\documentclass[]{siamonline1116}

\usepackage{lipsum}
\usepackage{amsfonts}
\usepackage{bbm}
\usepackage{graphicx}
\usepackage{epstopdf}
\usepackage{algorithmic}

\ifpdf
  \DeclareGraphicsExtensions{.eps,.pdf,.png,.jpg}
\else
  \DeclareGraphicsExtensions{.eps}
\fi

\numberwithin{theorem}{section}

\newcommand{\TheTitle}{Bayesian Updating and Uncertainty Quantification using Sequential Tempered MCMC with the Rank-One Modified Metropolis Algorithm}
\newcommand{\TheAuthors}{T. A. Catanach and J. L. Beck}

\headers{\TheTitle}{\TheAuthors}

\title{{\TheTitle}}

\author{
  Thomas A. Catanach\thanks{Extreme Scale Data Analytics, Sandia National Laboratories, Livermore, CA (\email{tacatan@sandia.gov}).}
  \and
  James L. Beck\thanks{Department of Computing and Mathematical Sciences, California Institute of Technology, Pasadena, CA \newline (\email{jimbeck@caltech.edu}).} 
}

\usepackage{amsopn}

\ifpdf
\hypersetup{
  pdftitle={\TheTitle},
  pdfauthor={\TheAuthors}
}
\fi

\begin{document}

\maketitle

\begin{abstract}
Bayesian methods are critical for quantifying the behaviors of systems. They capture our uncertainty about a system's behavior using probability distributions and update this understanding as new information becomes available. Probabilistic predictions that incorporate this uncertainty can then be made to evaluate system performance and make decisions. While Bayesian methods are very useful, they are often computationally intensive. This necessitates the development of more efficient algorithms. Here, we discuss a group of population Markov Chain Monte Carlo (MCMC) methods for Bayesian updating and system reliability assessment that we call Sequential Tempered MCMC (ST-MCMC) algorithms. These algorithms combine 1) a notion of tempering to gradually transform a population of samples from the prior to the posterior through a series of intermediate distributions, 2) importance resampling, and 3) MCMC. They are a form of Sequential Monte Carlo and include algorithms like Transitional Markov Chain Monte Carlo and Subset Simulation. We also introduce a new sampling algorithm called the Rank-One Modified Metropolis Algorithm (ROMMA), which builds upon the Modified Metropolis Algorithm used within Subset Simulation to improve performance in high dimensions. Finally, we formulate a single algorithm to solve combined Bayesian updating and reliability assessment problems to make posterior assessments of system reliability. The algorithms are then illustrated by performing prior and posterior reliability assessment of a water distribution system with unknown leaks and demands.

\end{abstract}

\section{Introduction}

Bayesian inference for system identification and rare event reliability analysis can both be formulated as Bayesian updating problems, which means that they can both be solved using the same algorithms \cite{Betz201872, CHIACHIO2014, VAKILZADEH20172, DIAZDELAO20171102}. In this work we consider Sequential Tempered Markov Chain Monte Carlo (ST-MCMC) algorithms for solving these updating problems. This family of algorithms, based on Sequential Monte Carlo, allows us to gradually transform the prior probability distribution describing the system's uncertainty to the updated posterior distribution that describes the system uncertainty conditioned on data or knowledge of a failure's occurrence. Previously, separate algorithms from this family have been used to solve these problems, such as TMCMC \cite{ching2007transitional} for posterior sampling in Bayesian system identification and Subset Simulation \cite{au2001estimation} for estimating prior probabilities of rare events. These algorithms share many commonalities and can be combined in the framework of ST-MCMC to enable full posterior probabilities of rare events to be estimated by a single algorithm, which has not been done previously in this framework. The unification of Bayesian updating and uncertainty quantification for posterior reliability problems has been considered before in \cite{beck2013prior, papadimitriou2001updating, beckau2002}, but has been significantly held back by inefficient MCMC methods that have made it generally computationally expensive. The development of a new MCMC sampler within ST-MCMC, the Rank One Modified Metropolis Algorithm presented in this work, allows for more efficient sampling of the posterior failure region than previous methods. Moreover, we find that the benefits of ST-MCMC and ROMMA are quite broad. Therefore, the contributions of this work are:

\begin{enumerate}
	\item Presenting a general framework for understanding Sequential Tempered MCMC algorithms like TMCMC and Subset Simulation
	\item Showing that this framework can be used to efficiently solve the posterior reliability problem while being robust to modeling uncertainty
	\item Introducing the Rank-One Modified Metropolis Algorithm to speed up sampling in ST-MCMC
\end{enumerate}

Typical MCMC methods rely on generating samples by sequentially evolving a Markov Chain which explores the posterior distribution and estimates expectations with respect to the posterior based upon the ergodicity of the chain. These single chain samplers are difficult to parallelize, tune, and adapt to complex posterior environments such as unidentifiable and locally identifiable models. This makes solving for the posterior failure probability difficult since the problem is often high dimensional and the posterior may be quite complex in shape \cite{beck1998updating1, beck1998updating2}.

Sequential Tempered MCMC methods are population-based methods that can more efficiently generate samples from complex posteriors since they evolve a population of chains which captures the global structure of the posterior, allowing for better adaptation. Examples of methods that fit in the framework of ST-MCMC include transitional/multilevel MCMC \cite{ching2007transitional, prudencio2012parallel}, AIMS \cite{beck2013asymptotically}, AlTar/CATMIP \cite{minson2013bayesian}, Subset Simulation \cite{au2001estimation}, and more generally, forms of Sequential Monte Carlo (SMC) \cite{kantas2014sequential, del2006sequential, jasra2011inference, chopin2002sequential}. These methods exploit parallelism by evolving a population of Markov chains simultaneously through a series of intermediate distribution levels until, as a population, they reach the final posterior. Since there are many chains sampling the final level, the mixing time of the Markov chain while sampling the ultimate posterior distribution is less relevant and so it can be much more efficient to use these methods. The intermediate levels also enable the algorithm to estimate the model evidence for solving model selection problems \cite{ching2007transitional, Neal2001} and rare event failure probabilities \cite{au2001estimation}.

We also introduce a better proposal method for the MCMC step in ST-MCMC to sample high dimensional distributions. Moving beyond standard Metropolis algorithms, like Random Walk Metropolis (RWM), which suffers from the curse of dimensionality, we adapt the Modified Metropolis Algorithm \cite{au2001estimation, zuev2011modified}, for use when the proposal distribution is a multivariate Gaussian with any covariance structure. This new algorithm, called the Rank-One Modified Metropolis Algorithm (ROMMA), performs a series of rank-one updates according to the prior distribution to form a proposal candidate. This means ROMMA can sample posterior distributions very effectively when they are significantly informed by the prior, which is common for many inference problems for physical systems. ROMMA avoids many of the high-dimensional scaling issues seen with RWM, particularly when the prior distribution has bounded support, since the proposed candidate is adapted to any prior constraint information.

The paper is organized as follows. First, we formulate the Bayesian updating and reliability problems in Section \ref{sec:bayes}. Then we discuss Sequential Tempered MCMC (ST-MCMC) algorithms and how they can be used to solve Bayesian inference and reliability problems in Section \ref{sec:stmcmc_alg}. Next, we introduce the ROMMA algorithm in Section \ref{sec:romma_alg}. Then we present experimental results of using ST-MCMC to solve a robust reliability assessment problem for estimating small failure probabilities for a water distribution system in Section \ref{sec:experiments}. Finally, we provide concluding remarks in Section \ref{sec:conclusions}. The appendix contains proofs, implementation details, and discussion about algorithm tuning.

\section{Bayesian Formulation}
\label{sec:bayes}

The Bayesian framework is a rigorous method for quantifying uncertainty using probability distributions. This philosophy is rooted in probability as a logic \cite{beck2010bayesian, cox1946probability, richard1961algebra, jaynes2003probability}. Within this framework, probability distributions are used to quantify uncertainty due to insufficient information, regardless of whether that information is believed to be unavailable (epistemic uncertainty) or is believed to not exist because of inherent randomness (aleatory uncertainty). This notion of uncertainty makes the Bayesian framework well suited for posing system identification problems, where postulated system models have uncertain, rather than random, mathematical structure and parameters. Therefore, we view system identification as updating a probability distribution that represents our beliefs about a system based on new information from system response data. This uncertainty description is then used to quantify the prediction of future system behaviors.

\subsection{Bayesian Inference}
\label{bayes_stat_sec}

Bayesian inference uses Bayes' theorem to update our uncertainty about a system using data, where uncertainty is quantified by assigning a probability function $p \left ( \right )$ to different system descriptions. The Bayesian inference formulation of parameter estimation and system identification is given in \cite{beck2010bayesian}: Given observation data $\mathcal{D}$ and assuming a system description in terms of a model class $\mathcal{M}$, find the posterior distribution $p \left (\theta \mid \mathcal{D} , \mathcal{M} \right )$ that represents the updated beliefs about a model parameter vector $\theta$ after incorporating the data. The data $\mathcal{D}$ is made up of measurements of the output $z_{1:n}$ and possibly measurements of the input $u_{1:n}$. The model class consists of (a) a set of predictive forward models, $p \left (z_{1:n} \mid u_{1:n}, \theta, \mathcal{M} \right )$, describing the plausibility of the predicted output $z_{1:n}$ for a parameter vector $\theta \in \Theta$, and (b) a prior distribution, $p \left (\theta \mid \mathcal{M} \right )$ over $\Theta$ representing our initial belief about the plausibility of different values of the model parameter vector $\theta$ and their corresponding forward models. To find the posterior distribution, Bayes' Theorem is used:

\begin{equation}
p \left (\theta \mid \mathcal{D} , \mathcal{M} \right ) = \frac{p \left (\mathcal{D} \mid \theta, \mathcal{M} \right )p \left (\theta \mid \mathcal{M} \right )}{p \left (\mathcal{D} \mid \mathcal{M} \right )}
\label{eq:bayes}
\end{equation}

\noindent The likelihood function, $p \left (\mathcal{D} \mid \theta, \mathcal{M} \right )$, denotes the probability of observing the data $\mathcal{D}$ according to the predictive model $p \left (z_{1:n} \mid u_{1:n}, \theta, \mathcal{M} \right )$ of the system with the measured input and output substituted into it. The normalizing factor in equation~\eqref{eq:bayes}, $p \left (\mathcal{D} \mid \mathcal{M} \right )$, is called the evidence (or marginal likelihood) for the model class $\mathcal{M}$ and is given by:

\begin{equation}
p \left (\mathcal{D} \mid \mathcal{M} \right ) = \int{p \left (\mathcal{D} \mid \theta, \mathcal{M} \right )p \left (\theta \mid \mathcal{M} \right ) d\theta}
\label{eq:bayesnorm}
\end{equation}

\noindent If multiple model classes exist to describe the uncertain behavior of the system, the evidence for each model class can be used to solve the corresponding model class selection problem \cite{beck2010bayesian}.

Typically, the model's prediction of the system's behavior has uncertain prediction accuracy, so we choose a probability model based on additive independent measurement and prediction errors, $\epsilon_i$, for the prediction of the measurement of the $i$th output $z_i = h_i \left ( \theta \right ) + \epsilon_i$. The stochastic predictive forward model then becomes

\begin{equation}
p \left (z_{1:n} \mid u_{1:n}, \theta, \mathcal{M} \right ) = \prod_{i=1}^n p \left ( \epsilon_i = z_i - h_i \left ( \theta \right ) \right )
\label{eq:indep_like}
\end{equation}

\noindent where $h_i \left ( \theta \right )$ is the model prediction equation for the $i$th output and $p \left ( \epsilon \right )$ is the PDF for the combined measurement and model prediction errors. With modern sensors, the measurement error will usually be quite small compared to the model prediction error. The deterministic prediction equation $h_i \left ( \theta \right )$ may be constructed from underlying physical principles while $p \left ( \epsilon \right )$ may be chosen based on a sensor model and Jaynes' principle of maximum information entropy \cite{beck2010bayesian, jaynes2003probability}.

We can broadly classify the posterior probability distributions from solving the inference problem for a model class into three types: globally identifiable, locally identifiable, and unidentifiable~\cite{beck1998updating1, beck1998updating2}. Globally identifiable model classes have a posterior distribution with a single peak around a unique maximum. Locally identifiable model classes have a posterior distribution with several separated peaks with approximately the same significance. Unidentifiable models have a manifold in the parameter space with approximately equally plausible parameter values. When the problem results in a locally identifiable or unidentifiable distribution, Bayesian methods are essential since they can capture the distribution unlike optimization-based methods. However, these problems are still challenging since it is often difficult to find and explore the peaks or the manifold of most plausible parameter values. This necessitates the development of better MCMC methods.

\subsection{Bayesian Uncertainty Quantification}

Taking a Bayesian perspective to studying complex systems enables scientists and engineers to quantitatively integrate all forms of uncertainty into their planning and decision making process using probability; for example, to make prior and posterior robust predictions about future system performance that take into account all available sources of uncertainty \cite{kennedyohaganUQ, chkrebtii2013bayesian, beck2013prior, najm2009uncertainty}. This is done by marginalizing over the collective modelling uncertainty and other sources of uncertainty.

Within this framework, we can assess the probability of a system in the future being in some failure domain $\mathcal{F}$ of unsatisfactory system performance that is defined as $\mathcal{F} = \left \{ \theta \in \mathbb{R}^n s.t. f\left ( \theta \right ) \geq 1 \right \}$ where $f\left ( \theta \right )$ is a performance failure function. For convenience, we now view the parameter vector $\theta$ as an expanded vector that captures both the modeling uncertainty in the system description and the uncertain future inputs. Therefore, the uncertainty for some of the components of $\theta$ will be described by the posterior in \eqref{eq:bayes} while for the remaining components it will be described by a prior distribution. When the system understanding only uses prior information about the model description of the system, $p \left ( \theta \mid \mathcal{M} \right )$, for a single model class $\mathcal{M}$, then we can define the prior failure probability \cite{beck2013prior, papadimitriou2001updating} as the expected value:

\begin{equation}
P \left ( \mathcal{F} \mid \mathcal{M} \right ) = \int \mathbbm{1} \left \{ \theta \in \mathcal{F} \right \} p \left ( \theta \mid \mathcal{M} \right ) d\theta
\label{eq:prior_fail}
\end{equation}

\noindent Once data $\mathcal{D}$ has been collected to better assess the model and state of the system, we can define the posterior robust failure probability \cite{beck2013prior, papadimitriou2001updating} as the expected value:

\begin{equation}
P \left ( \mathcal{F} \mid \mathcal{M}, \mathcal{D} \right ) = \int \mathbbm{1}\left \{ \theta \in \mathcal{F} \right \} p \left ( \theta \mid \mathcal{D}, \mathcal{M} \right ) d\theta
\label{eq:posterior_fail}
\end{equation}

\noindent where it is to be understood that some of the components of $\theta$ may not be informed by the data $\mathcal{D}$ and so the uncertainty in their values is controlled by the prior distribution. Using MCMC and importance sampling via ST-MCMC, we can provide estimates of the posterior failure probability. However, the MCMC method introduced in the Subset Simulation algorithm \cite{au2001estimation} for small prior failure probabilities (rare-event simulation) in \eqref{eq:prior_fail} requires modification to be efficient computationally in the posterior case \eqref{eq:posterior_fail} because the data induces significant correlations in $\theta$ that were not considered in the original algorithm.

\subsection{Solving Bayesian Updating Problems}

Bayesian inference and updating problems are usually analytically intractable \cite{beck2010bayesian, gelman2014bayesian} due to the required integration in \eqref{eq:bayesnorm}.  Common sampling methods for Bayesian inference are Markov Chain Monte Carlo (MCMC) methods~\cite{brooks2011handbook}, which do not require normalization of the posterior in \eqref{eq:bayes}. Generating samples through MCMC is computational intensive as often thousands to millions of model evaluations are needed to fully populate the high probability content of a complex posterior. While the central limit theorem implies that the estimate quality for the mean of a finite-variance stochastic variable scales independently of the dimension, given independent samples, MCMC methods produce correlated samples, which can introduce poor high dimensional scaling. As a result, solving updating problems using Bayesian methods is often prohibitively expensive for high dimensional or complex problems.

Recall that the basic idea of Monte Carlo estimation is to estimate expected values of $g\left (\theta \right )$ with respect to the posterior distribution by using a population of posterior samples $\theta_i$ for $i=1 \dots N$ as follows:

\begin{equation}
\mathbb{E} \left [ g\left (\theta \right ) \mid \mathcal{D} , \mathcal{M} \right ] = \int{g\left (\theta \right ) p \left (\theta \mid \mathcal{D}, \mathcal{M} \right ) d\theta} \approx \frac{1}{N} \sum_{i=1}^{N} g \left ( \theta_i \right )
\label{eq:bayesest}
\end{equation}

\noindent Assuming certain conditions hold, the quality of this estimate and its convergence can be assessed by the Markov chain central limit theorem \cite{geyer2011introduction}.

\begin{algorithm}
\caption{Metropolis-Hastings Algorithm}
\label{alg:mh_mcmc}
\begin{algorithmic}
\STATE{Define $Q \left ( \theta^\prime \mid \theta \right )$ as the proposal distribution for generating a candidate sample}
\STATE{Define $N_{steps}$ as the number of steps in the Markov chain}
\STATE{Initialize $\theta^0$}
\FOR{$i$ = 0 \TO $N_{steps}-1$}
\STATE{Draw the candidate $\theta_{i+1}^\prime \sim Q \left ( \theta_{i+1}^\prime \mid \theta_i\right )$}
\STATE{Compute the acceptance probability $\alpha \left (\theta_{i+1}^\prime \mid \theta_i \right )$ using equation \eqref{eq:aprob}}
\STATE{Draw $\eta \sim \mathcal{U} \left [0, 1 \right ]$}
\IF{$\eta < \alpha \left (\theta_{i+1}^\prime \mid \theta_i \right )$}
\STATE{Accept the candidate by setting $\theta_{i+1} = \theta_{i+1}^\prime$}
\ELSE
\STATE{Reject the candidate by setting $\theta_{i+1} = \theta_i$}
\ENDIF
\ENDFOR
\RETURN $\theta_0 \dots \theta_{N_{steps}}$
\end{algorithmic}
\end{algorithm}

\subsection{Basic Markov Chain Monte Carlo Algorithm}

The basis for many MCMC methods is the Metropolis-Hastings algorithm, which produces a Markov chain with a desired stationary distribution, $\pi \left ( \theta \right )$, by designing a transition kernel, $K \left ( \theta^\prime \mid \theta \right )$, such that the Markov chain is ergodic and reversible~\cite{geyer2011introduction, robert2011short}. Ergodicity ensures that the Markov chain has a unique stationary distribution, if it exists, while reversibility is a sufficient condition for the existence of a stationary distribution, $\pi \left ( \theta \right )$. Any proposal distribution $Q \left ( \theta^\prime \mid \theta \right )$ such that $Q \left ( \theta^\prime \mid \theta \right ) \neq 0$ for $Q \left ( \theta \mid \theta^\prime \right ) \neq 0$, can be used to construct such a $K \left ( \theta^\prime \mid \theta \right )$. Given the Markov chain is in state $\theta$, this is done by proposing a candidate sample $\theta^\prime$ according to $Q \left ( \theta^\prime \mid \theta \right )$ and then accepting the candidate $\theta^\prime$ with probability $\alpha$ given by:

\begin{equation}
\alpha \left (\theta^\prime \mid \theta \right ) = \min \left (1, \frac{\pi \left ( \theta^\prime \right )Q \left ( \theta \mid \theta^\prime \right )}{\pi \left ( \theta \right )Q \left ( \theta^\prime \mid \theta \right )} \right )
\label{eq:aprob}
\end{equation}

\noindent If the candidate is rejected, the initial sample $\theta$ is repeated. This leads to the Metropolis-Hastings (MH), Algorithm \ref{alg:mh_mcmc}. After settling into the stationary distribution, the resulting Markov chain produces samples from $\pi \left ( \theta \right )$ which are correlated.

The major challenge for the MH algorithm is designing the proposal distribution $Q$. A good proposal will cause the Markov chain to (1) converge quickly to the stationary distribution, that is, have a short burn-in time, and (2) have low correlation while sampling the stationary distribution. A common Metropolis-Hastings implementation is Random Walk Metropolis (RWM) in which $Q \left ( \theta^\prime \mid \theta \right )$ is a multivariate Gaussian distribution centered at $\theta$.

\subsection{Limitations of Traditional MCMC}

Generally, finding a proposal distribution that escapes the ``curse of dimensionality" is difficult because the high probability region of the posterior distribution lives on a low dimensional manifold in the high dimensional space. Therefore, it is important to find and sample this manifold efficiently. Even when starting on the manifold, randomly sampling the region around it without detailed knowledge of the structure will lead to a low acceptance rate of proposed samples in the Markov chain. Thus, if the proposal distribution is ill informed, very short steps are needed to ensure high acceptance rates. This leads to highly correlated samples. These types of posterior distributions are common in inverse problems for physical systems where the data is not sufficiently rich to detangle the complex relationships produced by the system.

Further, even for simple distributions without complicated geometry, RWM requires many model evaluations to produce a satisfactory sample population. Practitioners often run chains for hundreds of thousands or millions of iterations to ensure they have enough uncorrelated samples. For simple problems, the length of the chain required to decorrelate the samples scales linearly with the dimension of the problem \cite{roberts2001optimal}. Avoiding slow mixing by developing more efficient samplers is therefore critical for solving inference problems involving systems where evaluating the forward model is computationally intensive.

Finally, the standard MH algorithm is constrained by the fact that it forms a Markov chain. This means it requires sequential evaluation and it has a limited ability to adapt to its state without jeopardizing its reversibility and ergodicity. The sequential updating of the chain makes MH MCMC unsuitable for high performance computing because it cannot exploit parallelism. For efficient solution of computationally intensive inverse problems, algorithms are sought that exploit parallelism and are adaptive based upon global information.

\section{Sequential Tempered MCMC Algorithm}
\label{sec:stmcmc_alg}

\begin{figure}[t]
\centerline{\includegraphics[width=1.0\textwidth]{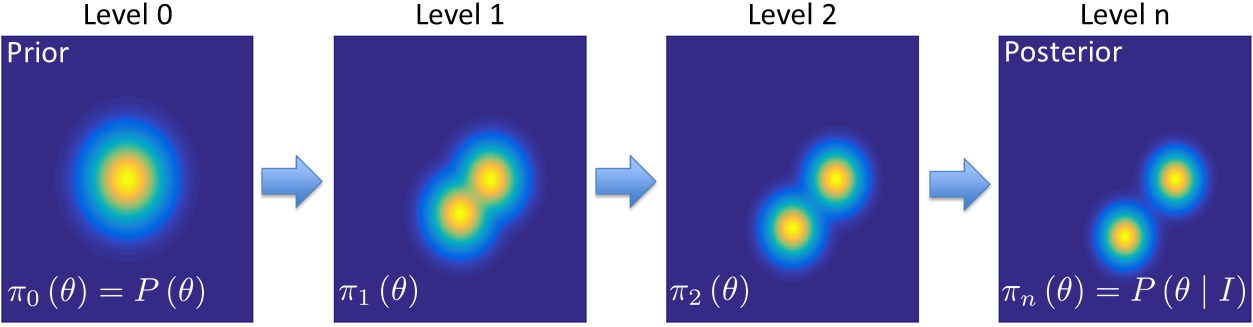}}
\caption{Illustration of a set of intermediate distributions which gradually transform a unimodal prior $P \left ( \theta \right )$ into a bimodal posterior $P \left ( \theta \mid I \right )$.}
\label{fig:anneal_ped}
\end{figure}

ST-MCMC methods provide many advantages over single chain methods because of the information and adaptation they gain through the population of samples and by tempering. The population aspect enables parallelism and the ability to capture and learn from the global structure of the posterior. The tempering enables the samples to gradually transition from the prior to the posterior. This means they better adapt to handle complicated distributions, like multi-modal distributions, without getting stuck sampling in one area when the chains have difficulty exploring the posterior. These methods have been shown to be very effective in many problems \cite{ching2007transitional, minson2013bayesian, au2001estimation, kantas2014sequential}. Thinking about all these algorithms within a single framework is helpful to implement and tune them, particularly for posterior robust uncertainty quantification.

ST-MCMC methods can be divided into three basic parts: tempering/annealing, importance resampling, and MCMC. The annealing step introduces intermediate distributions that gradually evolve the samples from the prior to the posterior, as illustrated in Figure \ref{fig:anneal_ped}. The importance resampling step discards relatively implausible samples and multiplies highly plausible samples to maintain and rebalance the samples with respect to the changes in the intermediate distributions from one level to the next. Then the MCMC step allows the samples to be perturbed in order to explore the intermediate distributions and to adjust to changes as the distributions evolve. This family of algorithms can be interpreted as a form of Sequential Monte Carlo as found in \cite{del2006sequential}, which provides the theoretical foundation of these algorithms.

\begin{algorithm}
\caption{General Sequential Tempered MCMC}
\label{alg:stmcmc}
\begin{algorithmic}
\STATE{Define prior distribution $p \left ( \theta \right )$}
\STATE{Define posterior distribution $p \left ( \theta \mid I \right )$} for data or information $I$
\STATE{Define $N$ as the number of samples in the population}
\STATE{Initialize the first intermediate distribution $\pi_0 \left ( \theta \right ) = p \left ( \theta \right )$}
\STATE{Draw the sample population $\theta_i^{\left ( 0 \right )} $ for $i = 1...N$ from $\pi_0 \left ( \theta \right )$}
\STATE{Set the level counter $k = 0$}
\WHILE{$\pi_k \left ( \theta \right ) \neq p \left ( \theta \mid I \right )$ }
\STATE{1) Increment the level counter $k = k+1$}
\STATE{2) Choose the next intermediate distribution $\pi_{k} \left ( \theta \right )$ based upon the sample population}
\STATE{3) Compute the importance weights for the population $\theta^{\left ( k-1 \right )}$ as $w \left (\theta_i^{\left ( k-1 \right )} \right ) = \frac{\pi_{k} \left ( \theta_i^{\left ( k-1 \right )} \right )}{\pi_{k-1} \left ( \theta_i^{\left ( k-1 \right )} \right )}$}
\STATE{4) Resample the population according to the normalized importance weights to find the initial level $k$ population $_0\theta^{\left ( k \right )}$}
\STATE{5) Adapt the MCMC proposal $Q_{k} \left (\hat \theta \mid \theta \right )$ using population statistics}
\STATE{6) Evolve samples $_0\theta^{\left ( k \right )}$ according to MCMC with respect to $\pi_k \left ( \theta \right )$ and proposal $Q_k \left ( \hat{\theta} \mid \theta \right )$ to get the final population $\theta^{\left ( k \right )}$}
\ENDWHILE
\RETURN $\theta_i^{\left ( k \right )}$ for $i = 1...N$
\end{algorithmic}
\end{algorithm}

A fully general ST-MCMC algorithm is described in Algorithm \ref{alg:stmcmc}. The choice for resampling, adaptation, and MCMC may vary. This algorithm begins with samples from the prior distribution $p \left ( \theta \right )$ and then evolves them to generate samples from the posterior distribution $p \left ( \theta \mid I \right )$, which is the distribution conditioned on additional data or information $I$. For each of the intermediate distributions $\pi_k \left ( \theta \right )$, the  previous level's population samples are weighted based upon their relative likelihood according to the initial and new distribution, $\frac{\pi_k \left ( \theta \right )}{\pi_{k-1} \left ( \theta \right )}$ and then resampled. Then each sample initiates a Markov chain which explores the intermediate distribution $\pi_k \left ( \theta \right )$ according to the MCMC process defined by the proposal distribution $Q_k \left ( \hat{\theta} \mid \theta \right )$. The next level proposal $Q_{k+1}$ and next intermediate distribution $\pi_{k+1}$ are typically adapted based upon information from the current sample population, with $\pi_{k+1}$ growing closer to the posterior as $k$ increases. Once the chains are sufficiently decorrelated from their initial seeds, the samples at the end of the Markov chains serve as the initial samples to start the weighting and resampling process for the next level. This process repeats until the samples are distributed according to the posterior distribution $p \left ( \theta \mid I \right )$, meaning that $\pi_{k}$ is $p \left ( \theta \mid I \right )$ for the final $k$. For a discussion for the general tuning and parameterizations of this algorithm, see the appendix \ref{sec:supp_tuning}.

\subsection{ST-MCMC for Bayesian Inference}

An approach to using ST-MCMC for Bayesian inference is found in the TMCMC \cite{ching2007transitional} and CATMIP \cite{minson2013bayesian} algorithms or in the context of an SMC algorithm as in \cite{kantas2014sequential}. A general implementation of these ideas is found in Algorithm \ref{alg:stmcmc_bayes}. When initializing the algorithm, the initial sample population $\theta^{\left ( 0 \right )}$ is drawn from the prior distribution. The number of samples in the population is typically fixed at all levels to be $N$. Parameters used for the algorithm are then initialized, such as the annealing factor $\beta$, level counter $k$, and parameters that define the proposal distribution $Q \left ( \hat{\theta} \mid \theta \right )$.

\begin{figure}[t]
\centerline{\includegraphics[width=1.0\textwidth]{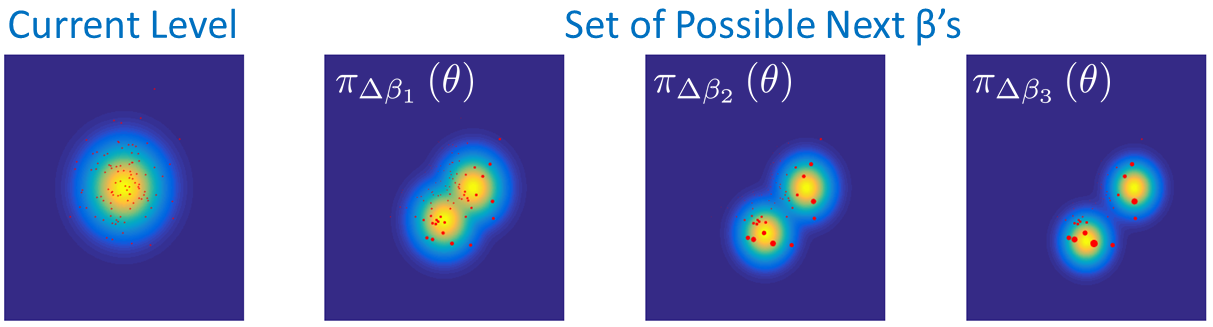}}
\caption{Illustration of finding $\Delta \beta$ that defines how much additional influence the data has in the next intermediate distribution level. Red dots indicate the samples and their size indicates their weight for three $\Delta \beta$ where $\Delta \beta_1 < \Delta \beta_2 < \Delta \beta_3$. If too large a $\Delta \beta$ step is made (e.g. $\Delta \beta_3$), only a few samples will have the majority of the weights, indicating that the samples poorly represent the distribution. If too small a $\Delta \beta$ step is made (e.g. $\Delta \beta_1$), the next distribution is too close to the current distribution making it an inefficient choice.}
\label{fig:beta_ped}
\end{figure}

At the beginning of each subsequent level $k$, the annealing factor $\beta_k$ is computed. The annealing factor controls the influence of the data at every level by gradually transitioning the level stationary distributions from the prior at $\beta_0 = 0$ to the posterior at $\beta_{final} = 1$. The increment $\Delta \beta$ at each level is chosen to ensure that the intermediate distributions are not too far apart, otherwise the sample population $\theta^{\left ( k-1 \right )}$ does a poor job representing the next level distribution $\pi_k \left ( \theta \right )$. This increment is controlled by looking at the degeneracy of the sample importance weights, which are weights that allow us to transform samples $\theta^{\left (k-1 \right )}$ from making estimates with respect to $\pi_{k-1} \left ( \theta \right )$ to $\pi_{k} \left ( \theta \right )$. This process is illustrated in Figure \ref{fig:beta_ped}. This weight function takes the form of $ w\left (\theta,  \Delta \beta \right ) = \frac{p \left (\mathcal{D} \mid \theta \right )^{\beta+\Delta \beta} \pi_0 \left ( \theta \right )}{p \left (\mathcal{D} \mid \theta \right )^{\beta} \pi_0 \left ( \theta \right )} = p \left (\mathcal{D} \mid \theta \right )^{\Delta \beta}$. The degeneracy in the weights is measured by computing their coefficient of variation $\text{COV} \left [ w\left (\theta^{\left (k-1 \right )},  \Delta \beta \right ) \right ]$ and trying to find a $\Delta \beta$ such that it is equal to some target threshold $\kappa^*$. We use the sample coefficient of variation and so we must solve for $\Delta \beta$ in

\begin{equation}
COV \left [ w\left (\theta^{\left ( k-1 \right )},  \Delta \beta \right ) \right ] = \frac{\sqrt{\frac{1}{N} \sum_{i=1}^{N} \left ( w\left (\theta_i^{\left ( k-1 \right )},  \Delta \beta \right ) - \frac{1}{N} \sum_{i=1}^{N} w\left (\theta_i^{\left ( k-1 \right )},  \Delta \beta \right )\right )^2}}{\frac{1}{N} \sum_{i=1}^{N} w\left (\theta_i^{\left ( k-1 \right )},  \Delta \beta \right )} = \kappa^*
\label{eq:cov_cal}
\end{equation}

\noindent This equation is typically solved using a bisector method since we have an upper and lower bound for $\Delta \beta$. Based upon the theory of importance sampling, the coefficient of variation is an estimate of the effective sample size (ESS), making it a good proxy for degeneracy \cite{mcbook}. Therefore, we set $\Delta \beta$ to have a certain target ESS in the sample population so that the population represents the intermediate level distribution well. The ESS is the equivalent number of independent samples that will provide the same variance estimate of a quantity of interest. For a more detailed discussion of the change of ESS during ST-MCMC, see \ref{sec:supp_tuning}.

\begin{algorithm}
\caption{ST-MCMC for Bayesian Updating}
\label{alg:stmcmc_bayes}
\begin{algorithmic}
\STATE{Define prior distribution $p \left ( \theta \right )$}
\STATE{Define posterior distribution $p \left ( \theta \mid \mathcal{D} \right )$} for system data $\mathcal{D}$
\STATE{Define $N$ as the number of samples in the population}
\STATE{Define $\kappa^*$ as the target COV (coefficient of variation) of the importance weights}
\STATE{Initialize the first intermediate distribution $\pi_0 \left ( \theta \right ) = p \left ( \theta \right )$}
\STATE{Draw the sample population $\theta_i^{\left ( 0 \right )} $ for $i = 1...N$ from $\pi_0 \left ( \theta \right )$}
\STATE{Set the level counter $k = 0$}
\STATE{Set the annealing parameter $\beta_0 = 0$}
\WHILE{$\beta_k < 1$ }
\STATE{1) Increment the level counter $k = k+1$}
\STATE{2) Define importance weights $ w\left (\theta_i^{\left ( k - 1 \right )},  \Delta \beta \right ) = p \left (\mathcal{D} \mid \theta_i^{\left ( k - 1 \right )} \right )^{\Delta \beta}$, for $i = 1...N$}
\STATE{3) Solve $ \text{COV} \left [ w\left (\theta^{\left ( k - 1 \right )},  \Delta \beta \right ) \right ] = \kappa^*$ for $\Delta \beta$. If $\beta_{k-1} + \Delta \beta > 1$ set $\Delta \beta = 1- \beta_{k-1}$}
\STATE{4) Find $k$\textsuperscript{th} level PDF: $\pi_k \left ( \theta \right ) = p\left (\mathcal{D} \mid \theta \right )^{\beta_k} p \left ( \theta \right )$} where $\beta_k = \beta_{k-1} + \Delta \beta$
\STATE{5) Set the importance weights for the population $\theta^{\left ( k-1 \right )}$ as $w\left (\theta_i^{\left ( k - 1 \right )},  \Delta \beta \right )$}
\STATE{6) Resample the population according to the normalized importance weights to find $N$ samples for the initial level $k$ population $_0\theta^{\left ( k \right )}$ based upon multinomial resampling}
\STATE{7) Adapt the MCMC proposal $Q_{k} \left (\hat \theta \mid \theta \right )$ using population statistics}
\STATE{8) For each sample in $_0\theta^{\left ( k \right )}$ evolve a Markov chain using MCMC with respect to $\pi_k \left ( \theta \right )$ and proposal $Q_k \left ( \hat{\theta} \mid \theta \right )$. The end of the chains return the final population $\theta^{\left ( k \right )}$. The chain length may be fixed or determined based upon chain correlation statistics.}
\ENDWHILE
\RETURN $\theta_i^{\left ( k \right )}$ for $i = 1...N$
\end{algorithmic}
\end{algorithm}

Once $\Delta \beta$ and $\beta_k = \beta_{k-1} + \Delta \beta$ are found, the sample population $\theta^{\left ( k-1 \right )}$ can be used to make expectation estimates with respect to the $k$\textsuperscript{th} level PDF $\pi_k \left ( \theta \right )$ using the normalized importance weights $\hat{w}^k$:

\begin{equation}
\hat{w}_i^k = \frac{w\left (\theta_i^{\left ( k-1 \right )},  \Delta \beta \right )}{\sum_{j=1}^N w\left (\theta_j^{\left ( k-1 \right )},  \Delta \beta \right )}
\label{eq:weights}
\end{equation}

The algorithm is now ready to produce the next level sample population, $_0\theta^{\left ( k \right )}$. Importance resampling produces an initial sample population that is asymptotically distributed according to $\pi_k \left ( \theta \right )$ for increasing population size. Multinomial resampling is often used, where each $_0\theta_i^{\left ( k \right )}$ is randomly picked from the samples $\theta^{\left ( k-1 \right )}$. The probability of choosing $\theta_j^{\left ( k-1 \right )}$ is $\hat{w}_j^k$. Because the new samples are a subset of the previous level population, there is added degeneracy. To add diversity, MCMC is used to evolve the sample population $_0\theta^{\left ( k \right )}$ according to $\pi_k \left ( \theta \right )$.

The MCMC step for ST-MCMC is defined by the type of proposal distribution used and the length of the chain. Both of these factors can significantly influence the performance of the algorithm as the chains must be allowed to evolve sufficiently to ensure that they explore the distribution $\pi_k \left ( \theta \right )$ and decorrelate from each other. Typically, within Metropolis-Hastings MCMC, a Gaussian proposal is used where the candidate $\hat{\theta} = \theta + \eta$, $\eta \sim \mathcal{N} \left ( 0, \sigma^2 \Sigma \right)$ and $\Sigma$ is the sample covariance matrix computed using the weighted samples while $\sigma^2$ is the scaling factor, which is adapted using the acceptance rate $\alpha_{k-1}$ of the sampler at the previous MCMC level. We use a feedback controller-based method to tune the scale factor and a theoretical method for choosing the chain length, both of which are discussed in the appendix \ref{sec:supp_tuning_acc}. Other methods besides RWM can be used as discussed in Section \ref{sec:romma_alg}.

The final sample population at level $k$ is $\theta^{\left ( k \right )}$ at the end of these chains. The algorithm then iterates until $\beta_k = 1$ in which case the final samples are from the posterior distribution.

\subsection{ST-MCMC for Estimating Prior Failure Probabilities}
\label{sec:stmcmc_ss}
Sequential Tempered MCMC can also be used to estimate small failure probabilities for reliability analysis as described in Algorithm \ref{alg:stmcmc_ss}. This algorithm as implemented is essentially a formulation of Subset Simulation \cite{au2001estimation}. Here, the intermediate distributions are intermediate failure domains for increasing threshold levels of a failure function $f \left ( \theta \right )$, where failure is defined when $f \left ( \theta \right ) \geq 1$. So by defining the intermediate failure region by $f \left ( \theta \right ) \geq \beta$ for some $\beta < 1$, this region contains the full failure region plus additional points. When $\beta = -\infty$ the distribution over the parameters is the prior distribution $p \left ( \theta \right )$ since no failure information has been incorporated. When $\beta = 1$, the distribution is now the posterior distribution $p \left ( \theta \mid \mathcal{F} \right )$ since it is the distribution conditioned on the failure event $\mathcal{F}$. By moving $\beta$ from $-\infty$ to $1$, we anneal down to the final failure region through a set of nested intermediate failure domains. With these prior, posterior, and intermediate distributions, we use ST-MCMC to find, sample, and estimate the probability contained in the failure region.

\begin{algorithm}
\caption{ST-MCMC for Estimating Failure Probabilities via Modified Subset Simulation}
\label{alg:stmcmc_ss}
\begin{algorithmic}
\STATE{Define prior distribution $p \left ( \theta \right )$}
\STATE{Define the failure function $f \left ( \theta \right )$ s.t. failure occurs when $f \left ( \theta \right ) \geq 1$}
\STATE{Define failure distribution $p \left ( \theta \mid \mathcal{F} \right ) = \frac{\mathbbm{1} \left \{ f \left ( \theta \right ) \geq 1 \right \} p \left ( \theta \right )}{P \left ( \mathcal{F} \right )}$}
\STATE{Define $N$ as the number of samples in the population}
\STATE{Define $\kappa$ as the sampling fraction}
\STATE{Initialize the first intermediate distribution $\pi_0 \left ( \theta \right ) = p \left ( \theta \right )$}
\STATE{Draw the sample population $\theta_i^{\left ( 0 \right )} $ for $i = 1...N$ from $\pi_0 \left ( \theta \right )$}
\STATE{Set the level counter $k = 0$}
\STATE{Set the failure parameter $\beta_0 = - \infty$}
\WHILE{$\beta_k < 1$ }
\STATE{1) Increment the level counter $k = k+1$}
\STATE{2) Solve $ \frac{1}{N} \sum_{i=1}^N \mathbbm{1} \left \{ f \left ( \theta_i^{k-1} \right ) > \beta_{k}\right \} = \kappa$ for $\beta_k$. If $\beta_{k} > 1$ set $\beta_{k} = 1$}
\STATE{3) Find $k$\textsuperscript{th} level PDF: $\pi_k \left ( \theta \right ) = \frac{\mathbbm{1} \left \{ f \left ( \theta \right ) \geq \beta_k \right \} p \left ( \theta \right )}{P \left ( \mathcal{F}_\beta \right )}$}
\STATE{4) Since the importance weights for the population $\theta^{\left ( k-1 \right )}$, $w\left (\theta_i^{\left ( k - 1 \right )}, \beta_k \right ) = \mathbbm{1} \left \{ f \left ( \theta_i^{k-1} \right ) \geq  \beta_{k} \right \} = 0$ or $1$, resample the population of $\kappa N$ failure samples $N$ times to find the initial level $k$ population $_0\theta^{\left ( k \right )}$}
\STATE{5) Adapt the MCMC proposal $Q_{k} \left (\hat \theta \mid \theta \right )$ using population statistics}
\STATE{6) For each sample in $_0\theta^{\left ( k \right )}$ evolve a Markov chain using MCMC with respect to $\pi_k \left ( \theta \right )$ and proposal $Q_k \left ( \hat{\theta} \mid \theta \right )$. The end of the chains return the final population $\theta^{\left ( k \right )}$. The chain length may be fixed or determined based upon chain correlation statistics.}
\ENDWHILE
\STATE{Estimate the failure probability $\hat p_f = \kappa^{k-1} \frac{1}{N}\sum_{i=1}^N \mathbbm{1} \left \{ f \left ( \theta_i^{k-1} \right ) \geq 1 \right \}$}
\RETURN $\hat p_f$ and $\theta_i^{\left ( k \right )}$ for $i = 1...N$
\end{algorithmic}
\end{algorithm}

The ST-MCMC method used to estimate small failure probabilities is simpler than that used for Bayesian updating since $p \left ( \theta \mid \mathcal{F}_{\beta} \right ) = \mathbbm{1}\left \{ \theta \in \mathcal{F}_{\beta} \right \}p \left ( \theta \right )/P \left ( \mathcal{F} \right )$. This means that the weights are either $1$ or $0$ depending if the sample is inside or outside, respectively, the failure region defined by $\beta$. Therefore, multinomial resampling is not needed. The next $\beta$ is determined in Step 2 of Algorithm \ref{alg:stmcmc_ss} by finding a $\beta_k$ such that a specified fraction, $\kappa$, of the samples satisfy $f \left ( \theta_i^{k-1} \right ) > \beta_k$. The fraction $\kappa$ controls the ESS of the sample population i.e. if $\kappa = \frac{1}{2}$ then the ESS is $\frac{N}{2}$. These samples are then replicated to regenerate the initial size $N$ of the sample population. The samples are then decorrelated and explore the intermediate failure domain distribution using MCMC.

It is important to note that Algorithm \ref{alg:stmcmc_ss} differs from the Subset Simulation algorithm presented in \cite{au2001estimation} in one aspect. In place of Step 4, the original Subset Simulation keeps the fraction $\kappa$ of the samples that satisfy $f \left (\theta_i^{k-1} \right ) > \beta_k$ and then evolves each of the $\kappa N$ chains $\frac{1}{\kappa}$ steps in Step 6 to recover a total of $N$ samples from the $k$th level failure PDF $\pi_k \left ( \theta \right )$. In contrast, we replicate the $\kappa N$ samples $\frac{1}{\kappa}$ times then evolve the resulting $N$ chains in parallel for a number of steps that is determined by how well the chain is mixing (Step 6). While this requires more function evaluations, it is very parallelizable and also more robust when long chains are needed to ensure decorrelation and better exploration of $\pi_k \left ( \theta \right )$.

\subsection{ST-MCMC for Estimating Posterior Failure Probabilities}
To estimate and sample the posterior failure region $p \left ( \theta \mid \mathcal{F}, \mathcal{D} \right )$, we can combine Algorithms \ref{alg:stmcmc_bayes} and \ref{alg:stmcmc_ss}. One approach is to use Algorithm \ref{alg:stmcmc_bayes} to transform samples from $p \left ( \theta \right )$ to samples from $p \left ( \theta \mid \mathcal{D} \right )$ and then use these samples to seed Algorithm \ref{alg:stmcmc_ss}, which then transforms samples from $p \left ( \theta \mid \mathcal{D} \right )$ to samples from $p \left ( \theta \mid \mathcal{F}, \mathcal{D} \right )$, yielding posterior failure samples. In this case, $p \left ( \theta \mid \mathcal{D} \right )$ is essentially the prior distribution for the failure assessment and the intermediate distributions are $\mathbbm{1}\left \{ \theta \in \mathcal{F}_{\beta} \right \}p \left ( \theta \mid \mathcal{D} \right )/P \left ( \mathcal{F}_{\beta} \right )$ where $\mathcal{F}_{\beta}$ is the intermediate failure domain. Therefore, it is important for the MCMC method to be able to efficiently explore $p \left ( \theta \mid \mathcal{D} \right )$, as we see in the experiments in Section \ref{sec:experiments} when we solve a posterior reliability analysis using this approach. Otherwise, Algorithm \ref{alg:stmcmc_ss} proceeds as described in Section \ref{sec:stmcmc_ss}.

Alternatively, different paths to transform samples from $p \left ( \theta \right )$ to $p \left ( \theta \mid \mathcal{F}, \mathcal{D} \right )$ could be used. The most efficient path will depend on the computational cost of computing the likelihood and failure functions, along with the level of difficulty of sampling the domains. This question of developing a strategy for finding the optimal path to minimize effort is left to future study.

\subsection{ST-MCMC for Integration}

One of the original motivations for the development of Sequential Tempered MCMC methods like TMCMC \cite{ching2007transitional} was to solve Bayesian model selection problems, for which population MCMC algorithms do well \cite{CALDERHEAD20094028}. Such problems can be very difficult to solve because they require computing the model evidence, that is, the likelihood of the data given the model class, $p \left ( \mathcal{D} \mid \mathcal{M} \right )$, which is given by the integral in~\eqref{eq:bayesnorm}. Similarly, when ST-MCMC is used for computing failure probabilities \cite{au2001estimation}, we must compute $P \left ( \mathcal{F} \mid \mathcal{M} \right )$, which can be expressed as integral~\eqref{eq:prior_fail}. The similarity of the model selection and failure probability estimation problem indicate that the same methods can be used to solve both problems. In this section, we describe such a method for computing failure probabilities using ST-MCMC. We discuss computing model evidences in the appendix to illustrate the similarities between the two methods, see \ref{sec:model_select}.

\subsubsection{Estimating Rare Event Failure Probability}
Computing the failure probability for a rare events is dicussed in \cite{au2001estimation}. In this case, $P \left ( \mathcal{F} \mid \mathcal{M} \right )$ is

\begin{equation}
P \left ( \mathcal{F} \mid \mathcal{M} \right ) = \int P \left ( \mathcal{F} \mid \theta, \mathcal{M}  \right ) p \left ( \theta \mid \mathcal{M} \right ) d\theta = \int \mathbbm{1}\left \{ \theta \in \mathcal{F} \right \} p \left ( \theta \mid \mathcal{M} \right ) d\theta
\label{eq:model_failure}
\end{equation}

\noindent This integral could be naively estimated using Monte Carlo sampling of the prior distribution $p \left ( \theta \mid \mathcal{M} \right )$. This estimate would be computationally inefficient when the probability of failure is small, since the probability of randomly generating a sample in the failure region is low. However, the intermediate levels of ST-MCMC solve this problem by decomposing the computation over the intermediate failure domains. The integral is then expressed as the product of Monte Carlo estimates of $s$ intermediate conditional failure probabilities:

\begin{equation}
\int \mathbbm{1}\left \{ \theta \in \mathcal{F} \right \} p \left ( \theta \mid \mathcal{M} \right ) d\theta = \prod_{k=1}^{s} \frac{\int \mathbbm{1}\left \{ \theta \in \mathcal{F}_{\beta_k} \right \} p \left ( \theta \mid \mathcal{M} \right ) d\theta}{\int \mathbbm{1}\left \{ \theta \in \mathcal{F}_{\beta_{k-1}} \right \} p \left ( \theta \mid \mathcal{M} \right ) d\theta} = \prod_{k=1}^{s} c_k
\label{eq:model_failure_prod}
\end{equation}

For each intermediate level, we can perform a fairly accurate Monte Carlo estimate between the previous level and the current level since these distributions are designed to be relatively close to each other in terms of the relative ESS (effective sample size) of samples coming from the previous level. Having a high ESS means Monte Carlo sampling will be effective. Noting that $\mathcal{F}_{\beta_k} \subset \mathcal{F}_{\beta_{k-1}}$, the Monte Carlo estimate for $c_k$ in~\eqref{eq:model_failure_prod} takes the form

\begin{equation}
\begin{split}
c_k &=  \int \mathbbm{1}\left \{ \theta \in \mathcal{F}_{\beta_k} \right \} \frac{\mathbbm{1}\left \{ \theta \in \mathcal{F}_{\beta_{k-1}} \right \} p \left ( \theta \mid \mathcal{M} \right )}{\int \mathbbm{1}\left \{ \theta' \in \mathcal{F}_{\beta_{k-1}} \right \} p \left ( \theta' \mid \mathcal{M} \right ) d\theta'} d\theta \\
&=  \int \mathbbm{1}\left \{ \theta \in \mathcal{F}_{\beta_k} \right \} p \left ( \theta \mid \mathcal{F}_{\beta_{k-1}}, \mathcal{M} \right )d\theta \\
&\approx \frac{1}{N} \sum_{i=1}^N \mathbbm{1}\left \{ \theta_i^{\left ( k - 1\right )} \in \mathcal{F}_{\beta_k} \right \}
\end{split}
\label{eq:model_failure_mc_est}
\end{equation}

\noindent where $\theta_i^{\left ( k - 1\right )} \sim p \left ( \theta \mid \mathcal{F}_{\beta_{k-1}}, \mathcal{M} \right )$.  Then, if the fraction of intermediate failure samples at each level of the algorithm is set to $\kappa$, the total estimate of the failure probability becomes

\begin{equation}
P \left ( \mathcal{F} \mid \mathcal{M} \right ) \approx \kappa^{\left ( s-1 \right )} \left ( \frac{1}{N} \sum_{i=1}^N \mathbbm{1}\left \{ \theta_i^{\left ( s - 1\right )} \in \mathcal{F} \right \} \right )
\label{eq:model_failure_final}
\end{equation}

By breaking up the estimation of the failure probability into a number of levels, the number of model evaluations needed to check whether a system configuration leads to failure scales sub-linearly with the inverse failure probability $p^{-1}$. Normal Monte Carlo requires $O \left ( p^{-1} \right )$ samples to estimate the failure probability while ST-MCMC algorithms require $O \left ( \log{p^{-1}}  \right )$ since it requires approximately $\log_{\kappa^{-1}} p^{-1}$ levels to reach the failure domain.

\section{Rank-One Modified Metropolis Algorithm}
\label{sec:romma_alg}

The most computationally intensive step of ST-MCMC is evolving the population of Markov chains during the MCMC step. This is particulary significant for high dimensional inference problems where exploring the posterior distribution is challenging. For certain types of problems where the posterior distribution is informed by the prior, integrating prior information in the proposal can reduce the computational cost. This avoids the wasted computational effort of computing the likelihood function when the candidate was rejected mostly due to the influence of prior information and not the likelihood. The Modified Metropolis Algorithm (MMA), developed in \cite{au2001estimation}, does this under the assumption that the proposal distribution is a Gaussian with independent variables and that the prior has independent components. The Rank-One Modified Metropolis Algorithm (ROMMA) presented in this work generalizes the MMA algorithm to any prior distribution and any Gaussian proposal distribution.

In general, ROMMA provides speed ups when the distribution $p \left ( \theta \mid I \right )$ conditioned on information or data $I$ is still significantly informed by the prior distribution $p \left ( \theta \right )$. There are two common situations that lead to these types of posteriors. The first case is where the prior enforces a constraint such as a prior inequality constraint on model parameter. The second case is inference problems where the data is only rich enough to inform the parameter distribution along certain directions, also known as active subspaces \cite{constantine2015active}. This typically occurs for unidentifiable model classes which have a posterior distribution with a manifold of approximately equally plausible parameter values \cite{EQE:EQE122}. Both of these situations are common in posterior reliability problems for complex models of physical systems.

\subsection{Modified Metropolis Algorithm}
The Modified Metropolis Algorithm was developed in \cite{au2001estimation} to overcome the curse of dimensionality that comes from sampling high dimensional posteriors when estimating small failure probabilities. This algorithm originally assumed that the posterior is the product of the prior and an indicator function but it can be expanded to the more general Bayesian inference setting as in Algorithm \ref{alg:og_mm_mcmc}. A proof of its reversibility is given in \ref{sec:proof_mma}. The algorithm assumes that the prior distribution $p \left (\theta \right)$ has independent components such that $p \left (\theta \right) = \prod_{j=1}^{N_d} p \left (\theta_j \right)$, which is a common assumption for many prior reliability problems. In order to evolve the Markov chain that samples the posterior $p \left (\theta \mid \mathcal{D} \right)$, the authors break it up into a two-step proposal. The first step of the proposal deals with the prior and can be done component-wise. The second step deals with the likelihood $p \left (\mathcal{D} \mid \theta \right)$, which is done by evaluating the forward model. By separating out the prior and evolving it component-wise, this algorithms avoids the poor dimensional scaling introduced by the prior. This is particularly important for priors with bounded support because they often have a significant impact on the posterior. However, the independence assumptions for the prior and proposal distributions pose a significant drawback for applying MMA to general Bayesian updating problems where there is significant correlation induced by the data.

\begin{algorithm}
\caption{Modified Metropolis Algorithm}
\label{alg:og_mm_mcmc}
\begin{algorithmic}
\STATE{Define $D$  as a diagonal positive definite proposal covariance matrix}
\STATE{Define $N_d$ as the number of components of the vector $\theta$}
\STATE{Define $N_{steps}$ as the number of steps in the Markov chain}
\STATE{Initialize $\theta^0$}
\FOR{$i$ = 0 \TO $N_{steps}-1$}
\STATE{Draw $\xi \sim \mathcal{N} \left (0, I_{N_d} \right )$}
\FOR{$j$ = 1 \TO $N_{d}$}
\STATE{Compute the component update $\hat{\theta}_j = \theta_j^i + D_j \xi_j$}
\STATE{Draw $\zeta \sim \mathcal{U} \left [0, 1 \right ]$}
\IF{$\zeta > \frac{p \left (\hat{\theta}_j \right )}{p \left (\theta_j^i \right )}$}
\STATE{Reject the update by setting $\xi_j  = 0$}
\ENDIF
\ENDFOR
\STATE{Compute the candidate $\hat{\theta} = \theta^i + D \xi$}
\STATE{Draw $\eta \sim \mathcal{U} \left [0, 1 \right ]$}
\IF{$\eta < \frac{p \left (\mathcal{D} \mid \hat{\theta} \right )}{p \left (\mathcal{D} \mid \theta^i \right )}$}
\STATE{Accept the candidate $\theta^{i+1} = \hat{\theta}$}
\ELSE
\STATE{Reject the candidate $\theta^{i+1} = \theta^{i}$}
\ENDIF
\ENDFOR
\RETURN $\theta^{0} \dots \theta^{N_{steps}}$
\end{algorithmic}
\end{algorithm}

\subsection{Rank-One Modified Metropolis Algorithm (ROMMA)}

\begin{algorithm}
\caption{Rank-One Modified Metropolis Algorithm (ROMMA)}
\label{alg:romma_mcmc}
\begin{algorithmic}
\STATE{Define $S = \sqrt{\Sigma}$ as the square root of the proposal covariance $\Sigma$}
\STATE{Define $P_+ = I$ as the matrix for the forward parameter ordering}
\STATE{Define $P_- = Flip \left ( I \right )$ as the matrix for the reverse parameter ordering}
\STATE{Define $N_d$ as the number of components of the vector $\theta$}
\STATE{Define $N_{steps}$ as the number of steps in the Markov chain}
\STATE{Initialize $\theta^0$}
\FOR{$i$ = 0 \TO $N_{steps}-1$}
\STATE{Draw $\xi \sim \mathcal{N} \left (0, I_{N_d} \right )$}
\STATE{Draw $\eta_1 \sim \mathcal{U} \left [0, 1 \right ]$}
\IF{$\eta_1 < \frac{1}{2}$}
\STATE{Choose the forward ordering $P = P_+$}
\ELSE
\STATE{Choose the reverse ordering $P = P_-$}
\ENDIF
\STATE{Compute the transformed components $R = P S P^T$}
\STATE{Set $\hat{\theta} = \theta^i$}
\FOR{$j$ = 1 \TO $N_{d}$}
\STATE{Compute rank-one update $\tilde{\theta} = \hat{\theta} + P R_j \xi_j$}
\STATE{Draw $\eta_2 \sim \mathcal{U} \left [0, 1 \right ]$}
\IF{$\eta_2 < \frac{p \left (\tilde{\theta} \right )}{p \left (\hat{\theta} \right )}$}
\STATE{Accept the rank one update $\hat{\theta}  = \tilde{\theta}$}
\ENDIF
\ENDFOR
\STATE{Draw $\eta_3 \sim \mathcal{U} \left [0, 1 \right ]$}
\IF{$\eta_3 < \frac{p \left (\mathcal{D} \mid \hat{\theta} \right )}{p \left (\mathcal{D} \mid \theta^i \right )}$}
\STATE{Accept the candidate $\theta^{i+1} = \hat{\theta}$}
\ELSE
\STATE{Reject the candidate $\theta^{i+1} = \theta^{i}$}
\ENDIF
\ENDFOR
\RETURN $\theta^{0} \dots \theta^{N_{steps}}$
\end{algorithmic}
\end{algorithm}

We develop a similar two-step proposal process for a more general setting where the proposal and prior may not correspond to independent variables. In particular, we study the case of a multivariate Gaussian proposal and a general prior. The key idea is that instead of thinking of the algorithm as a set of component-wise proposals, think of it as a set of linearly independent rank one proposals. By employing this algorithm, we can significantly reduce the number of forward model evaluations, which provides a significant speed up. The tradeoff is that this algorithm requires an increased number of prior evaluations, which scales linearly with dimension, and it is sensitive to the proposal covariance used to generate the rank-one updates. However, when used as part of ST-MCMC, the covariance structure and scaling can be well estimated.

ROMMA for MCMC is described in Algorithm \ref{alg:romma_mcmc}. In this algorithm, the correlation structure in the Gaussian proposal is handled by computing the matrix square root, $S$, of the proposal covariance, $\Sigma$; however, in principle, any matrix decomposition may be used. We also need two permutation matrices, $P_+$ and $P_-$, where $P_+$ is the identity matrix and corresponds to performing the rank-one updates in the forward direction while $P_- = Flip \left ( I \right )$ corresponds to reversing or flipping the indices of the variables and performing the updates in the reverse direction. Using these two permutation matrices is necessary to produce a reversible sampler.

Then, for each step in the Markov chain, we initialize the candidate $\hat{\theta}$ to be the current sample $\theta^i$ and randomly choose the permutation $P$ to be the forward or reverse ordering with equal probability. Based upon the choice of the permutation, the transformed matrix square root $R$ is formed. The $i$\textsuperscript{th} column of $R$, $R_i$, will be the $i$\textsuperscript{th} rank one update. Finally, we draw a random standard Normal vector $\xi$, as when generating a zero-mean multivariate Gaussian with transformed covariance $P \Sigma P^T$ using $S\xi \sim \mathcal{N} \left (0,P \Sigma P^T \right)$.

Iterating through all of the $N_d$ rank-one updates, we construct a proposed candidate $\tilde{\theta}$ based upon the current rank-one update vector $R_i$, as $\tilde{\theta} = \hat{\theta} + R_i \xi_i$. We then compute the ratio of the priors, $\frac{p \left (\tilde{\theta} \right )}{p \left ( \hat{\theta} \right )}$ and choose whether to accept or reject the proposed rank one change according to a Metropolis step. If the component is rejected, $\hat{\theta}$ remains the same, else $\hat{\theta}$ is updated to $\tilde{\theta}$. These two steps are performed for all rank-one updates until we reach the final $\hat{\theta}$. This set of rank-one proposals can be thought of as evolving the Markov chain according to the prior since the prior distribution would be the invariant distribution of this Markov chain in the absence of the likelihood evaluation step that follows.

After choosing $\hat{\theta}$, we then perform a Metropolis step to accept or reject the entire vector $\hat{\theta}$ according to only the likelihood $p \left (\mathcal{D} \mid \hat{\theta} \right )$. Thus, we compute the ratio, $\frac{p \left (\mathcal{D} \mid \hat{\theta} \right )}{p \left (\mathcal{D} \mid \theta^i \right )}$, of the likelihood for the candidate and current parameter vectors. If the sample is accepted, then $\theta^{i+1} = \hat{\theta}$, else $\theta^{i+1} = \theta^{i}$.

\section{Example: Reliability of a Water Distribution Network}
\label{sec:experiments}

Assessing the performance reliability of water distribution networks is important given the increasing age of many of these infrastructure systems that leads to component degradation and failure \cite{scheidegger2015statistical, maskit2014leakage}. We consider first identifying leak positions and intensity within a water distribution system and then making robust predictions about the reliability of the distribution system given uncertainty in demand, leak position, and leak intensity. The Bayesian leak detection problem was previously considered in \cite{poulakis2003leakage} but the posterior reliability problem has not been addressed before. This style of posterior reliability problem has been formulated in \cite{beck2013prior} but, in general, it remains computationally intractable using existing methods. This test problem has been designed to have many parameters whose values are not significantly informed by the data, which makes the problem reflect many physical systems. Using ST-MCMC and ROMMA, we are able to solve this problem significantly faster than ST-MCMC approaches based upon Random Walk Metropolis (RWM) or the Modified Metropolis Algorithm (MMA), with similar accuracy as judged by looking at the failure estimates and posterior marginal distributions.

\subsection{Test System}

\begin{figure}[t]
\centerline{\includegraphics[width=1.0\textwidth]{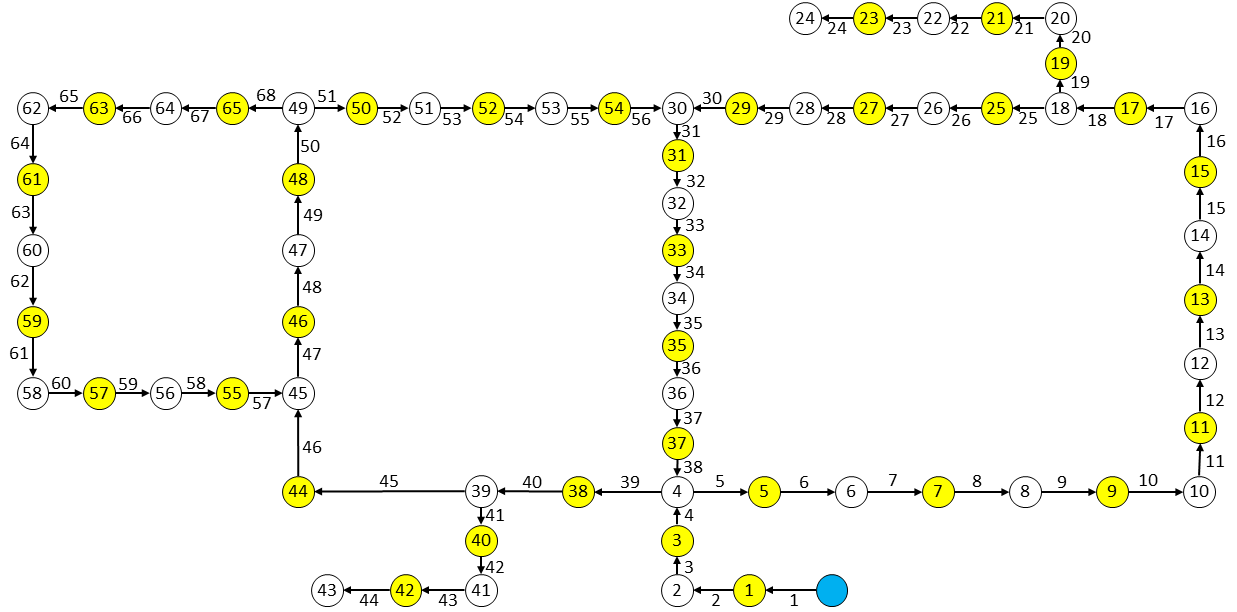}}
\caption{Network model of the Hanoi water distribution test problem. The blue node indicates the reservoir, yellow nodes represent possible leaks in pipes, and white nodes represent user loads.}
\label{fig:hanoi}
\end{figure}

We consider the Hanoi water distribution model, Figure \ref{fig:hanoi}, found in \cite{WRCR:WRCR5014, cunha1999water} and assume it is operating under steady state conditions. The network model considers frictional losses in the pipes and pressure-dependent leak rates. The static hydraulic system model is determined by solving mass conservation equations at the nodes and energy conservation along loops. Mass conservation at each node $n$ of the system is captured by

\begin{equation}
\sum_{i \in I_n} Q_i - \sum_{j \in O_n} Q_j = S_n
\label{eq:mass_conservation}
\end{equation}

\noindent where $I_n$ and $O_n$ are the sets of in-flow and out-flow pipes to node $n$, respectively. $Q_i$ is the flow rate of pipe $i$ and $S_n$ is the demand or leak rate at node $n$. In this model we treat leaks as nodes where the demand is pressure dependent according to the hydraulic head $H_n$ at the node so that $S_n = c_n \sqrt{H_n}$ if $n$ is a leakage node.

Second, energy conservation is enforced for each loop $L_i$:

\begin{equation}
\sum_{k \in L_i} \Delta H_k = 0
\label{eq:energy_conservation}
\end{equation}

\noindent where the change in the hydraulic head from the beginning to the end of pipe $k$, $\Delta H_k$, is captured by the Hazen-Williams equation expressing fictional losses:

\begin{equation}
\Delta H_k = H_{s,k} - H_{e,k} = Q_k | Q_k |^{\beta-1} \frac{w l_k}{C_k^\beta D_k^\gamma}
\label{eq:dhead}
\end{equation}

\noindent Here, $H_{s,k}$ and $H_{e,k}$ are the hydraulic head at the start and end of the pipe, respectively. $l_k$ is the pipe length and $D_k$ is its diameter. There are four fixed model parameters: the numerical conservation constant, $w=10.5088$, the pipe roughness, $C_k = 130$, and the regression coefficients, $\beta = 1.85$ and $\gamma = 4.87$. For leaks at an unknown position along a pipe, we parameterize the location of that leak node $n$ along pipe $k$ using $\delta_k \in \left [0,1 \right]$ such that the length of the pipe between the source and the leak is $\delta_k l_k$. The ``source'' direction of each pipe is defined in the model but this does not constrain the flows since the flow can either be positive or negative. Therefore, we can compute the hydraulic head at the leak on pipe k as

\begin{equation}
H_{leak,k} = H_{s,k} - Q_k | Q_k |^{\beta-1} \frac{w \delta_k l_k}{C_k^\beta D_k^\gamma}
\label{eq:leak_head}
\end{equation}

The combined equations that describe the mass conservation, energy conservation, and leaks is solved using Newton's method to find the vector $Q$ of flows along the pipes and the vector $H$ of hydraulic heads at the nodes. This approach follows standard techniques in the water distribution community as described in  \cite{cunha1999water}.

The Hanoi network in \cite{cunha1999water} is a reservoir fed system with 31 nodes and 34 pipes, leading to 34 possible leaks. Therefore, the network state is parameterized by 31 nodal loads and 34 leak sizes and positions, leading to 99 parameters in the model parameter vector $\theta$. The hydraulic head at the reservoir is fixed at $100$ m. The network description and the physics captured by Equations \eqref{eq:mass_conservation} - \eqref{eq:leak_head} define a model class $\mathcal{M}$ used in the analysis. The prior distribution on the nodal demands are modeled as a multiplicative factor on a nominal high value that follows a Gaussian distribution with mean $0.75$ and standard variation $0.15$. The leak sizes have an exponential prior with mean $0.002$, while the leak position has a uniform prior over the length of the pipe. Our choice of an exponential leak size prior means that most of the leaks will be small but a few large leaks are possible. For more details, see \ref{sec:supp_exp}.

In this test system, failure is defined as being unable to provide a minimum level of hydraulic head at each node, which in this example is set to $30$ m. The head will be influenced by the uncertain demand and leak properties. This operations constraint is mapped to the failure function $f \left ( \theta \right ) \geq 1$ by defining $f \left ( \theta \right ) = 2 - \frac{min_n H_{d,n} \left ( \theta \right )}{30}$. Here, $H_{d,n}$ are the heads for each of the demand nodes. The prediction of the hydraulic head for failure estimation is treated as deterministic and does not include model prediction uncertainty. This could be included at the cost of a more complex failure model.

The investigation is divided into three phases: 1) prior failure estimation where the failure probability is assessed based upon the prior uncertainty in the nodal demands and the leak properties, 2) leak identification based upon pressure data observed from the network under different known loading conditions, and 3) posterior failure probably estimation based upon the prior uncertainty on the demand and posterior uncertainty on the leak properties. For the posterior analysis, we consider two cases: Case 1, where failure is likely because of a large leak in a sensitive part of the network and Case 2, where failure is unlikely because of limited damage to sensitive areas of the network.

For each of these investigations, we compare ST-MCMC based on ROMMA to ST-MCMC using MMA and RWM. The methods use $N = 1024$ chains and their chain lengths at each level are determined by evolving the chains until a correlation target, $\rho = 0.6$, is reached. For inference (Algorithm \ref{alg:stmcmc_bayes}), the correlation is assessed as the maximum correlation of a parameter to its starting value in the chains. For failure probability estimation (Algorithm \ref{alg:stmcmc_ss}), since multimodality is common to insure the chains explore the likelihood levels of each mode, the correlation of the log posterior likelihood is used. For these algorithms, the next intermediate distribution is chosen such that the effective sample size is approximately $\frac{N}{2}$. For inference this is done by setting $\kappa^* = 1$ in Algorithm \ref{alg:stmcmc_bayes}. For failure probability estimation this is done by setting $\kappa = \frac{1}{2}$ in Algorithm \ref{alg:stmcmc_ss}. For a more detailed discussion of these choices see \ref{sec:supp_tuning}.

The sampling based uncertainty estimates for Subset Simulation from \cite{ZUEV2012283} are used to capture sampling uncertainty in terms of variance but they do not capture any bias due to the chains not mixing sufficiently to adequately decorrelate and explore the intermediate distributions. The quoted uncertainties are the sampling standard deviations.

\subsection{Prior Failure Estimation}
\label{sec:prior_fail}
\begin{figure}[t]
\centerline{\includegraphics[width=1.0\textwidth]{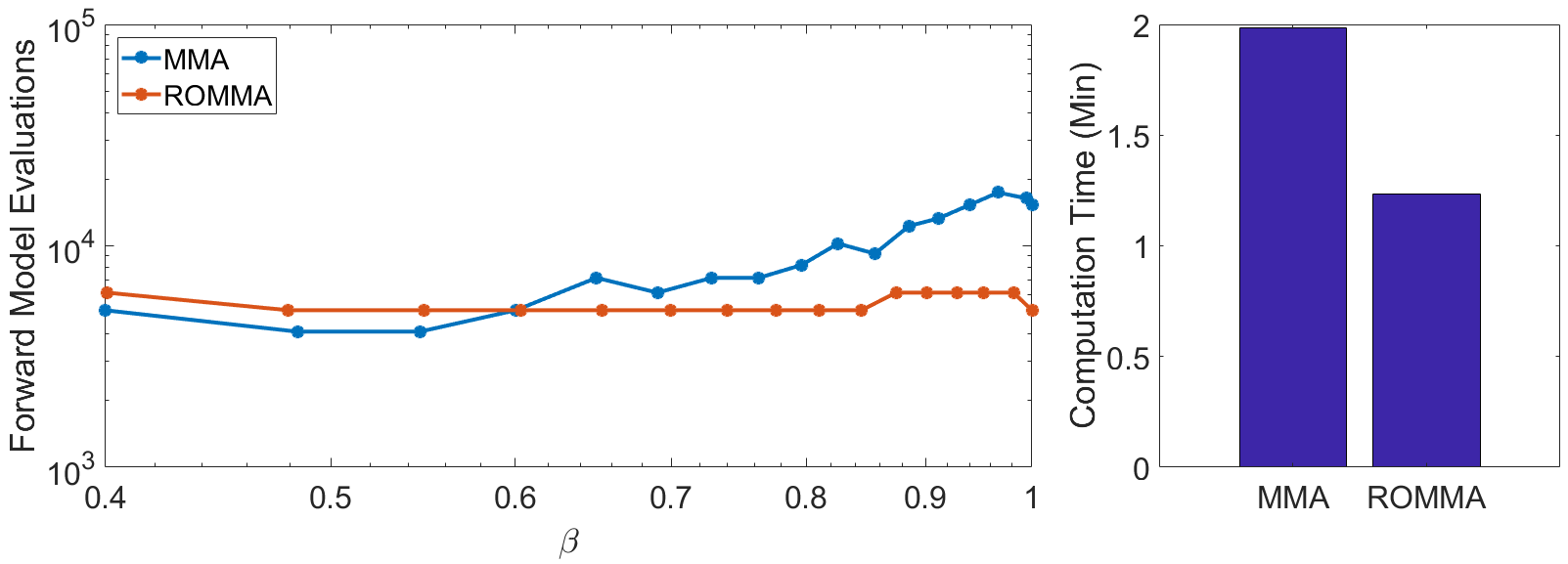}}
\caption{Comparison of performance while solving the prior failure probability estimation problem}
\label{fig:prior_fail_times}
\end{figure}

The estimation of the prior failure probability $P \left (\mathcal{F} \right )$ for systems has been extensively studied and the Subset Simulation algorithm using Modified Metropolis is generally quite effective. Here, the prior uncertainty for nodal demand, leak size, and leak position is considered, yielding 99 uncertain parameters. Two ST-MCMC algorithms were used that are similar to Subset Simulation, one based upon MMA and one based upon ROMMA. The two algorithms had the same tuning parameters so their expected accuracy should be similar. The estimated prior failure probabilities were $1.3 \pm 0.17 \times 10^{-5}$  (MMA) and $1.8 \pm 0.23 \times 10^{-5}$ (ROMMA). The prior failure probability was found to be $1.54 \pm 0.12 \times 10^{-5}$ based upon $10^7$ Monte Carlo samples. We see in Figure \ref{fig:prior_fail_times} that ROMMA improves on the efficiency of MMA for solving this problem since it takes only 60\% of the time. This improvement is significant but the full power of ROMMA comes when solving the posterior problem, since handling the data-induced correlation becomes much more important.

The fact that ROMMA better handles correlation can be seen in Figure \ref{fig:prior_fail_times} by the fact that the number of model evaluations needed at each level to reduce the correlation in the chains to a desired threshold does not significantly change as $\beta$ increases. This increase is seen in MMA since it does not efficiently handle the correlation in the failure domain as $\beta \rightarrow 1$.

\subsection{CASE 1: Leak Identification}

\begin{figure}[t]
\centerline{\includegraphics[width=1.0\textwidth]{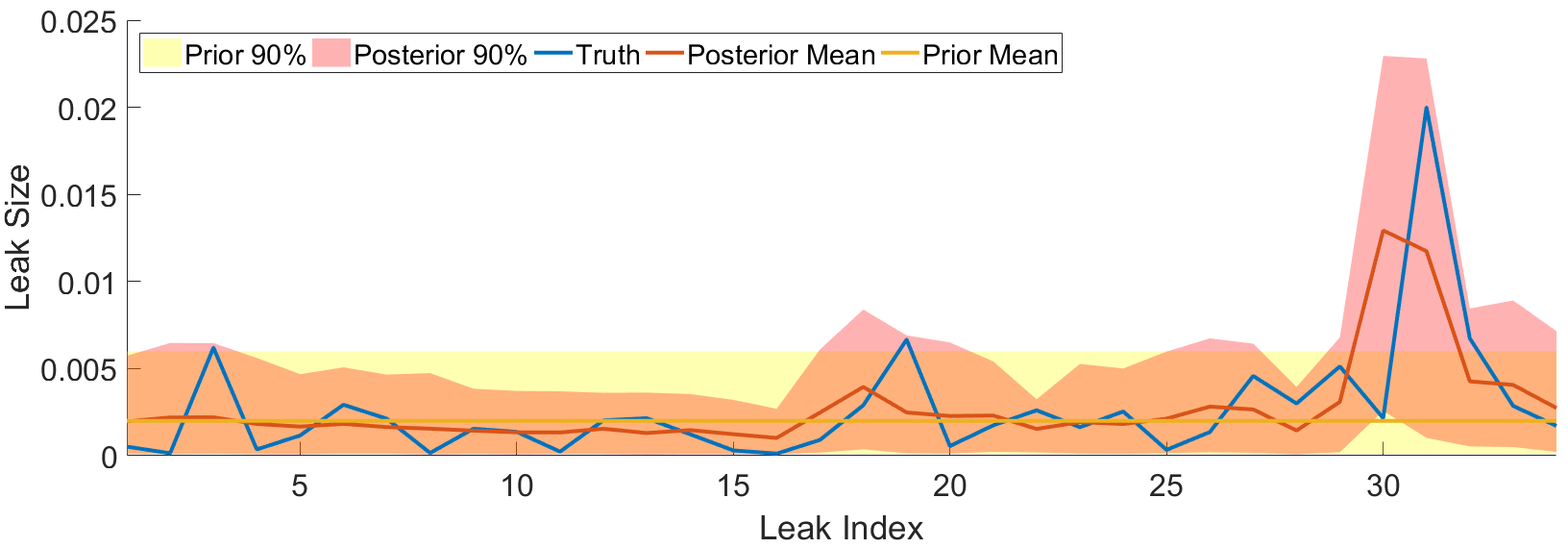}}
\caption{Case 1: Posterior mean leak size and the 90\% credibility interval compared to the prior mean and 90\% credibility interval. The true values of the parameters are in blue. We can see that many of the parameters are informed by the data since there is a large leak.}
\label{fig:c1_post_leak_size}
\end{figure}

\begin{figure}[t]
\centerline{\includegraphics[width=1.0\textwidth]{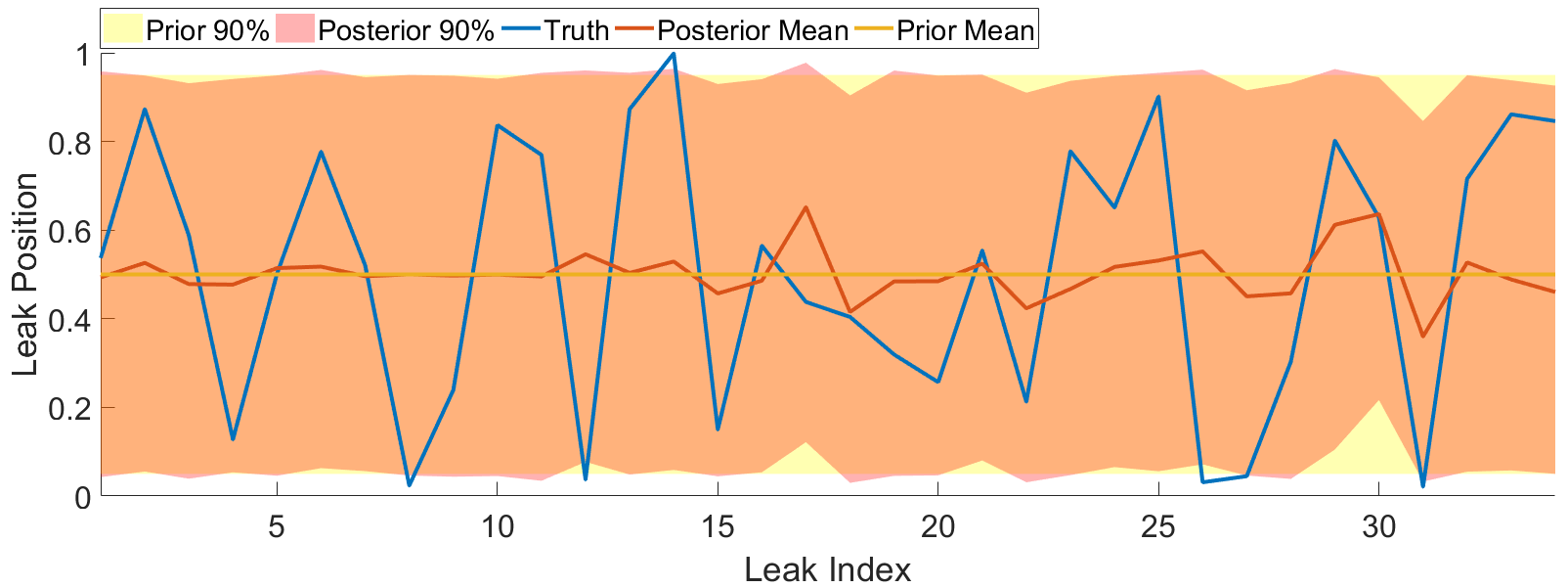}}
\caption{Case 1: Posterior mean leak position and the 90\% credibility interval compared to the prior mean and 90\% credibility interval. The true values of the parameters are in blue. We can see that the posterior appears to be very close to the prior.}
\label{fig:c1_post_leak_pos}
\end{figure}

\begin{figure}[t]
\centerline{\includegraphics[width=1.0\textwidth]{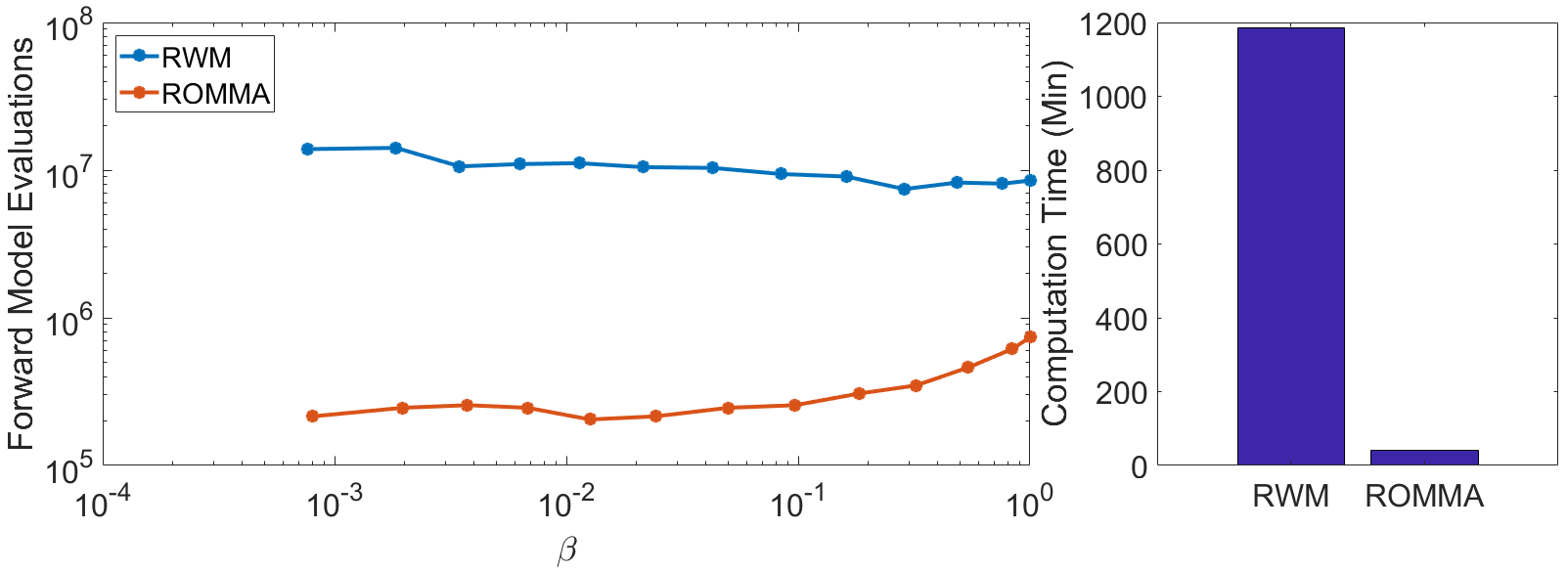}}
\caption{Case 1: Comparison of performance while solving the posterior leak detection problem}
\label{fig:c1_post_times}
\end{figure}

The leak identification problem seeks to estimate the size and position of leaks given observations under an arbitrary set of nodal demands. For the first experiment, data $\mathcal{D}$ was generated from the water distribution system subject to a large leak in a critical section of the system. This data corresponds to noisy observations of the hydraulic head at all demand nodes under 10 known demand conditions. Therefore, 10 model evaluations are needed for every likelihood evaluation. The observation uncertainty was modeled as Gaussian with $\sigma = 1$m. Using this data, the leak sizes and positions are inferred using two ST-MCMC algorithms based upon Algorithm \ref{alg:stmcmc_bayes}, standard RWM and ROMMA, to sample from the posterior $p \left ( \theta \mid \mathcal{D} \right )$. The means and Bayesian credibility intervals of the leak size and position posterior parameters are shown in Figures \ref{fig:c1_post_leak_size} - \ref{fig:c1_post_leak_pos}. We see that many of the parameters are not significantly informed by the data and that they are similar to their prior distributions. This is particularly true for the leak position variables. Further, the data implies some significant correlations between some parameters, but most parameters remain uncorrelated as in the prior. Along many directions the posterior is very similar to the prior, enabling ROMMA to give very high performance since it explicitly integrates the prior information into the proposal and can also handle the directions where correlation is imposed by the data.
Indeed, Figure \ref{fig:c1_post_times} shows that ROMMA enables much higher computational performance, requiring significantly fewer model evaluations. The performance gains of ROMMA do decrease relative to RWM as $\beta$ increases since the intermediate distribution $p \left ( \mathcal{D} \mid \theta \right )^{\beta} p \left ( \theta \right )$ gets farther from the prior $p \left ( \theta \right )$.

Finally, the distributions of posterior samples from ST-MCMC with RWM and ROMMA are practically the same, as we would expect from a converged MCMC algorithm. The comparisons of the posterior parameter marginal distributions and the posterior parameter correlations for ROMMA and RWM can be seen in the appendix, Figures \ref{fig:c1_post_corr_comp} - \ref{fig:c1_post_pos_marg}.

\subsection{CASE 1: Posterior Failure Estimation}

\begin{figure}[t]
\centerline{\includegraphics[width=1.0\textwidth]{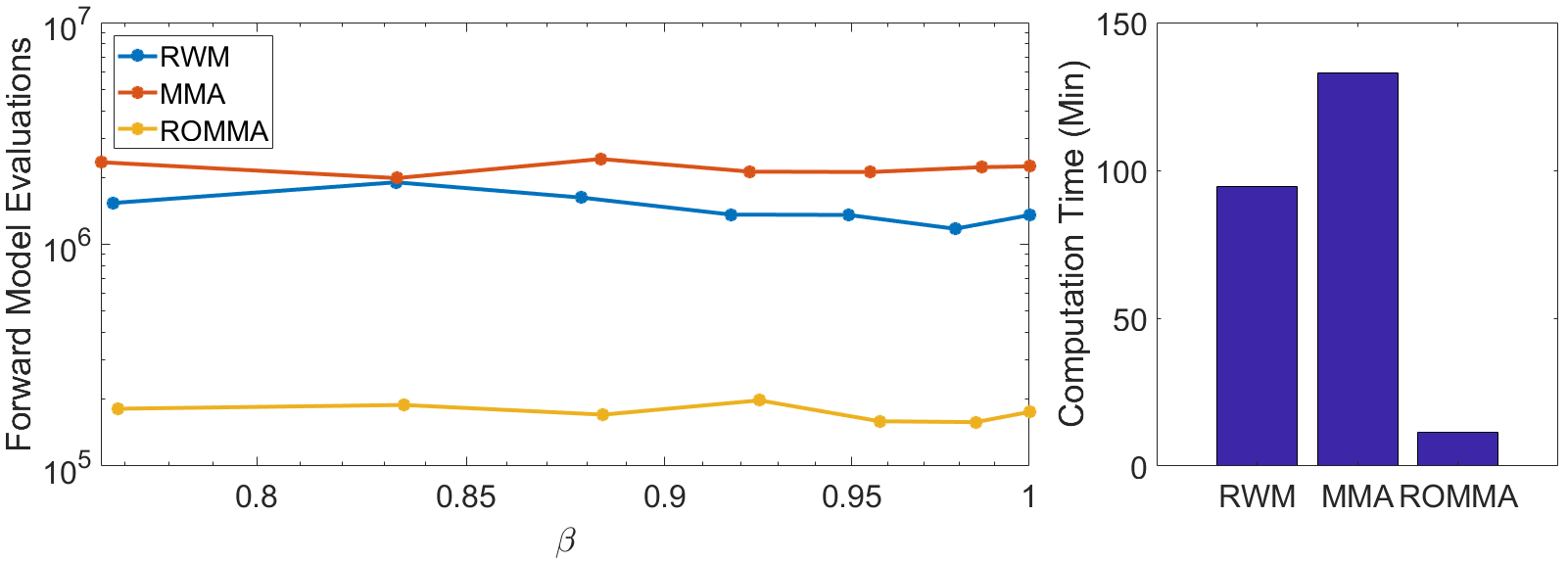}}
\caption{Case 1: Performance comparison while solving for the posterior failure probability.}
\label{fig:c1_post_fail_times}
\end{figure}

In this case, since there is a large leak in a critical part of the network, failure is likely. We compute the failure probability $P \left ( \mathcal{F} \mid \mathcal{D} \right )$ using ST-MCMC with Algorithm \ref{alg:stmcmc_ss}, based upon RWM, MMA, and ROMMA. In Figure \ref{fig:c1_post_fail_times}, we see that ROMMA outperforms both RWM and MMA significantly. MMA requires smaller steps because it cannot effectively handle the correlation introduced by the data (see Figure \ref{fig:c1_fail_corr_comp} for parameter correlations), leading to slower sampling. RWM also requires smaller steps because of the high dimensionality of the problem where it is hard for large random steps to stay in the high probability region of the posterior. The three algorithms are in good agreement with their posterior failure estimates. ROMMA estimates the failure probability as $9.9 \pm 0.79 \times 10^{-3}$, RWM estimates it as $9.0 \pm 0.73 \times 10^{-3}$, and MMA as $11.0 \pm 0.87 \times 10^{-3}$.

\begin{figure}[t]
\centerline{\includegraphics[width=1.0\textwidth]{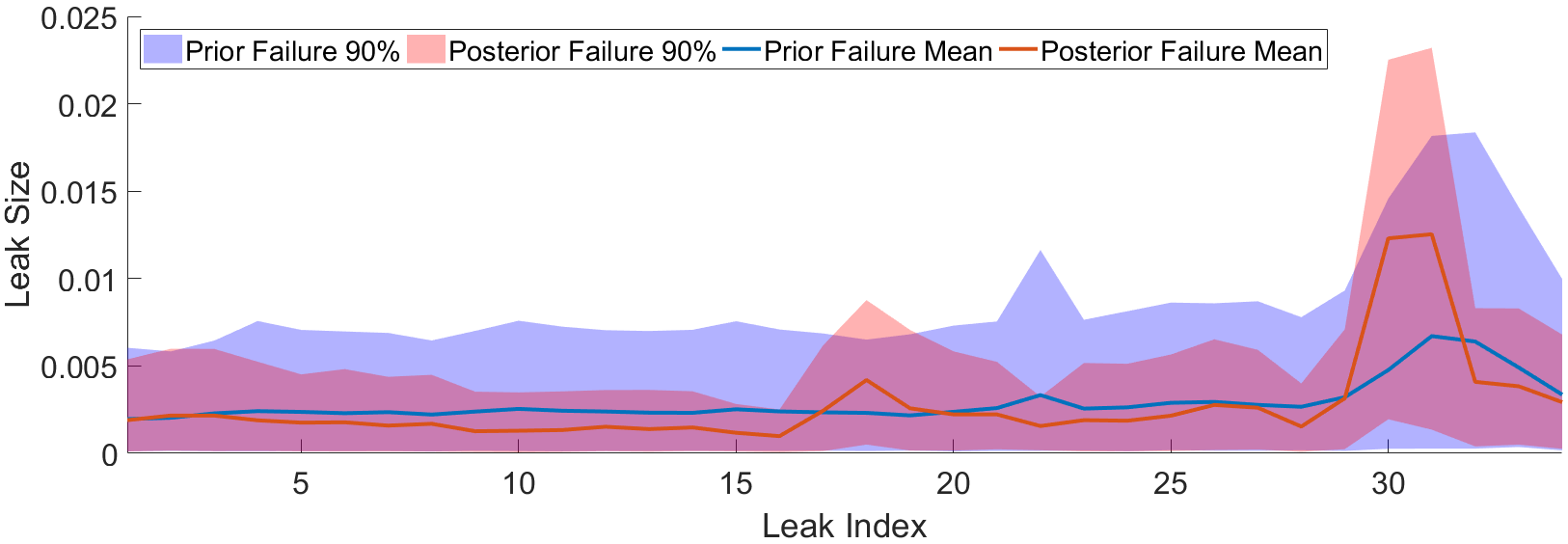}}
\caption{Case 1: Posterior failure mean leak size and the 90\% credibility interval compared to the prior failure mean and 90\% credibility interval. Because there is a large leak in a sensitive area of the network, the prior and posterior failure domains are qualitatively similar.}
\label{fig:c1_fail_leak_size}
\end{figure}

\begin{figure}[t]
\centerline{\includegraphics[width=1.0\textwidth]{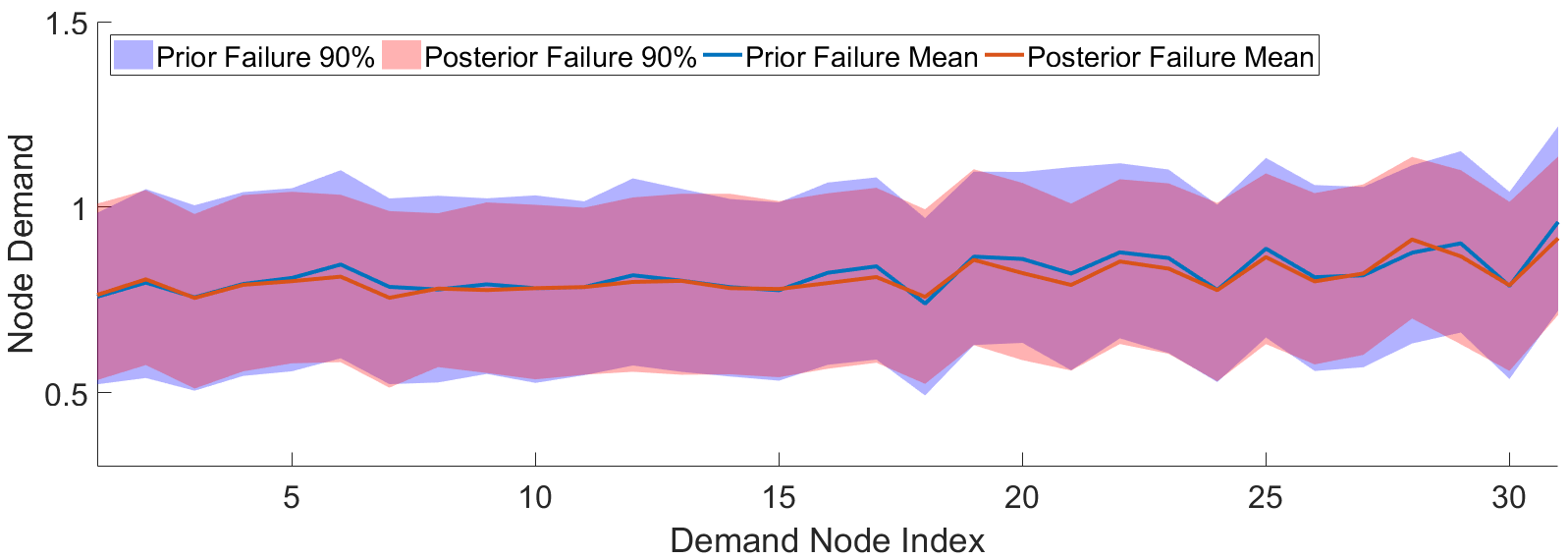}}
\caption{Case 1: Posterior failure mean demand and the 90\% credibility interval compared to the prior failure mean and 90\% credibility interval. Because there is a large leak in a sensitive area of the network, the prior and posterior failure domains are qualitatively similar.}
\label{fig:c1_fail_demands}
\end{figure}

Figures \ref{fig:c1_fail_leak_size} - \ref{fig:c1_fail_demands} compare the structure of the prior and posterior failure distributions $p \left ( \theta \mid \mathcal{F} \right )$ and $p \left ( \theta \mid \mathcal{F}, \mathcal{D} \right )$ for the leak size and demand parameters. We see that the large leak in the system matches well with the prior failure distribution so qualitatively the prior and posterior failure regions look very similar. Further, there is really only one failure mode present so the distribution is unimodal, making it work well with our algorithm tuning theory and implementation. For a full look at the posterior failure marginal distributions and correlations for the three algorithms, see the appendix, Figures \ref{fig:c1_fail_corr_comp} - \ref{fig:c1_fail_demand_marg}.

\subsection{CASE 2: Leak Identification}

\begin{figure}[t]
\centerline{\includegraphics[width=1.0\textwidth]{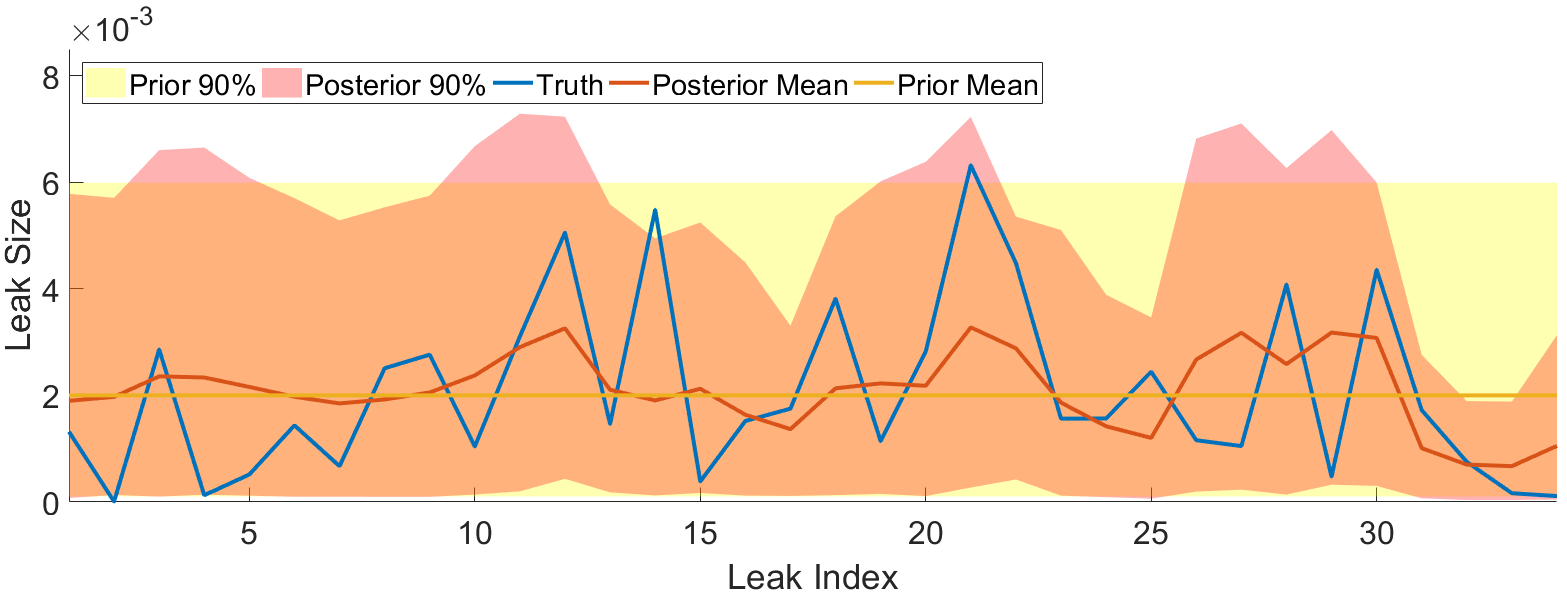}}
\caption{Case 2: Posterior mean leak size and the 90\% credibility interval compared to the prior mean and 90\% credibility interval. The true values of the parameters are in blue. We can see that a only a few of the parameters have been significantly informed by the data.}
\label{fig:post_leak_size}
\end{figure}

\begin{figure}[t]
\centerline{\includegraphics[width=1.0\textwidth]{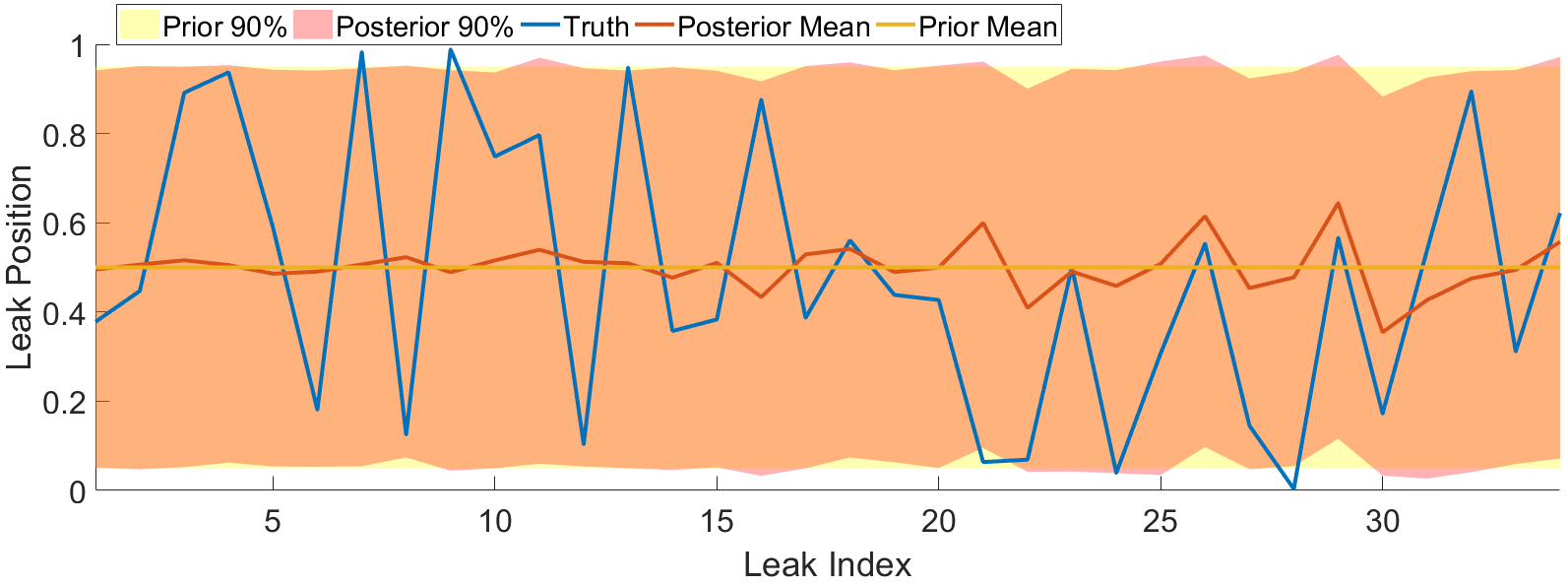}}
\caption{Case 2: Posterior mean leak position and the 90\% credibility interval compared to the prior mean and 90\% credibility interval. The true values of the parameters are in blue. We can see that the posterior appears to be very close to the prior.}
\label{fig:post_leak_pos}
\end{figure}

\begin{figure}[t]
\centerline{\includegraphics[width=1.0\textwidth]{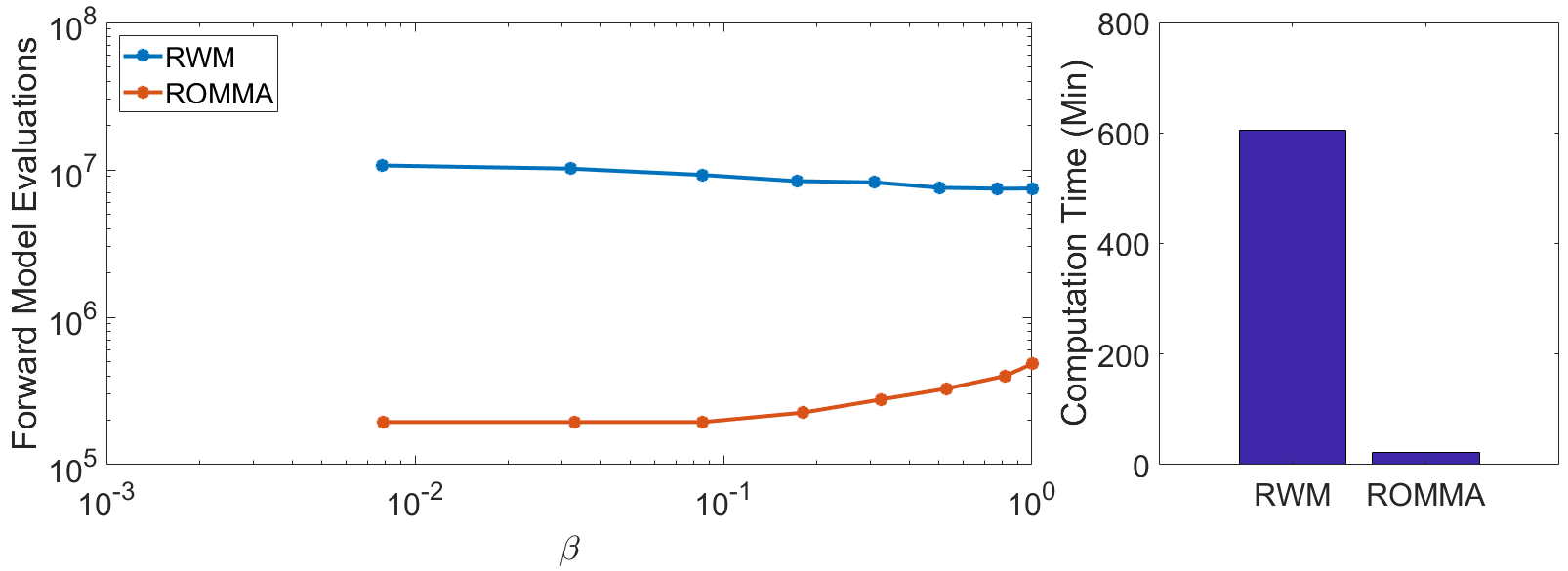}}
\caption{Case 2: Comparison of performance while solving the posterior leak detection problem}
\label{fig:post_times}
\end{figure}

Similar to Case 1, leak sizes and positions are inferred from noisy observations of the hydraulic head under different conditions. However, in this case there is no catastrophic leak. The true leak parameters used to generate the data can be seen in Figures \ref{fig:post_leak_size} - \ref{fig:post_leak_pos}. Again, when solving this identification problem, we compare a ST-MCMC method that uses RWM to one using ROMMA. Figure \ref{fig:post_times} shows a considerable speed up where ROMMA take only 3\% of the time to solve the problem than RWM, giving it about a thirty times speed up. While both RWM and ROMMA use a tuned proposal covariance, because ROMMA samples the prior very efficiently through its rank-one proposal in the first part of the algorithm, it is much more efficient than RWM. RWM must take significantly smaller proposal steps than ROMMA to maintain an appropriate acceptance rate.

The posterior is still significantly influenced by the prior since the majority of leaks are small and close to the inequality constraint that their flow rate must be non-negative. Further, the data is not rich enough to really inform the posterior and constrain all the leak sizes and positions any more than the prior already does, although it does introduce a few significant parameter correlations. We do see from Figures \ref{fig:post_leak_size} - \ref{fig:post_leak_pos} that only the leak size parameters that influence the system have been significantly identified. Figures \ref{fig:fail_corr_comp} - \ref{fig:post_pos_marg} in the appendix gives a more complete view of the parameter correlations and marginal distributions of the posterior. ROMMA excels in this environment and samples very efficiently. However, again we see in Figure \ref{fig:post_times} that the number of model evaluations needed for ROMMA increases as $\beta$ increases since the posterior is moving further away from the prior as the data is being integrated and some of the parameters are being constrained.

\subsection{CASE 2: Posterior Failure Estimation}

\begin{figure}[t]
\centerline{\includegraphics[width=1.0\textwidth]{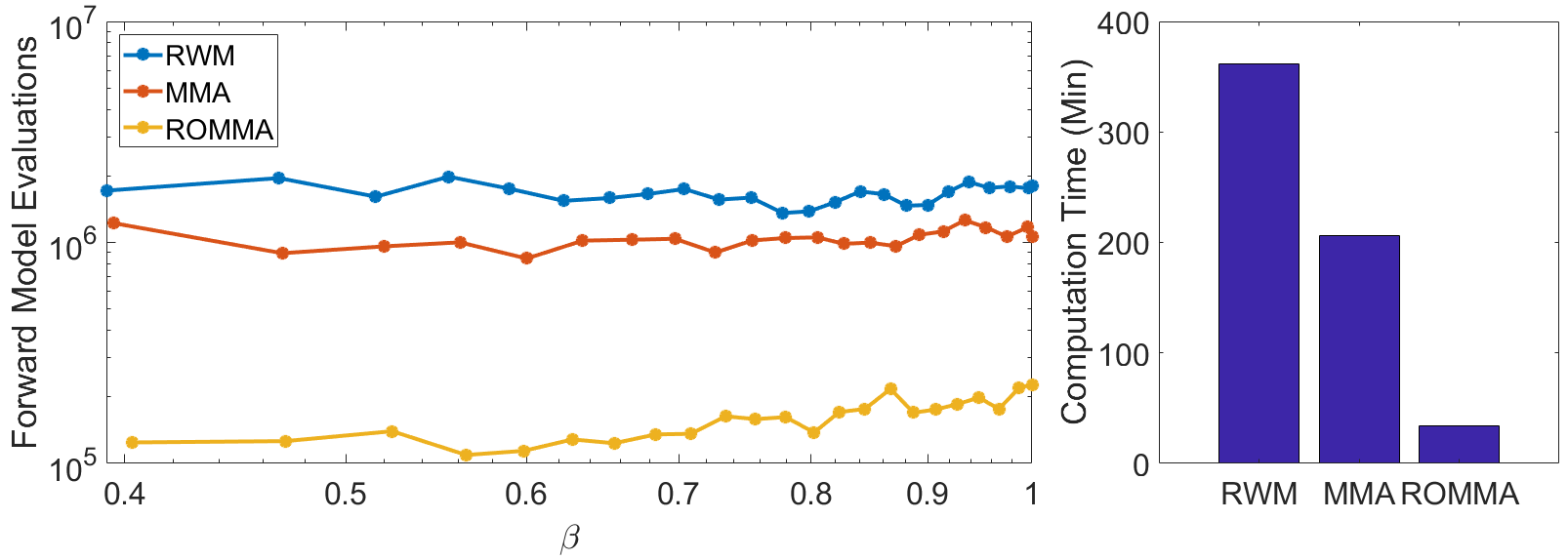}}
\caption{Case 2: Performance comparison while solving for the posterior failure probability.}
\label{fig:post_fail_times}
\end{figure}

When we combine the two ST-MCMC algorithms to estimate the posterior robust failure probability, we still see significant speed ups in Figure \ref{fig:post_fail_times} using ROMMA instead of RWM or MMA. The latter require taking much smaller step sizes than ROMMA, causing higher computational cost. The algorithms give the posterior failure probability to be $1.5 \pm 0.22 \times 10^{-7}$ for ROMMA, $1.1 \pm 0.16 \times 10^{-7}$ for RWM, and $3.9 \pm 0.56 \times 10^{-7}$ for MMA. Therefore, the posterior information about the leaks indicates that the system is significantly less likely to fail than when only considering prior information about the leaks (Section \ref{sec:prior_fail}). Estimating the very small posterior failure probability based upon Monte Carlo samples would be very computationally challenging since the estimate needs to incorporate the data through importance sampling, which would be very inefficient. Using ROMMA with 16384 samples instead of 1024 samples gives the failure probability estimate $3.1 \pm 0.11 \times 10^{-7}$. All these estimates agree within an order of magnitude, but their variation reflects that the quality of the estimate degrades for very small failure probabilities with complex failure domains.

The high fidelity ROMMA simulation with 16384 samples shows that the posterior failure domain is tri-modal, corresponding to failures due to large loads on nodes 24, 43, or 64 in Figure \ref{fig:hanoi}. This tri-modality means that using the covariance estimate from the data will lead to a suboptimal proposal mechanism. This could contribute to the variation in the failure probability estimates. The algorithms generally agree on the marginal distributions over the failure domain. The means and confidence intervals are seen in Figures \ref{fig:fail_leak_size}-\ref{fig:fail_demands} and the marginals are seen in the supplemental information in the appendix, Figures \ref{fig:fail_size_marg}-\ref{fig:fail_demand_marg}.

\begin{figure}[t]
\centerline{\includegraphics[width=1.0\textwidth]{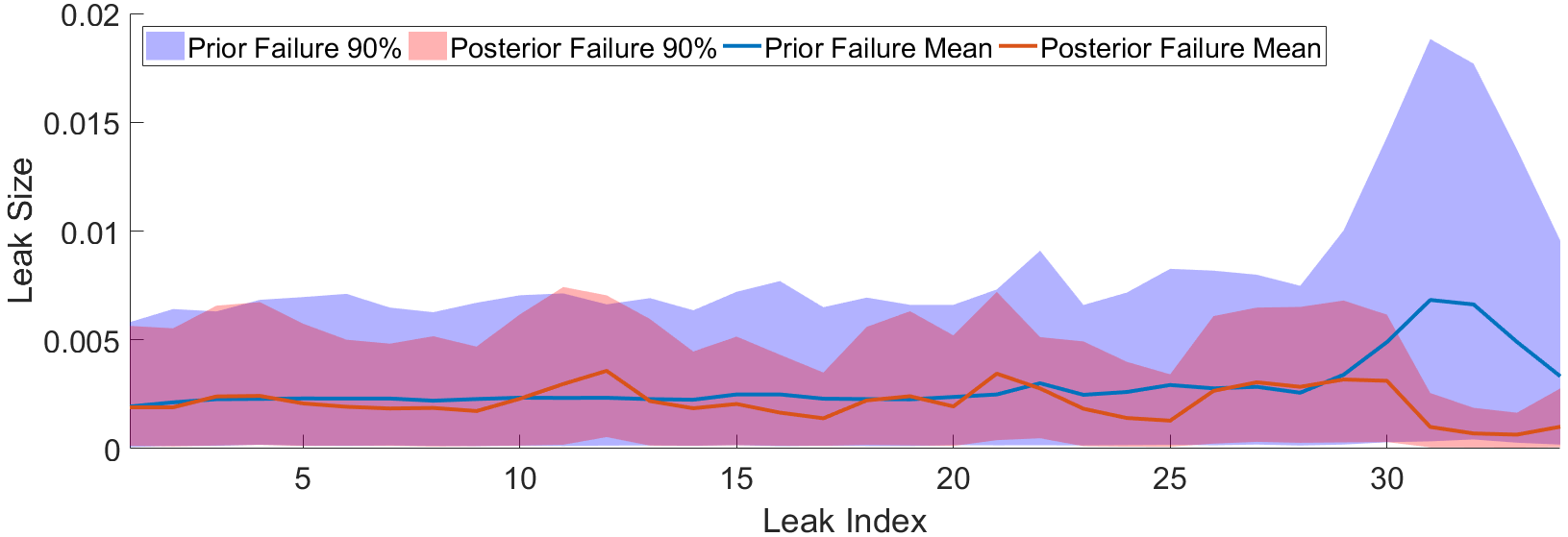}}
\caption{Case 2: Posterior failure mean leak size and the 90\% credibility interval compared to the prior failure mean and 90\% credibility interval. In the prior failure domain, failures typically results from large leaks on pipes 31 and 32. However, once data has been introduced these large leaks do not agree with the observations, so the posterior failure domains finds alterative points of failure which are overall much more rare.}
\label{fig:fail_leak_size}
\end{figure}

\begin{figure}[t]
\centerline{\includegraphics[width=1.0\textwidth]{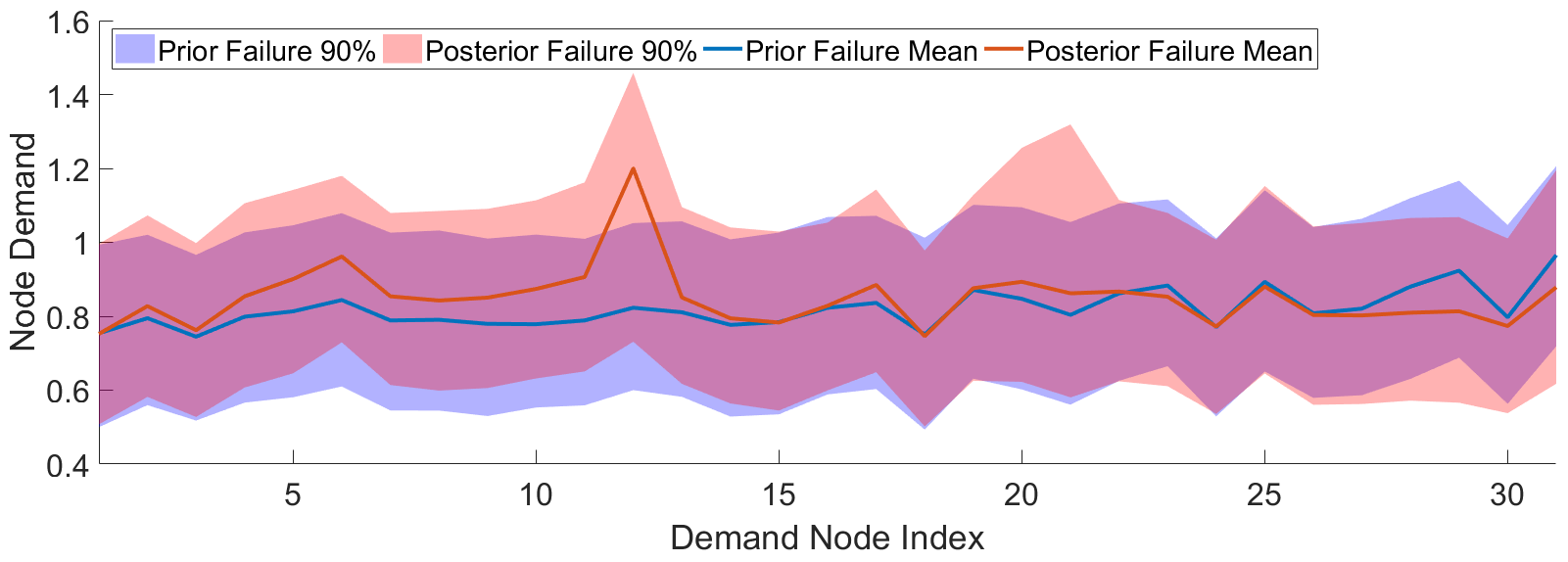}}
\caption{Case 2: Posterior failure mean demand and the 90\% credibility interval compared to the prior failure mean and 90\% credibility interval. Since large leaks in sensitive areas are unlikely according to the data, the posterior failure domain is characterized by higher demands in certain sensitive areas compared to the prior.}
\label{fig:fail_demands}
\end{figure}

\section{Concluding Remarks}
\label{sec:conclusions}

In this work we explore using Sequential Tempered MCMC (ST-MCMC) with the proposed Rank-One Modified Metropolis Algorithm (ROMMA) to solve problems in Bayesian updating and failure estimation arising in system identification, where we wish to update our understanding of a system given data, and in reliability assessment, where given our current understanding of a system we want to estimate the probability of failure based on some performance quantity. Our main contributions are as follows:

\begin{enumerate}
	\item Presenting a general framework for understanding Sequential Tempered MCMC algorithms like TMCMC and Subset Simulation which have been used for solving Bayesian updating problems and reliability problems separately in the past.
	\item Showing that this framework can be used to solve posterior reliability problems efficiently and robustly with respect to modeling uncertainty by combining multiple algorithms within the class of ST-MCMC algorithms.
	\item Introducing the Rank-One Modified Metropolis Algorithm to speed up sampling in ST-MCMC for high-dimensional distributions with inequality constraints.
\end{enumerate}

ST-MCMC combines tempering, importance resampling, and MCMC into a single algorithm that gradually transforms a population of samples from being distributed according to the prior to being distributed from the data-updated posterior. These methods gain efficiency because they can be easily parallelized and because they can adapt through the tempering process and learn from global population information. Further, these methods can be used to efficiently estimate Bayesian model evidence and failure probabilities for systems. ST-MCMC type algorithms have been used separately to solve the Bayesian inference problem and the prior failure probability assessment problem but in this work, we combine them to solve the joint posterior failure probability problem. We demonstrate efficient estimation of the prior and posterior failure probabilities for the reliability of a water distribution network subject to uncertain user demand and uncertain leak conditions. This high-dimensional problem reflects realistic complex systems where the data is often uninformative about many of the parameters. ROMMA with ST-MCMC is shown to perform very well under these conditions.

The efficiency for solving the posterior reliability problem is achieved by speeding up the MCMC sampling step within ST-MCMC using ROMMA, which builds upon the Modified Metropolis Algorithm (MMA) introduced in the original Subset Simulation reliability algorithm to allow scaling to high dimensional spaces. Unlike MMA, ROMMA does not require the prior and proposal distributions to be expressed in terms of independent stochastic variables, making it much more suited to posterior estimation that often involves high correlation induced by the data. ROMMA especially speeds up problems where the prior distribution significantly informs the posterior, as is the case where the prior enforces certain constraints or where the data only informs the posterior along a certain parameter subspace. This performance gain comes by first sequentially updating a candidate sample in the chain along rank-one directions using prior information and only then accepting or rejecting the candidate based upon the likelihood of the data.

There are many future opportunities for expanding this work. First, ST-MCMC and ROMMA could be further accelerated by making better approximations of the global structure of the intermediate distributions based upon the sample population to inform the MCMC proposal step to speed up sampling. Second, the path taken for transforming the prior to the posterior failure region by ST-MCMC could be better optimized to minimize the computational effort. Finally, the underlying theory of ST-MCMC needs to be further explored, as discussed in the appendix, to better inform its tuning for some desired level of performance, particularly for  approximations of integrals, such as estimating the failure probability.

\section*{Acknowledgments}

The authors wish to acknowledge the Department of Energy Computational Science Graduate Fellowship and the John von Neumann Fellowship granted to the first author. Sandia National Laboratories is a multimission laboratory managed and operated by National Technology \& Engineering Solutions of Sandia, LLC, a wholly owned subsidiary of Honeywell International Inc., for the U.S. Department of Energy’s National Nuclear Security Administration under contract DE-NA0003525. SAND no. 2018-3857 J

\bibliographystyle{siamplain}
\bibliography{references}

\pagebreak
\appendix
\section{Discussion of Algorithm Implementation and Tuning}
\label{sec:supp_tuning}

\subsection{General ST-MCMC Tuning}

\subsubsection{Effective Sample Size}

The number of effective samples is a common measure used in sampling problems like MCMC and importance sampling. Because of sample weights or sample correlations, the estimate quality does not necessary behave the same as if the estimate was made using independent samples. Thus, the Effective Sample Size (ESS), given by:

\begin{equation}
N_{ess} = \frac{var \left ( \theta \right )}{var \left ( \hat{\mu} \left (\theta \right ) \right )}
\label{eq:ESS_DEF}
\end{equation}

estimates the size of the population of independent samples that would give the same variance of the estimate $\hat{\mu} \left (\theta \right )$ as the weighted or correlated sample population. For example, in the case of a simple mean estimate, $\hat{\mu} \left (\theta \right ) = \frac{1}{N} \sum_{i=1}^N \theta_i$. In order to estimate the ESS, for Sequential Tempered MCMC methods, we must consider the effects of weighting the samples in the Importance Sampling step, performing the resampling, and evolving the population during the MCMC step. In general, weighting and resampling reduce the ESS while the MCMC step increases the ESS.

An approximation of the evolution of the ESS for ST-MCMC is presented in \cite{catanach2017computational}. This evolution of the effective sample size of the population is a function of the target coefficient of variation $\kappa$ and correlation $\rho$ between the start and end of the Markov chains at level $\left (k + 1\right )$:

\begin{equation}
n_{k+1} \approx n_k \frac{N}{\left (N - 1 \right) \left (1 + \kappa^2 \right) \rho^2 + n_k}
\label{eq:total_relation}
\end{equation}

\noindent If the population size is large and the COV $\kappa$ and correlation $\rho$ targets are constant for all steps, we can find a condition for the existence of a non-zero stationary number of effective samples by finding the fixed point of~\eqref{eq:total_relation} as the number of levels increases:

\begin{equation}
\rho^2 < \frac{1}{1+\kappa^2}
\label{eq:non_zero_ess_cond}
\end{equation}

\noindent If this condition holds, then an asymptotic expression for the effective number of samples is

\begin{equation}
N_{ESS} \approx N \left [1 - \left ( 1 + \kappa^2 \right )\rho^2 \right ]
\label{eq:non_zero_ess}
\end{equation}

\begin{figure}[h]
\centerline{\includegraphics[width=0.75\textwidth]{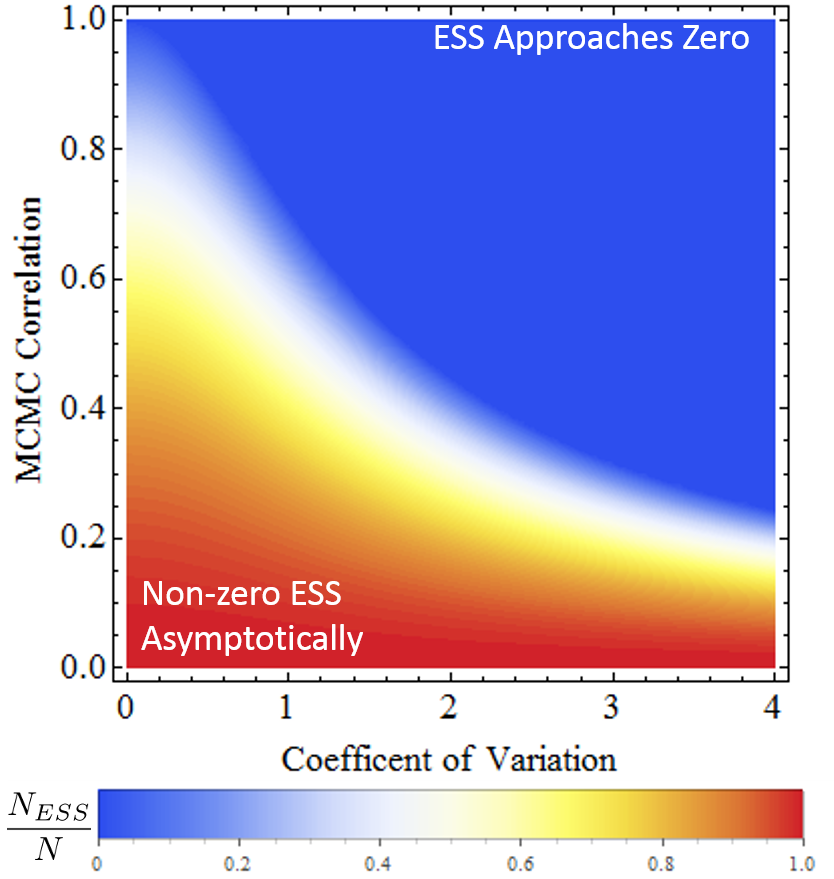}}
\caption{Asymptotically, with respect to the number of levels, the ratio of the effective numbers of samples to total samples is determined by the choices of the target coefficient of variation and MCMC correlation.}
\label{fig:ml_learning}
\end{figure}

In general, this analysis gives us a guide for setting the target COV and correlation to obtain a satisfactory number of samples from the posterior distribution. The region where learning is possible (i.e. the ESS will be non-zero at the end of all the levels) is found in Figure \ref{fig:ml_learning}. For example, if $\kappa = 1$, then $N_{ESS} = N \left ( 1 - 2 \rho^2 \right )$ and it requires $\rho < \frac{1}{\sqrt{2}}$ for learning.

\subsubsection{Measure of Correlation}

Finding a good measure of the correlation of a multivariate sample population is important to insure that the sample population converges to the correct distribution. A simple measure is to look at the component-wise correlations of each of the parameters. This technique is commonly used when estimating the autocorrelation of the Markov Chain for more typical effective sample size analyses for MCMC.

\hfill \break
\indent However, even if the parameter-wise correlations are small, there might be large correlations in some transformed set of coordinates. One strategy to mitigate this issue is to use Canonical Correlation Analysis (CCA) \cite{Hardle2007}. CCA is a technique to efficiently find the direction and magnitude of maximum correlation between two populations, i.e., find vectors $a$ and $b$ to maximize $\text{corr} \left (a^T \theta_0, b^T \theta_1 \right )$. By minimizing the canonical correlation, we can insure the correlation target is achieved. This approach was used when solving the Bayesian inference problem for the leak identification.

When the posterior distribution is multimodal, as is often the case for posterior reliability problems like Case 2, looking at the correlation of the components will cause the algorithm to be impractically slow. This is because it is very hard for the Markov Chain to traverse the multiple components which means the correlation will always be high. In this case, the rebalancing of samples between the components will be achieved by the importance resampling step so the most important factor is decorrelating the sample weights and making sure the chain explores the mode it is confined too. Therefore we use as correlation the log likelihood of the start and end of each chain.

\subsection{ROMMA Acceptance Rate Tuning}
\label{sec:supp_tuning_acc}
The scaling for the spread of a MCMC proposal distribution is typically tuned by trying to find a scaling factor that achieves an acceptance rate target. In ROMMA, because there are multiple Metropolis steps, finding the appropriate definition of the acceptance rate is non-trivial. For example, having the function that relates the scaling factor to the acceptance rate be monotonic is important for many of the tuning algorithms to achieve a target acceptance rate that corresponds to a scaling factor that is neither too large or too small and thus induces low correlation. However, the acceptance rate in the second part of the ROMMA algorithm does not have this property. When the scale factor is small, the second Metropolis step acceptance rate is high and generally decreases. Then once the scale factor gets sufficiently large, the acceptance rate starts to increase again since most of the rank one proposals in the first step of the algorithm are now getting rejected. This causes the bifurcation of the correlation with respect to the acceptance rate.

An alternative definition of the acceptance rate is to look at the acceptance rate for a specific rank one component. This means the probability that a specific component proposal is accepted during step one of the ROMMA algorithm and also accepted as part of the combined candidate in step 2. Since there are multiple rank one components, we take the minimum acceptance rate among all of them. This quantity is generally monotonic since as the scale factor grows very large, the higher acceptance rate in step 2 is balanced by the higher rejection rate in step 1. Since it is monotonic, it is a much better tuning mechanism to find a scaling factor that leads to low correlation.

The Gaussian proposal distribution in ST-MCMC, $\hat{\theta} \sim \mathcal{N} \left (\theta, \sigma^2 \Sigma \right )$, was tuned over the levels of the ST-MCMC algorithm. The proposal covariance was chosen to be a scaled version of the sample population covariance, $\Sigma$ where the scaling, $\sigma$, was adapted using a feedback controller to get a desired acceptance rate. The feedback controller is described in Algorithm \ref{alg:feedback}. The motivation and derivation of this feedback controller for MCMC can be found in \cite{catanach2017computational}. Based upon \cite{rosenthal2011optimal, roberts2001optimal}, $\alpha^* = 0.234$, and $G$ was chosen to be $2.1$ based upon \cite{catanach2017computational}.

\begin{algorithm}
\caption{MCMC Feedback Controller}
\label{alg:feedback}
\begin{algorithmic}
\STATE{Define $\alpha^*$ as the desired acceptance rate}
\STATE{Define $G$ as the feedback gain}
\STATE{Initialize scaling factor $\sigma_1$}
\FOR{$k$ = 1 \TO $N_k$}
\STATE{Get the acceptance rate,$\alpha_k$, for the MCMC using $\sigma_k$ at level k of ST-MCMC}
\STATE{Update the scale factor $\sigma_{k+1} = \sigma_{k}\exp \left [ G \left (\alpha_k - \alpha^* \right )\right ]$}
\ENDFOR
\end{algorithmic}
\end{algorithm}

\section{ST-MCMC for Model Selection}
\label{sec:model_select}
The model evidence is the high dimensional normalization factor in Bayes' Theorem for $\theta$ that MCMC was developed to avoid computing. The evidence can be thought of as the expectation of the probability of the data with respect to the prior distribution generated by the model class $\mathcal{M}$ and it is important for computing the posterior probability of $\mathcal{M}$:

\begin{equation}
P \left ( \mathcal{M} \mid \mathcal{D} \right ) = \frac{p \left ( \mathcal{D} \mid \mathcal{M} \right ) P \left ( \mathcal{M} \right ) }{p \left ( \mathcal{D} \right ) } \propto  \left ( \int p \left ( \mathcal{D} \mid \theta, \mathcal{M}  \right ) p \left ( \theta \mid \mathcal{M} \right ) d\theta \right ) P \left ( \mathcal{M} \right )
\label{eq:model_evidence}
\end{equation}

This integral could be naively estimated using Monte Carlo sampling of the prior distribution $p \left ( \theta \mid \mathcal{M} \right )$. This estimate would be very computationally inefficient when the data is informative, since the high probability content of the prior may be very far from the high probability content of the posterior. However, the intermediate levels of ST-MCMC enable us to address this problem by decomposing the evidence computation over the intermediate levels \cite{Neal2001, ching2007transitional}. This can be thought of as thermodynamic integration as in \cite{CALDERHEAD20094028}. Let $c_k$ denote the ratio of the evidences for the intermediate levels $k$ and $k-1$ of the $s$ levels, then:

\begin{equation}
\int p \left ( \mathcal{D} \mid \theta, \mathcal{M}  \right ) p \left ( \theta \mid \mathcal{M} \right ) d\theta = \prod_{k=1}^{s} \frac{\int p \left ( \mathcal{D} \mid \theta, \mathcal{M}  \right )^{\beta_k} p \left ( \theta \mid \mathcal{M} \right ) d\theta}{\int p \left ( \mathcal{D} \mid \theta, \mathcal{M}  \right )^{\beta_{k-1}} p \left ( \theta \mid \mathcal{M} \right ) d\theta} = \prod_{k=1}^{s} c_k
\label{eq:evidence_level}
\end{equation}

\noindent where $\beta_0 = 0$ and $\beta_s = 1$. For each intermediate level, we can perform a fairly accurate Monte Carlo estimate between the previous level and the current level since these distributions are designed to be relatively close to each other in terms of the relative effective sample size of samples coming from the previous level. Having a high ESS means Monte Carlo sampling will be effective. This leads to the Monte Carlo estimate:

\begin{equation}
c_k = \int p \left ( \mathcal{D} \mid \theta, \mathcal{M}  \right )^{\Delta \beta_k} \frac{p \left ( \mathcal{D} \mid \theta, \mathcal{M}  \right )^{\beta_{k-1}} p \left ( \theta \mid \mathcal{M} \right )}{\prod_{j=1}^{k-1}c_j} d\theta \approx \frac{1}{N} \sum_{i=1}^N p \left ( \mathcal{D} \mid \theta^{\left ( k-1 \right )}_{i}, \mathcal{M}  \right )^{\Delta \beta_k}
\label{eq:evidence_level_mc_est}
\end{equation}

This integral can be thought of as the evidence provided by level $k$ where $p \left ( \mathcal{D} \mid \theta, \mathcal{M}  \right )^{\Delta \beta_k}$ is the data likelihood added by level $k$ and $\frac{p \left ( \mathcal{D} \mid \theta, \mathcal{M}  \right )^{\beta_{k-1}} p \left ( \theta \mid \mathcal{M} \right )}{\prod_{j=1}^{k-1}c_j}$ is the prior for level $k$. The combined estimate of the level model evidences $c_k$ provides an asymptotically unbiased estimate of the total model evidence.
\vfill
\pagebreak
\hfill

\section{Proofs of MMA and ROMMA Reversibility}

Reversibility is a sufficient condition for the existence of a stationary distribution, $\pi \left ( \theta \right )$, that satisfies the detailed-balance condition:

\begin{equation}
\pi \left ( \theta \right ) K \left ( \theta^\prime \mid \theta \right ) = \pi \left ( \theta^\prime \right ) K \left ( \theta \mid \theta^\prime \right )
\label{eq:detail_balance}
\end{equation}

\noindent This sufficient condition means that any transition kernel may be chosen to maintain the stationary distribution $\pi \left ( \theta \right )$, as long as the reversibility condition~\eqref{eq:detail_balance} holds. Further, the composition of kernels which have the same invariant distribution, $\pi \left ( \theta \right )$, also has $\pi \left ( \theta \right )$ as its invariant distribution \cite{geyer2011introduction}. This method can be used to create non-reversible Markov chains with the correct stationary distribution.

\subsection{Proof of MMA Reversibility}
\label{sec:proof_mma}

The MMA MCMC Markov process step from $\theta$ to $\hat{\theta}$, with transition distribution denoted by $Q \left (\hat{\theta} \mid \theta \right)$, forms a reversible Markov chain whose invariant measure is the posterior distribution $p \left( \theta \mid \mathcal{D} \right ) \propto p \left ( \mathcal{D} \mid \theta \right ) \pi \left ( \theta \right )$. For this algorithm we assume that the proposal and prior distributions have independent components i.e. $P \left (\hat{\theta} \mid \theta \right) = \prod_{i=1}^N P_i \left (\hat{\theta}_i \mid \theta_i \right)$ and $\pi \left ( \theta \right) = \prod_{i=1}^N \pi_i \left ( \theta_i \right)$.

\begin{theorem}
Reversibility: $p \left ( \mathcal{D} \mid \hat{\theta} \right ) \pi \left ( \hat{\theta} \right ) Q \left (\theta \mid \hat{\theta} \right)= p \left ( \mathcal{D} \mid \theta \right ) \pi \left ( \theta \right ) Q \left (\hat{\theta} \mid \theta \right)$
\end{theorem}

\begin{proof}

Let $P \left ( \theta \rightarrow \hat{\theta}\right )$ denote the probability density that describes moving from $\theta$ to $\hat{\theta}$ under the Markov chain proposal, then the transition density from $\theta$ to $\hat{\theta}$, $Q \left (\hat{\theta} \mid \theta \right)$, is:

\begin{equation}
\begin{split}
Q \left (\hat{\theta} \mid \theta \right) &= \min \left ( \frac{p \left (\mathcal{D} \mid \hat{\theta} \right )}{p \left (\mathcal{D} \mid \theta \right )}, 1\right ) P \left ( \theta \rightarrow \hat{\theta}\right )\\
&= \min \left ( \frac{p \left (\mathcal{D} \mid \hat{\theta} \right )}{p \left (\mathcal{D} \mid \theta \right )}, 1\right ) \prod_{i=1}^N P_i \left (\theta_i \rightarrow \hat{\theta}_i \right)
\label{eq:mma_qprop}
\end{split}
\end{equation}

We define $\hat{\theta}_i$ to be the i\textsuperscript{th} component of the candidate $\hat{\theta}$ and $P_i \left ( \hat{\theta}_{i-1} \rightarrow \hat{\theta}_i \right )$ as the full transition probability according to both the i\textsuperscript{th} component update and the Metropolis accept/reject step. The component-wise proposals $q_i$ is introduced in Algorithm \ref{alg:og_mm_mcmc}. Each factor in this product can be express in two different ways that depend on whether the candidate was accepted or rejected at the i\textsuperscript{th} step:

\begin{equation}
\begin{split}
P_i \left ( \theta_{i} \rightarrow \hat{\theta}_i \right ) &\\
&=\begin{cases}
\min \left ( \frac{\pi \left ( \hat{\theta}_i \right ) q_i \left ( \hat{\theta}_i \rightarrow \theta_i \right )}{\pi \left (\theta_i \right ) q_i \left ( \theta_i \rightarrow \hat{\theta}_i \right )}, 1\right ) q_i \left ( \theta_{i} \rightarrow \hat{\theta}_i \right ) & \theta_i \neq \hat{\theta}_{i} \\
\int_{\tilde{\theta}} \left (1 - \min \left (\frac{\pi \left ( \tilde{\theta} \right ) q_i \left ( \hat{\theta}_i \rightarrow \theta_i \right )}{\pi \left (\theta_{i} \right ) q_i \left ( \theta_i \rightarrow \hat{\theta}_i \right )}, 1\right ) \right ) q_i \left ( \theta_{i} \rightarrow \tilde{\theta} \right ) d\tilde{\theta} &  \theta_i = \hat{\theta}_{i}\\
\label{eq:mma_ppropi}
\end{cases}
\end{split}
\end{equation}

 This leads to the ratio:

\begin{equation}
\begin{split}
\frac{P_i \left ( \hat{\theta}_{i} \rightarrow \theta_{i} \right )}{P_i \left ( \theta_i \rightarrow \hat{\theta}_i \right )}&\\\\
&=\begin{cases}
 \frac{\min \left ( \frac{\pi \left ( \theta_i \right ) q_i \left ( \theta_i \rightarrow \hat{\theta}_i \right )}{\pi \left ( \hat{\theta}_i \right ) q_i \left ( \hat{\theta}_i \rightarrow \theta_i \right )}, 1\right ) q_i \left ( \hat{\theta}_{i} \rightarrow \theta_i \right )}{\min \left ( \frac{\pi \left ( \hat{\theta}_i \right ) q_i \left ( \hat{\theta}_1 \rightarrow \theta_i \right )}{\pi \left (\theta \right ) q_i \left ( \theta_i \rightarrow \hat{\theta}_i \right )}, 1\right ) q_i \left ( \theta_{i} \rightarrow \hat{\theta}_i \right )} & \hat{\theta}_i \neq \theta_{i} \\\\
 \frac{\int_{\tilde{\theta}} \left (1 - \min \left (\frac{\pi \left (\theta_{i} \right ) q_i \left ( \theta_i \rightarrow \hat{\theta}_i \right )}{\pi \left ( \tilde{\theta} \right ) q_i \left ( \hat{\theta}_i \rightarrow \theta_i \right )}, 1\right ) \right ) q_i \left ( \tilde{\theta}_{i} \rightarrow \theta_i \right ) d\tilde{\theta}}{\int_{\tilde{\theta}} \left (1 - \min \left (\frac{\pi \left ( \tilde{\theta} \right ) q_i \left ( \hat{\theta}_i \rightarrow \theta_i \right )}{\pi \left (\theta_{i} \right ) q_i \left ( \theta_i \rightarrow \hat{\theta}_i \right )}, 1\right ) \right ) q_i \left ( \theta_{i} \rightarrow \tilde{\theta} \right ) d\tilde{\theta}}&  \hat{\theta}_i = \theta_{i}
\end{cases}\\\\
&=\begin{cases}
\frac{\pi \left ( \theta_{i} \right )}{\pi \left (\hat{\theta}_{i} \right )} & \hat{\theta}_i \neq \theta_i \\\\
1 &  \hat{\theta}_i = \theta_i
\end{cases}\\\\
&=\frac{\pi \left ( \theta_{i} \right )}{\pi \left (\hat{\theta}_{i} \right )}
\label{eq:mma_ppropirat}
\end{split}
\end{equation}

Therefore, we can put these results together to find:

\begin{equation}
\begin{split}
\frac{P \left ( \hat{\theta} \rightarrow \theta \right )}{P \left ( \theta \rightarrow \hat{\theta} \right )} &= \prod_{i=1}^{N} \frac{P_i \left ( \hat{\theta}_{i} \rightarrow \theta_{i} \right )}{P \left ( \theta_{i} \rightarrow \hat{\theta}_i \right )}\\\\
&= \prod_{i=1}^{N} \frac{\pi \left ( \theta_i \right )}{\pi \left (\hat{\theta}_{i} \right )}\\\\
&= \frac{\pi \left ( \theta \right )}{\pi \left (\hat{\theta}\right )}
\label{eq:mma_pproprat}
\end{split}
\end{equation}

Substituting this result into the Markov chain transition probability ratio $\frac{Q \left (\theta \mid \hat{\theta} \right)}{Q \left (\hat{\theta} \mid \theta \right)}$, we can prove the reversibility of the Markov chain with respect to the posterior distribution:

\begin{equation}
\begin{split}
&\frac{Q \left (\theta \mid \hat{\theta} \right)}{Q \left (\hat{\theta} \mid \theta \right)} = \frac{\min \left ( \frac{p \left (\mathcal{D} \mid \theta \right )}{p \left (\mathcal{D} \mid \hat{\theta} \right )}, 1\right ) P \left ( \hat{\theta} \rightarrow \theta \right )}{\min \left ( \frac{p \left (\mathcal{D} \mid \hat{\theta} \right )}{p \left (\mathcal{D} \mid \theta \right )}, 1\right ) P \left ( \theta \rightarrow \hat{\theta} \right )}\\\\
&= \frac{p \left (\mathcal{D} \mid \theta \right )}{p \left (\mathcal{D} \mid \hat{\theta} \right )}\frac{P \left ( \hat{\theta} \rightarrow \theta \right )}{P \left ( \theta \rightarrow \hat{\theta} \right )}\\\\
&= \frac{p \left (\mathcal{D} \mid \theta \right )}{p \left (\mathcal{D} \mid \hat{\theta} \right )}\frac{\pi \left ( \theta \right )}{\pi \left (\hat{\theta}\right )}
\label{eq:mma_qrat}
\end{split}
\end{equation}

\begin{equation}
\begin{split}
&\frac{Q \left (\theta \mid \hat{\theta} \right)}{Q \left (\hat{\theta} \mid \theta \right)} = \frac{p \left (\mathcal{D} \mid \theta \right )}{p \left (\mathcal{D} \mid \hat{\theta} \right )}\frac{\pi \left ( \theta \right )}{\pi \left (\hat{\theta}\right )}\\\\
&\implies p \left ( \mathcal{D} \mid \hat{\theta} \right ) \pi \left ( \hat{\theta} \right ) Q \left (\theta \mid \hat{\theta} \right)= p \left ( \mathcal{D} \mid \theta \right ) \pi \left ( \theta \right ) Q \left (\hat{\theta} \mid \theta \right)
\label{eq:mma_final_rev}
\end{split}
\end{equation}

\end{proof}

\subsection{Proof of ROMMA Reversibility}
\label{sec:proof_romma}

The ROMMA MCMC Markov process step from $\theta$ to $\hat{\theta}$, with transition distribution denoted by $Q \left (\hat{\theta} \mid \theta \right)$, forms a reversible Markov chain whose invariant measure is the posterior distribution $p \left( \theta \mid \mathcal{D} \right ) \propto p \left ( \mathcal{D} \mid \theta \right ) \pi \left ( \theta \right )$.

\begin{theorem}
Reversibility: $p \left ( \mathcal{D} \mid \hat{\theta} \right ) \pi \left ( \hat{\theta} \right ) Q \left (\theta \mid \hat{\theta} \right)= p \left ( \mathcal{D} \mid \theta \right ) \pi \left ( \theta \right ) Q \left (\hat{\theta} \mid \theta \right)$
\end{theorem}

\begin{proof}

Let $P \left ( \theta \rightarrow \hat{\theta}\right )$ denote the probability density that describes moving from $\theta$ to $\hat{\theta}$ under the Markov chain proposal, then the transition density from $\theta$ to $\hat{\theta}$, $Q \left (\hat{\theta} \mid \theta \right)$, is:

\begin{equation}
\begin{split}
Q \left (\hat{\theta} \mid \theta \right) &= \min \left ( \frac{p \left (\mathcal{D} \mid \hat{\theta} \right )}{p \left (\mathcal{D} \mid \theta \right )}, 1\right ) P \left ( \theta \rightarrow \hat{\theta}\right )\\
&= \min \left ( \frac{p \left (\mathcal{D} \mid \hat{\theta} \right )}{p \left (\mathcal{D} \mid \theta \right )}, 1\right )\left [  \frac{1}{2} P \left ( \theta \rightarrow \hat{\theta} \mid P_+ \right ) + \frac{1}{2} P \left ( \theta \rightarrow \hat{\theta} \mid P_- \right ) \right ]\\
\label{eq:qprop}
\end{split}
\end{equation}

We define $\hat{\theta}_i$ to be the i\textsuperscript{th} intermediate evolution step of the candidate $\hat{\theta}$ and $\tilde{\theta}_i$ to be the i\textsuperscript{th} proposal step under the i\textsuperscript{th} rank one update and $P \left ( \hat{\theta}_{i-1} \rightarrow \hat{\theta}_i \mid P \right )$ as the full transition probability according to both the i\textsuperscript{th} rank one update and the Metropolis accept/reject step:

\begin{equation}
\begin{split}
P \left ( \hat{\theta}_0 = \theta \rightarrow \hat{\theta}_N = \hat{\theta} \mid P \right ) &= \prod_{i=1}^{N} P \left ( \hat{\theta}_{i-1} \rightarrow \hat{\theta}_i \mid P \right )\\
\label{eq:pprop}
\end{split}
\end{equation}

Each factor in this product can be express in two different ways that depend on whether the candidate was accepted or rejected at the i\textsuperscript{th} step:

\begin{equation}
\begin{split}
P \left ( \hat{\theta}_{i-1} \rightarrow \hat{\theta}_i \mid P \right ) &\\
&=\begin{cases}
\min \left ( \frac{\pi \left ( \tilde{\theta}_i \right )}{\pi \left (\hat{\theta}_{i-1} \right )}, 1\right ) P \left ( \hat{\theta}_{i-1} \rightarrow \tilde{\theta}_i \mid P \right ) & \hat{\theta}_i \neq \hat{\theta}_{i-1} \\
\int_{\tilde{\theta}} \left (1 - \min \left (\frac{\pi \left ( \tilde{\theta} \right )}{\pi \left (\hat{\theta}_{i-1} \right )}, 1\right ) \right ) P \left ( \hat{\theta}_{i-1} \rightarrow \tilde{\theta} \mid P \right ) d\tilde{\theta} &  \hat{\theta}_i = \hat{\theta}_{i-1}\\
\label{eq:ppropi}
\end{cases}
\end{split}
\end{equation}

If we assume the structure of the rank one proposals introduced in Algorithm \ref{alg:romma_mcmc}, i.e. $\tilde{\theta}_i = \hat{\theta}_{i-1} + P R_i \xi_i$, we find:

\begin{equation}
\begin{split}
P \left ( \hat{\theta}_{i-1} \rightarrow \hat{\theta}_i \mid P \right ) &\\\\
&=\begin{cases}
\min \left ( \frac{\pi \left ( \hat{\theta}_{i-1} + P \vec{R}_i \xi_i \right )}{\pi \left (\hat{\theta}_{i-1} \right )}, 1\right ) P \left (\xi_i \mid P \right ) & \hat{\theta}_i \neq \hat{\theta}_{i-1} \\\\
\int_{\xi} \left (1 - \min \left (\frac{\pi \left ( \hat{\theta}_{i-1} + P \vec{R}_i \xi \right )}{\pi \left (\hat{\theta}_{i-1} \right )}, 1\right ) \right ) P \left ( \xi \mid P \right ) d\tilde{\theta} &  \hat{\theta}_i = \hat{\theta}_{i-1}
\label{eq:ppropxi}
\end{cases}
\end{split}
\end{equation}

\hfill \break
\indent The key insight into proving reversibility is that the rank one update $P \vec{R}_i \mid P_+$ is the same as $P \vec{R}_{N-i+1} \mid P_-$ so we can undo all the updates from $\theta \rightarrow \hat{\theta}$ update using the ordering implied by $P_+$ by applying the reverse ordering $P_-$ or vice versa. This leads to:

\begin{equation}
\begin{split}
\frac{P \left ( \hat{\theta}_{i} \rightarrow \hat{\theta}_{i-1} \mid P_- \right )}{P \left ( \hat{\theta}_{i-1} \rightarrow \hat{\theta}_i \mid P_+ \right )}&\\\\
&=\begin{cases}
\frac{\min \left ( \frac{\pi \left ( \hat{\theta}_{i} - P_- \vec{R}_{N-i+1} \xi_i \right )}{\pi \left (\hat{\theta}_{i} \right )}, 1\right ) P \left (-\xi_i \mid P_- \right )}{\min \left ( \frac{\pi \left ( \hat{\theta}_{i-1} + P_+ \vec{R}_i \xi_i \right )}{\pi \left (\hat{\theta}_{i-1} \right )}, 1\right ) P \left (\xi_i \mid P_+ \right )} & \hat{\theta}_i \neq \hat{\theta}_{i-1} \\\\
\frac{\int_{\xi} \left (1 - \min \left (\frac{\pi \left ( \hat{\theta}_{i} - P_- \vec{R}_{N-i+1} \xi \right )}{\pi \left (\hat{\theta}_{i} \right )}, 1\right ) \right ) P \left ( -\xi \mid P_- \right ) d\tilde{\theta}}{\int_{\xi} \left (1 - \min \left (\frac{\pi \left ( \hat{\theta}_{i-1} + P_+ \vec{R}_i \xi \right )}{\pi \left (\hat{\theta}_{i-1} \right )}, 1\right ) \right ) P \left ( \xi \mid P_+ \right ) d\tilde{\theta}} &  \hat{\theta}_i = \hat{\theta}_{i-1}
\end{cases}\\\\
&=\begin{cases}
\frac{\pi \left ( \hat{\theta}_{i-1} \right )}{\pi \left (\hat{\theta}_{i} \right )} & \hat{\theta}_i \neq \hat{\theta}_{i-1} \\\\
\frac{\pi \left ( \hat{\theta}_{i-1} \right )}{\pi \left (\hat{\theta}_{i} \right )} = 1 &  \hat{\theta}_i = \hat{\theta}_{i-1}
\end{cases}\\\\
&=\frac{\pi \left ( \hat{\theta}_{i-1} \right )}{\pi \left (\hat{\theta}_{i} \right )}
\label{eq:ppropirat}
\end{split}
\end{equation}

Therefore, we can put these results together to find:

\begin{equation}
\begin{split}
\frac{P \left ( \hat{\theta}_0 = \hat{\theta} \rightarrow \hat{\theta}_N = \theta \mid P_- \right )}{P \left ( \hat{\theta}_0 = \theta \rightarrow \hat{\theta}_N = \hat{\theta} \mid P_+ \right )} &= \prod_{i=1}^{N} \frac{P \left ( \hat{\theta}_{i} \rightarrow \hat{\theta}_{i-1} \mid P_- \right )}{P \left ( \hat{\theta}_{i-1} \rightarrow \hat{\theta}_i \mid P_+ \right )}\\\\
&= \prod_{i=1}^{N} \frac{\pi \left ( \hat{\theta}_{i-1} \right )}{\pi \left (\hat{\theta}_{i} \right )}\\\\
&= \frac{\pi \left ( \theta \right )}{\pi \left (\hat{\theta}\right )}
\label{eq:pproprat}
\end{split}
\end{equation}

Substituting this result into the Markov chain transition probability ratio $\frac{Q \left (\theta \mid \hat{\theta} \right)}{Q \left (\hat{\theta} \mid \theta \right)}$, we can prove the reversibility of the Markov chain:

\begin{equation}
\begin{split}
&\frac{Q \left (\theta \mid \hat{\theta} \right)}{Q \left (\hat{\theta} \mid \theta \right)} = \frac{\min \left ( \frac{p \left (\mathcal{D} \mid \theta \right )}{p \left (\mathcal{D} \mid \hat{\theta} \right )}, 1\right )\left [  \frac{1}{2} P \left ( \hat{\theta} \rightarrow \theta \mid P_- \right ) + \frac{1}{2} P \left ( \hat{\theta} \rightarrow \theta \mid P_+ \right ) \right ]}{\min \left ( \frac{p \left (\mathcal{D} \mid \hat{\theta} \right )}{p \left (\mathcal{D} \mid \theta \right )}, 1\right )\left [  \frac{1}{2} P \left ( \theta \rightarrow \hat{\theta} \mid P_+ \right ) + \frac{1}{2} P \left ( \theta \rightarrow \hat{\theta} \mid P_- \right ) \right ]}\\\\
&= \frac{p \left (\mathcal{D} \mid \theta \right )}{p \left (\mathcal{D} \mid \hat{\theta} \right )}\frac{P \left ( \hat{\theta} \rightarrow \theta \mid P_- \right ) +  P \left ( \hat{\theta} \rightarrow \theta \mid P_+ \right )}{P \left ( \theta \rightarrow \hat{\theta} \mid P_+ \right ) + P \left ( \theta \rightarrow \hat{\theta} \mid P_- \right )}\\\\
&= \frac{p \left (\mathcal{D} \mid \theta \right )}{p \left (\mathcal{D} \mid \hat{\theta} \right )}\left [ \frac{P \left ( \hat{\theta} \rightarrow \theta \mid P_- \right )}{P \left ( \theta \rightarrow \hat{\theta} \mid P_+ \right ) + P \left ( \theta \rightarrow \hat{\theta} \mid P_- \right )} + \frac{P \left ( \hat{\theta} \rightarrow \theta \mid P_+ \right )}{P \left ( \theta \rightarrow \hat{\theta} \mid P_- \right ) + P \left ( \theta \rightarrow \hat{\theta} \mid P_+ \right )}\right ]\\\\
&= \frac{p \left (\mathcal{D} \mid \theta \right )}{p \left (\mathcal{D} \mid \hat{\theta} \right )}\left [ \frac{P \left ( \hat{\theta} \rightarrow \theta \mid P_- \right )}{P \left ( \theta \rightarrow \hat{\theta} \mid P_+ \right )}\frac{1}{1 + \frac{P \left ( \theta \rightarrow \hat{\theta} \mid P_- \right )}{P \left ( \theta \rightarrow \hat{\theta} \mid P_+ \right )}} + \frac{P \left ( \hat{\theta} \rightarrow \theta \mid P_+ \right )}{P \left ( \theta \rightarrow \hat{\theta} \mid P_- \right )}\frac{1}{1 + \frac{P \left ( \theta \rightarrow \hat{\theta} \mid P_+ \right )}{P \left ( \theta \rightarrow \hat{\theta} \mid P_- \right )}}\right ]\\\\
&= \frac{p \left (\mathcal{D} \mid \theta \right )}{p \left (\mathcal{D} \mid \hat{\theta} \right )}\frac{\pi \left ( \theta \right )}{\pi \left (\hat{\theta}\right )}\left [ \frac{1}{1 + \frac{P \left ( \theta \rightarrow \hat{\theta} \mid P_- \right )}{P \left ( \theta \rightarrow \hat{\theta} \mid P_+ \right )}} + \frac{1}{1 + \frac{P \left ( \theta \rightarrow \hat{\theta} \mid P_+ \right )}{P \left ( \theta \rightarrow \hat{\theta} \mid P_- \right )}}\right ]\\\\
&= \frac{p \left (\mathcal{D} \mid \theta \right )}{p \left (\mathcal{D} \mid \hat{\theta} \right )}\frac{\pi \left ( \theta \right )}{\pi \left (\hat{\theta}\right )}
\label{eq:qrat}
\end{split}
\end{equation}

\begin{equation}
\begin{split}
&\frac{Q \left (\theta \mid \hat{\theta} \right)}{Q \left (\hat{\theta} \mid \theta \right)} = \frac{p \left (\mathcal{D} \mid \theta \right )}{p \left (\mathcal{D} \mid \hat{\theta} \right )}\frac{\pi \left ( \theta \right )}{\pi \left (\hat{\theta}\right )}\\\\
&\implies p \left ( \mathcal{D} \mid \hat{\theta} \right ) \pi \left ( \hat{\theta} \right ) Q \left (\theta \mid \hat{\theta} \right)= p \left ( \mathcal{D} \mid \theta \right ) \pi \left ( \theta \right ) Q \left (\hat{\theta} \mid \theta \right)
\label{eq:final_rev}
\end{split}
\end{equation}

\end{proof}

\pagebreak[4]
\hfill

\section{Supplementary Experiment Information}
\label{sec:supp_exp}

\subsection{Water Distribution System Description Tables}
\hfill
\begin{table}[!h]
\begin{center}
    \begin{tabular}{ | l | l | l | p{1cm} |}
    \hline
    Node Index & Demand ($m^3/h$) \\ \hline
    $2$      & 890  \\ \hline
    $4$      & 850  \\ \hline
    $6$      & 130  \\ \hline
    $8$      & 725  \\ \hline
    $10$      & 1005 \\ \hline
    $12$      & 1350 \\ \hline
    $14$      & 550  \\ \hline
    $16$      & 525  \\ \hline
    $18$     & 525  \\ \hline
    $20$     & 500  \\ \hline
    $22$     & 560  \\ \hline
    $24$     & 940  \\ \hline
    $26$     & 615  \\ \hline
    $28$     & 280  \\ \hline
    $30$     & 310  \\ \hline
    $32$     & 865  \\ \hline
    $34$     & 1345 \\ \hline
    $36$     & 60   \\ \hline
    $39$     & 1275 \\ \hline
    $41$     & 930  \\ \hline
    $43$     & 485  \\ \hline
    $45$     & 1045 \\ \hline
    $47$     & 820  \\ \hline
    $49$     & 170  \\ \hline
    $51$     & 900  \\ \hline
    $53$     & 370  \\ \hline
    $56$     & 290  \\ \hline
    $58$     & 360  \\ \hline
    $60$     & 360  \\ \hline
    $62$     & 105  \\ \hline
    $64$     & 805  \\ \hline

    \end{tabular}
\end{center}
\caption{Parameters describing the reference node demand used in the model of the Hanoi water distribution network when generating demand conditions. The indices correspond to notes in Figure \ref{fig:hanoi}}
\label{table:demand}
\end{table}
\vfill
\pagebreak

\begin{table}[!h]
\begin{center}
    \begin{tabular}{ | l | l | l | p{1cm} |}
    \hline
    Pipe Index & Length (m) & Diameter (m) \\ \hline
    $1$      & 100  & 1.016 \\ \hline
    $3$      & 1350 & 1.016\\ \hline
    $5$      & 900  & 1.016\\ \hline
    $7$      & 1150 & 1.016\\ \hline
    $9$      & 1450 & 1.016\\ \hline
    $11$     & 450  & 1.016\\ \hline
    $13$     & 850  & 1.016\\ \hline
    $15$     & 850  & 1.016\\ \hline
    $17$     & 800  & 0.762\\ \hline
    $19$     & 950  & 0.762\\ \hline
    $21$     & 1200 & 0.762\\ \hline
    $23$     & 3500 & 0.6096\\ \hline
    $25$     & 800  & 0.4064\\ \hline
    $27$     & 500  & 0.4064\\ \hline
    $29$     & 550  & 0.3048\\ \hline
    $31$     & 2730 & 0.4064\\ \hline
    $33$     & 1750 & 0.508\\ \hline
    $35$     & 800  & 0.6096\\ \hline
    $37$     & 400  & 0.6096\\ \hline
    $38$     & 2200 & 1.016\\ \hline
    $40$     & 1500 & 0.508\\ \hline
    $42$     & 500  & 0.3048\\ \hline
    $44$     & 2650 & 1.016\\ \hline
    $46$     & 1230 & 0.762\\ \hline
    $48$     & 1300 & 0.762\\ \hline
    $50$     & 850  & 0.508\\ \hline
    $52$     & 300  & 0.3048\\ \hline
    $54$     & 750  & 0.3048\\ \hline
    $55$     & 1500 & 0.4064\\ \hline
    $57$     & 2000 & 0.4064\\ \hline
    $59$     & 1600 & 0.3048\\ \hline
    $61$     & 150  & 0.3048\\ \hline
    $63$     & 860  & 0.4064\\ \hline
    $65$     & 950  & 0.508\\ \hline

    \end{tabular}
\end{center}
\caption{Parameters describing the pipes used in the model of the Hanoi water distribution network. The indices correspond to notes in Figure \ref{fig:hanoi}}
\label{table:pipes}
\end{table}

\pagebreak
\hfill
\subsection{Computational Setup}

The computational results in this paper were computed using a mobile workstation with a 2.7 GHz four core Intel Core i7-4800MQ with 16 GB of RAM. The simulations were run using MATLAB.

\section{Supplemental Experimental Result Plots}
\subsection{Case 1: Posterior Leak Correlation Plots}
\hfill
\begin{figure}[!h]
\centerline{\includegraphics[width=1.0\textwidth]{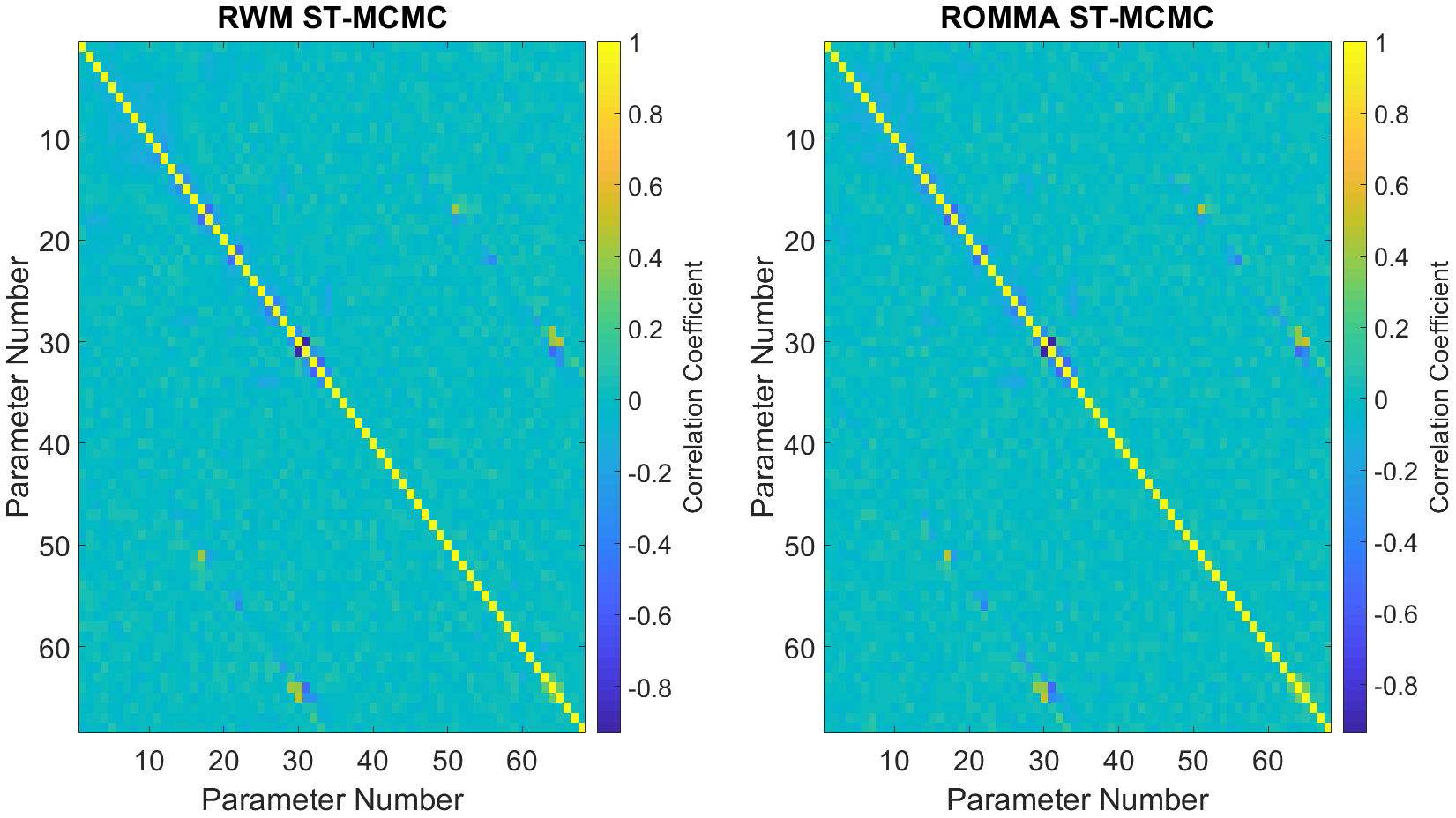}}
\caption{Case 1: Posterior correlation comparison between RWM and ROMMA}
\label{fig:c1_post_corr_comp}
\end{figure}
\vfill
\pagebreak
\hfill
\subsection{Case 1: Posterior Marginal Distributions Plots}
\hfill
\begin{figure}[!h]
\centerline{\includegraphics[width=1.0\textwidth]{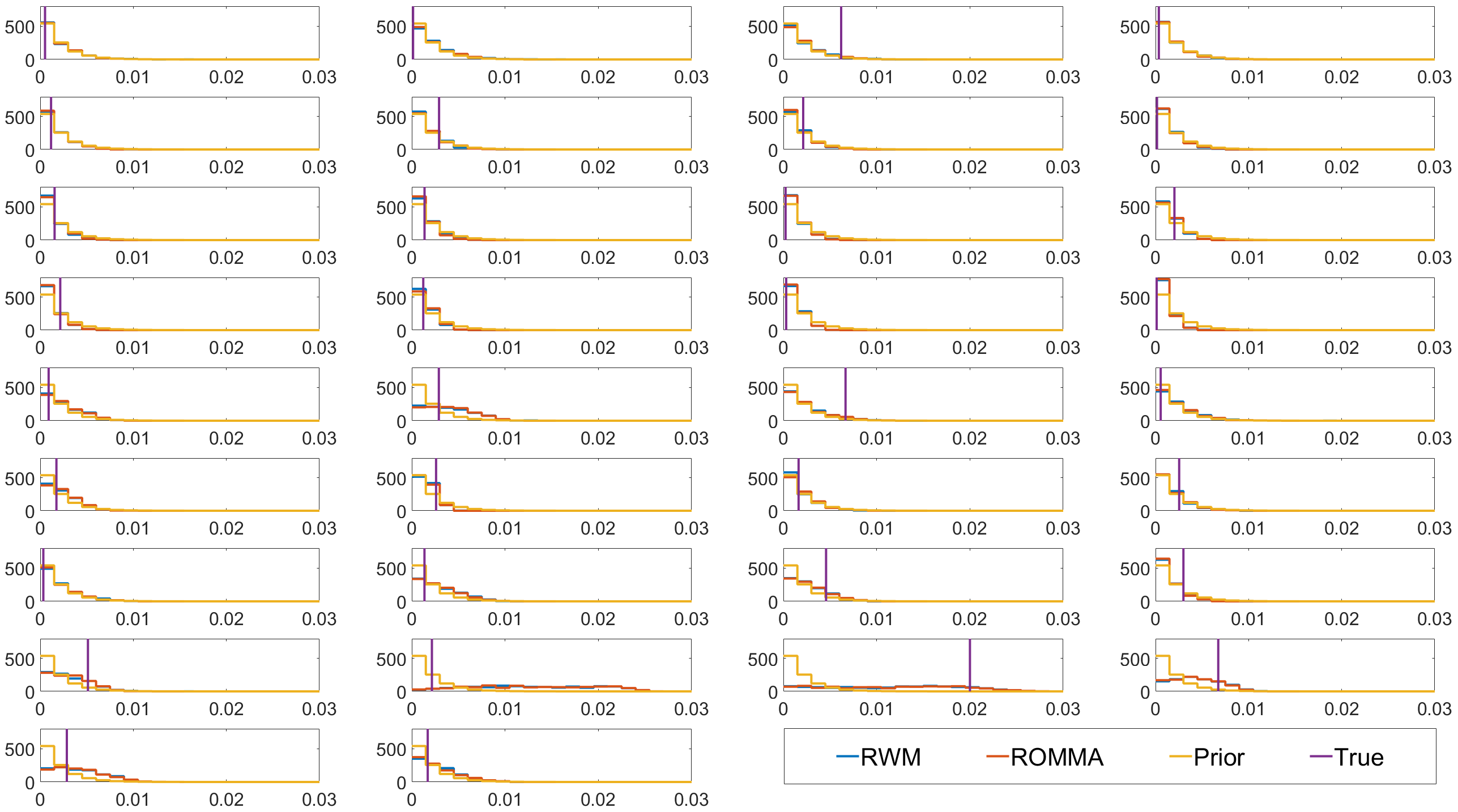}}
\caption{Case 1: Posterior marginal distributions for the leak size parameters computed using RWM and ROMMA. They are compared to the prior marginal and the true value.}
\label{fig:c1_post_size_marg}
\end{figure}
\vfill
\pagebreak
\hfill
\begin{figure}[!h]
\centerline{\includegraphics[width=1.0\textwidth]{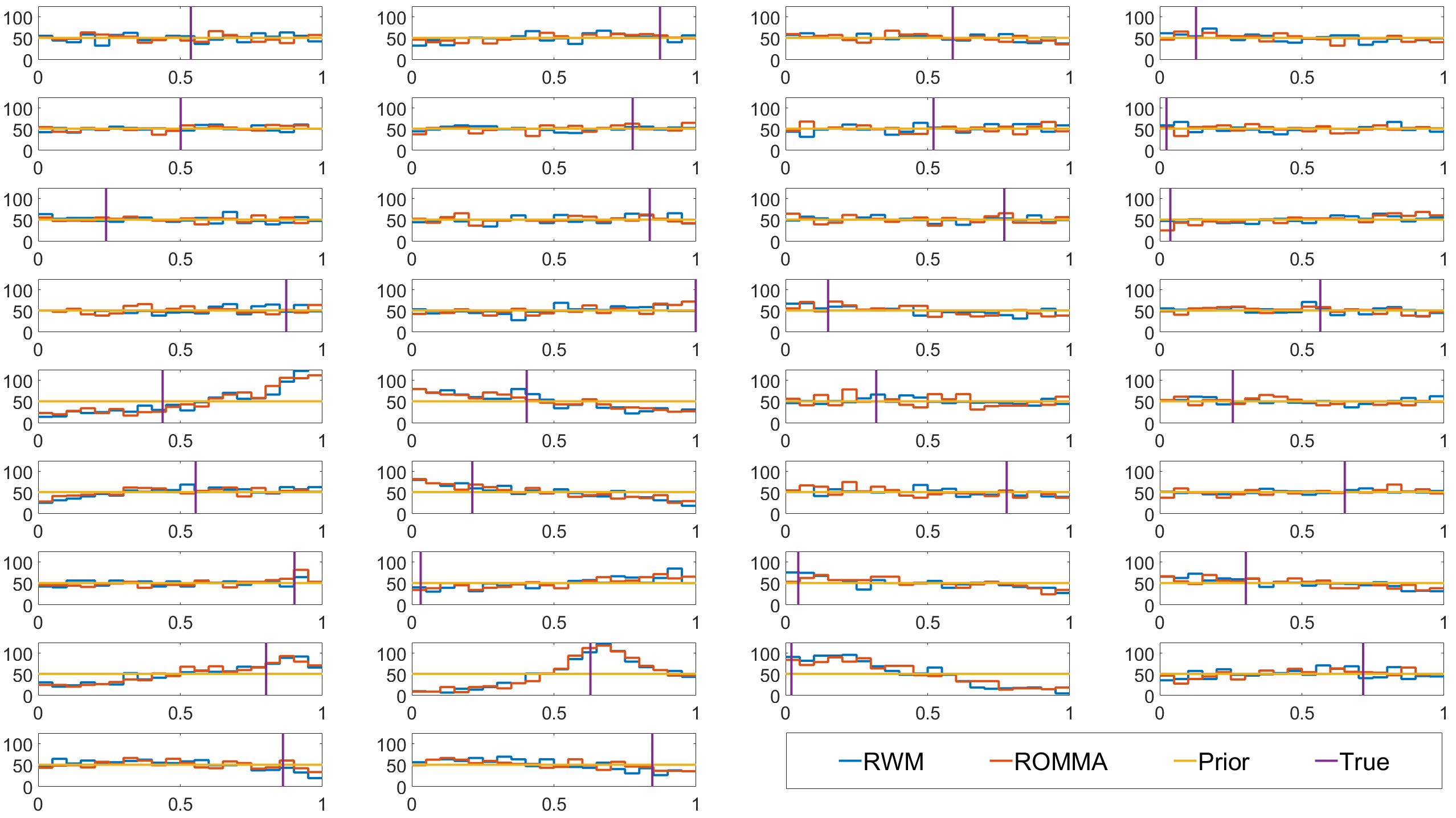}}
\caption{Case 1: Posterior marginal distributions for the leak position parameters computed using RWM and ROMMA. They are compared to the prior marginal and the true value.}
\label{fig:c1_post_pos_marg}
\end{figure}
\vfill
\pagebreak
\hfill
\subsection{Case 1: Posterior Failure Correlation Plots}
\hfill
\begin{figure}[!h]
\centerline{\includegraphics[width=1.0\textwidth]{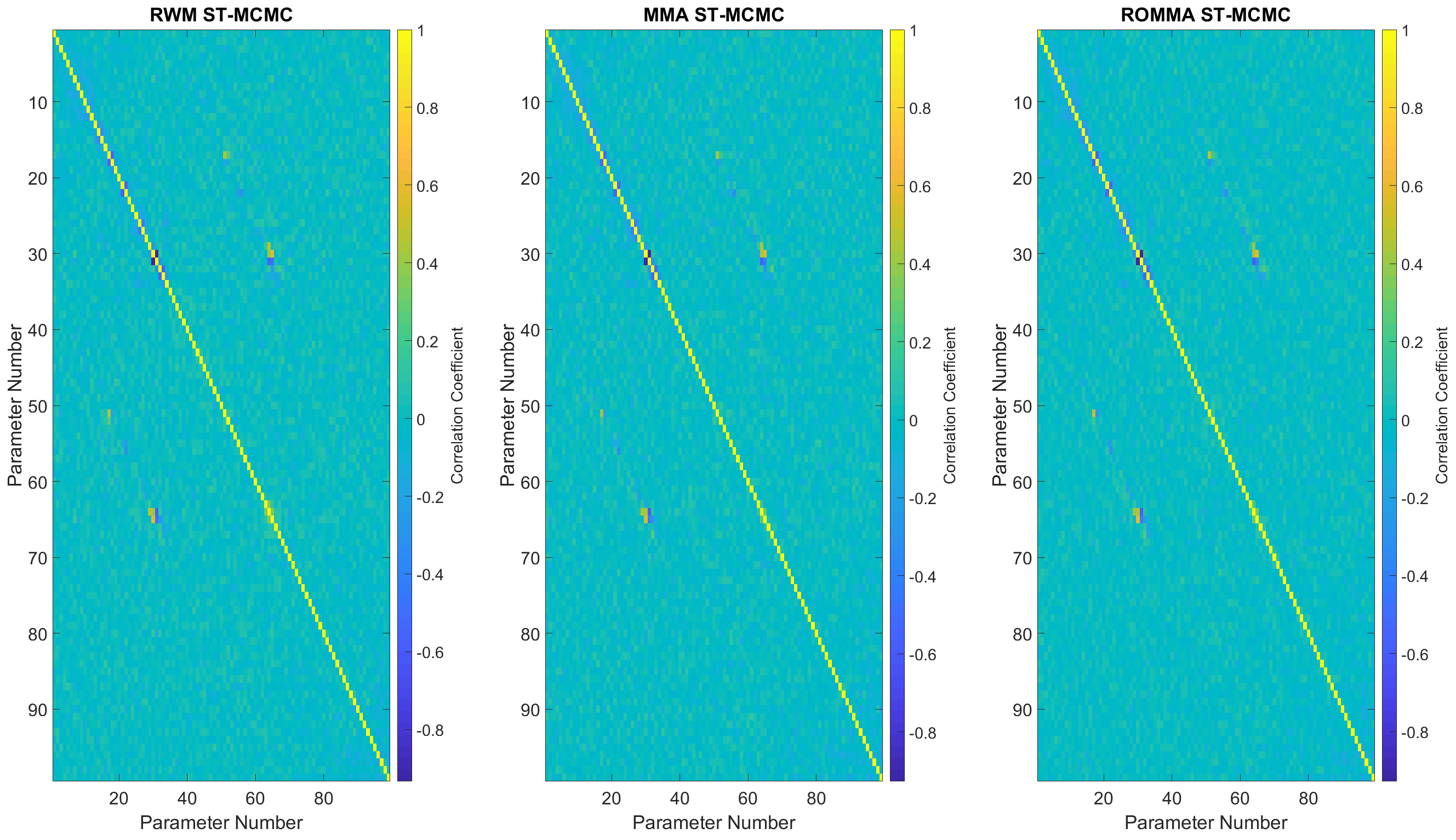}}
\caption{Case 1: Posterior failure correlation comparison between RWM, MMA, and ROMMA}
\label{fig:c1_fail_corr_comp}
\end{figure}
\vfill
\pagebreak
\hfill
\subsection{Case 1: Posterior Failure Marginal Distributions Plots}
\hfill
\begin{figure}[!h]
\centerline{\includegraphics[width=1.0\textwidth]{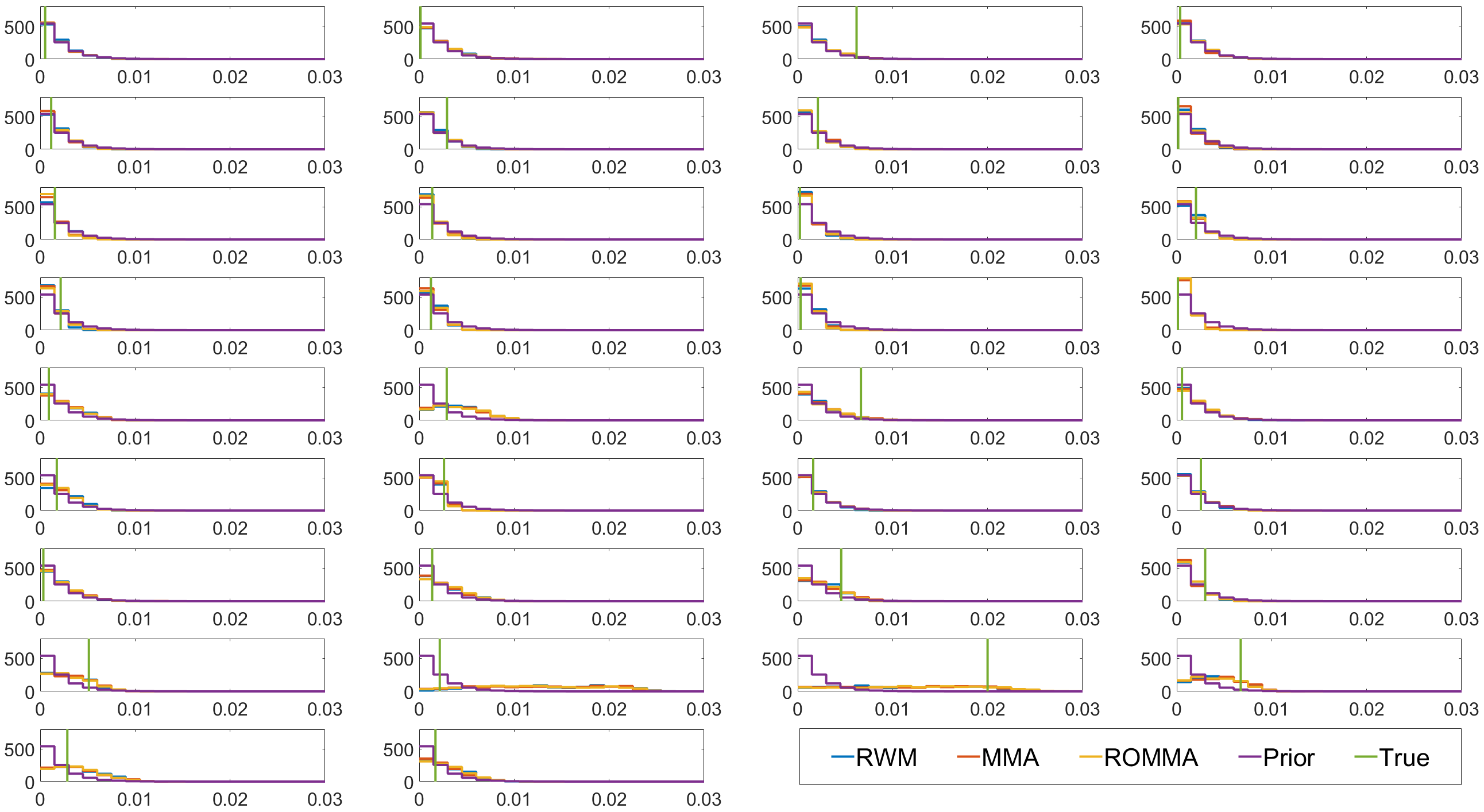}}
\caption{Case 1: Posterior failure marginal distributions for the leak size parameters computed using RWM, MMA, and ROMMA. They are compared to the prior marginal and the true value.}
\label{fig:c1_fail_size_marg}
\end{figure}

\vfill
\pagebreak
\hfill

\begin{figure}[!h]
\centerline{\includegraphics[width=1.0\textwidth]{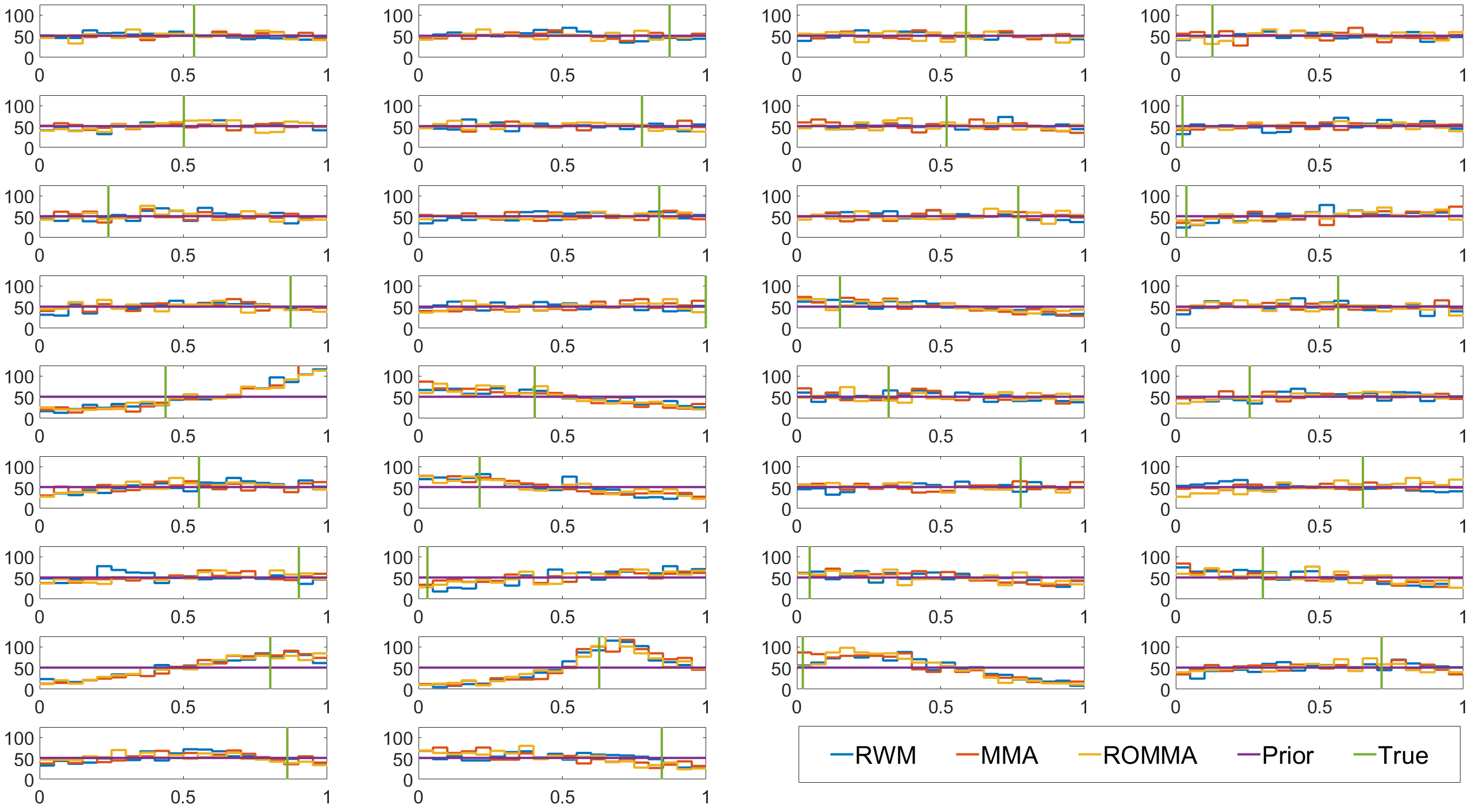}}
\caption{Case 1: Posterior failure marginal distributions for the leak position parameters computed using RWM, MMA, and ROMMA. They are compared to the prior marginal and the true value.}
\label{fig:c1_fail_pos_marg}
\end{figure}

\vfill
\pagebreak
\hfill

\begin{figure}[!h]
\centerline{\includegraphics[width=1.0\textwidth]{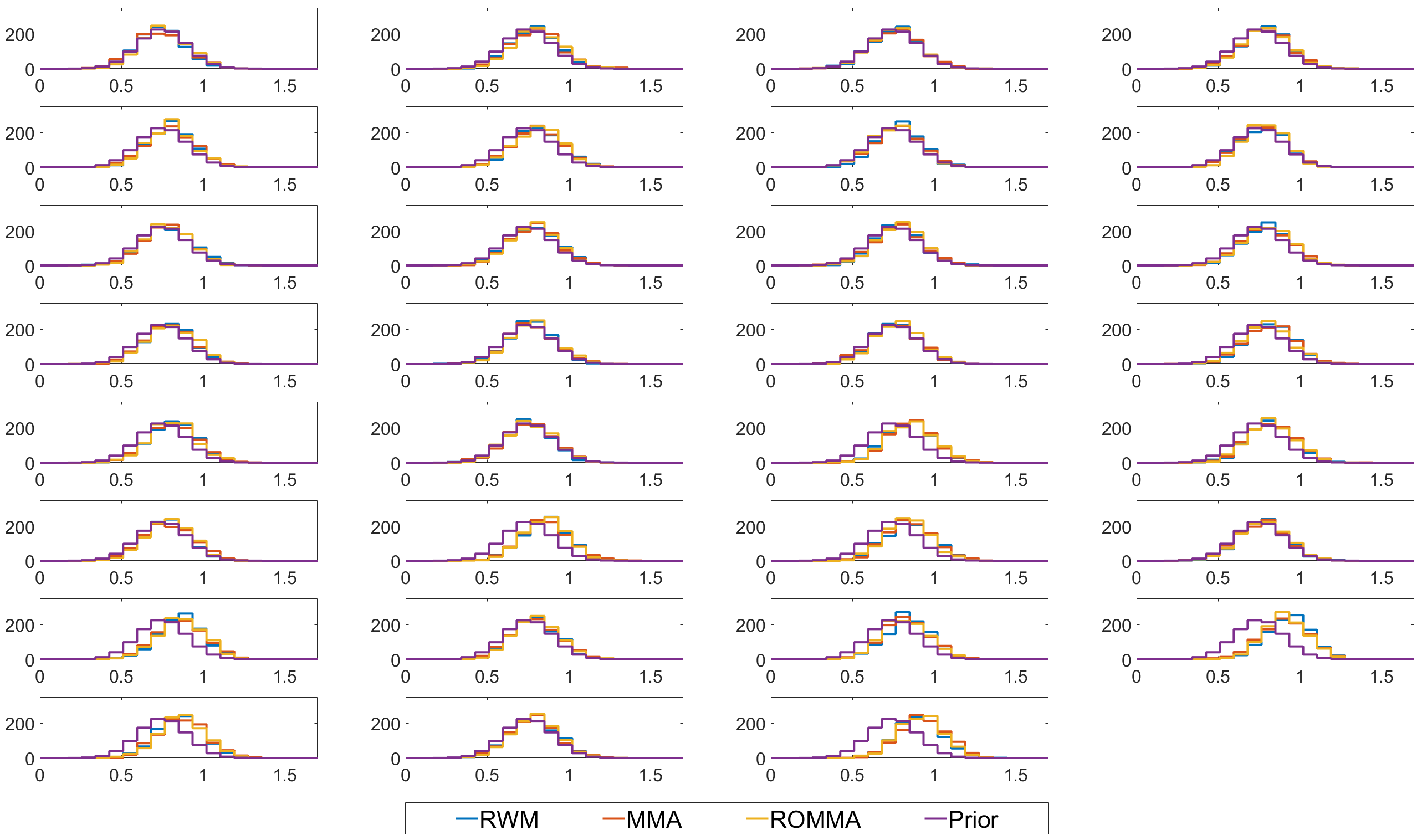}}
\caption{Case 1: Posterior failure marginal distributions for the nodal demand parameters computed using RWM, MMA, and ROMMA. They are compared to the prior marginal.}
\label{fig:c1_fail_demand_marg}
\end{figure}

\vfill
\pagebreak
\hfill
\subsection{Case 2: Posterior Leak Correlation Plots}
\hfill
\begin{figure}[!h]
\centerline{\includegraphics[width=1.0\textwidth]{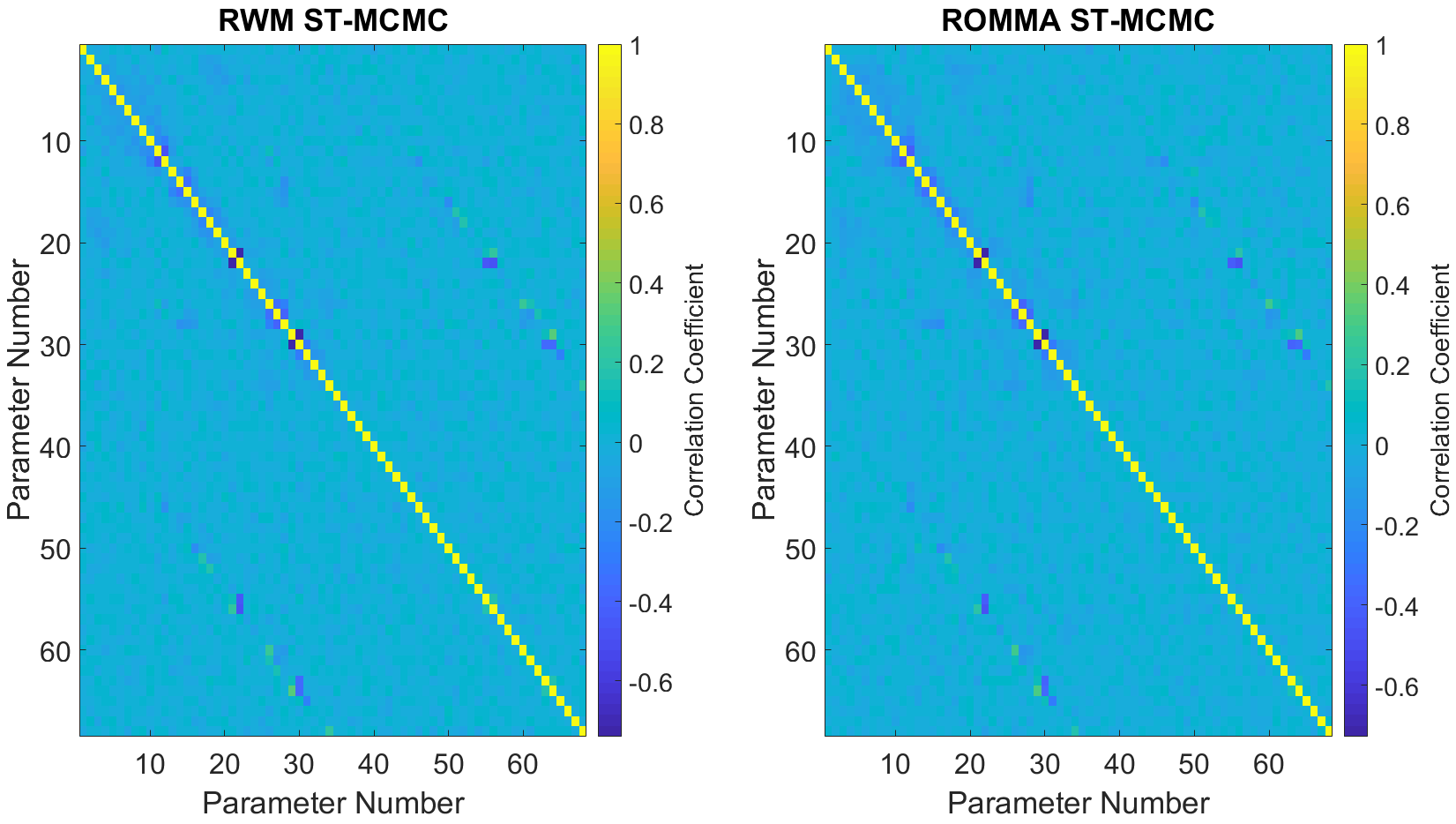}}
\caption{Case 2: Posterior correlation comparison between RWM and ROMMA}
\label{fig:post_corr_comp}
\end{figure}
\vfill
\pagebreak
\hfill
\subsection{Case 2: Posterior Marginal Distributions Plots}
\hfill
\begin{figure}[!h]
\centerline{\includegraphics[width=1.0\textwidth]{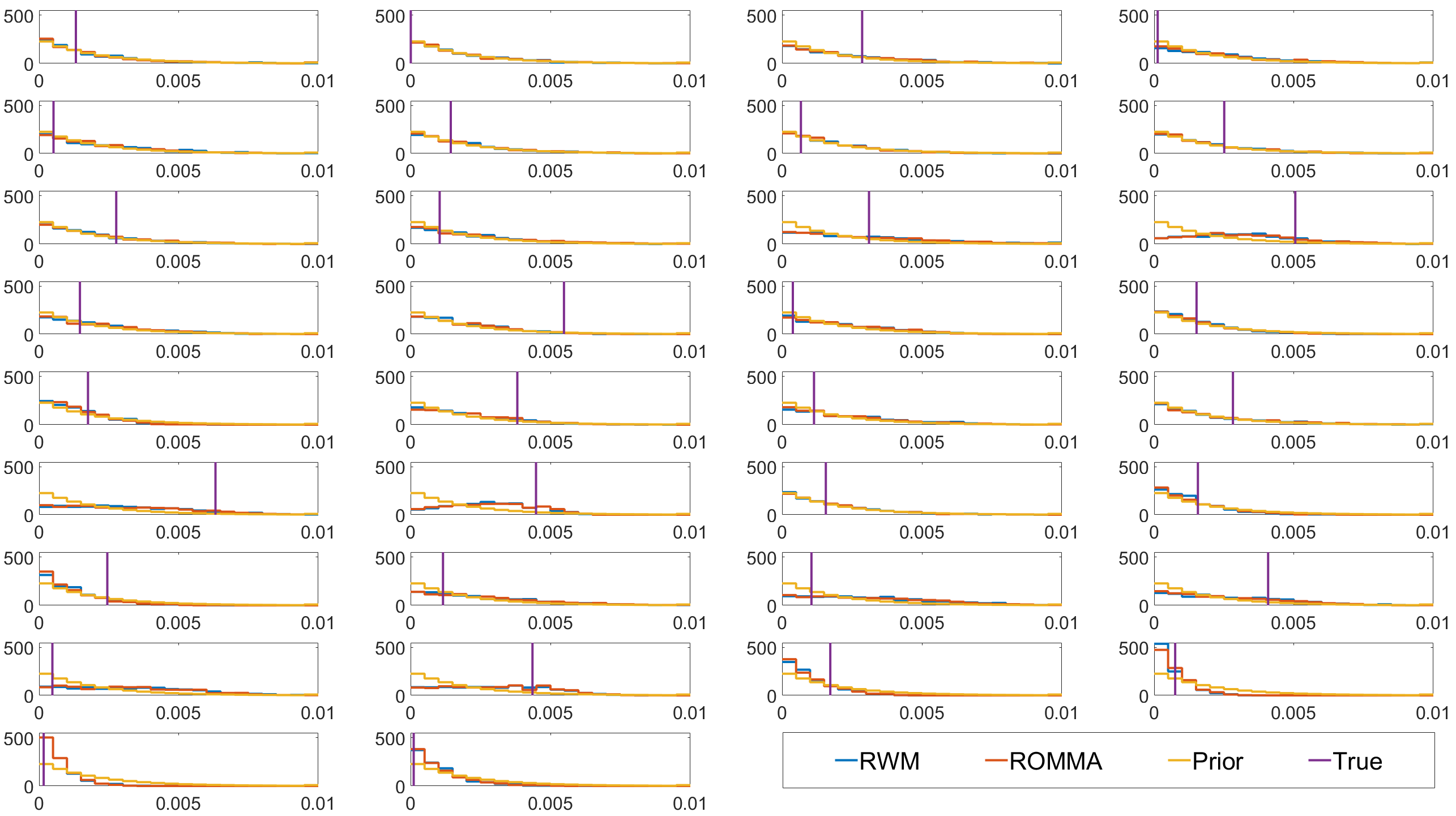}}
\caption{Case 2: Posterior marginal distributions for the leak size parameters computed using RWM and ROMMA. They are compared to the prior marginal and the true value.}
\label{fig:post_size_marg}
\end{figure}
\vfill
\pagebreak
\hfill
\begin{figure}[!h]
\centerline{\includegraphics[width=1.0\textwidth]{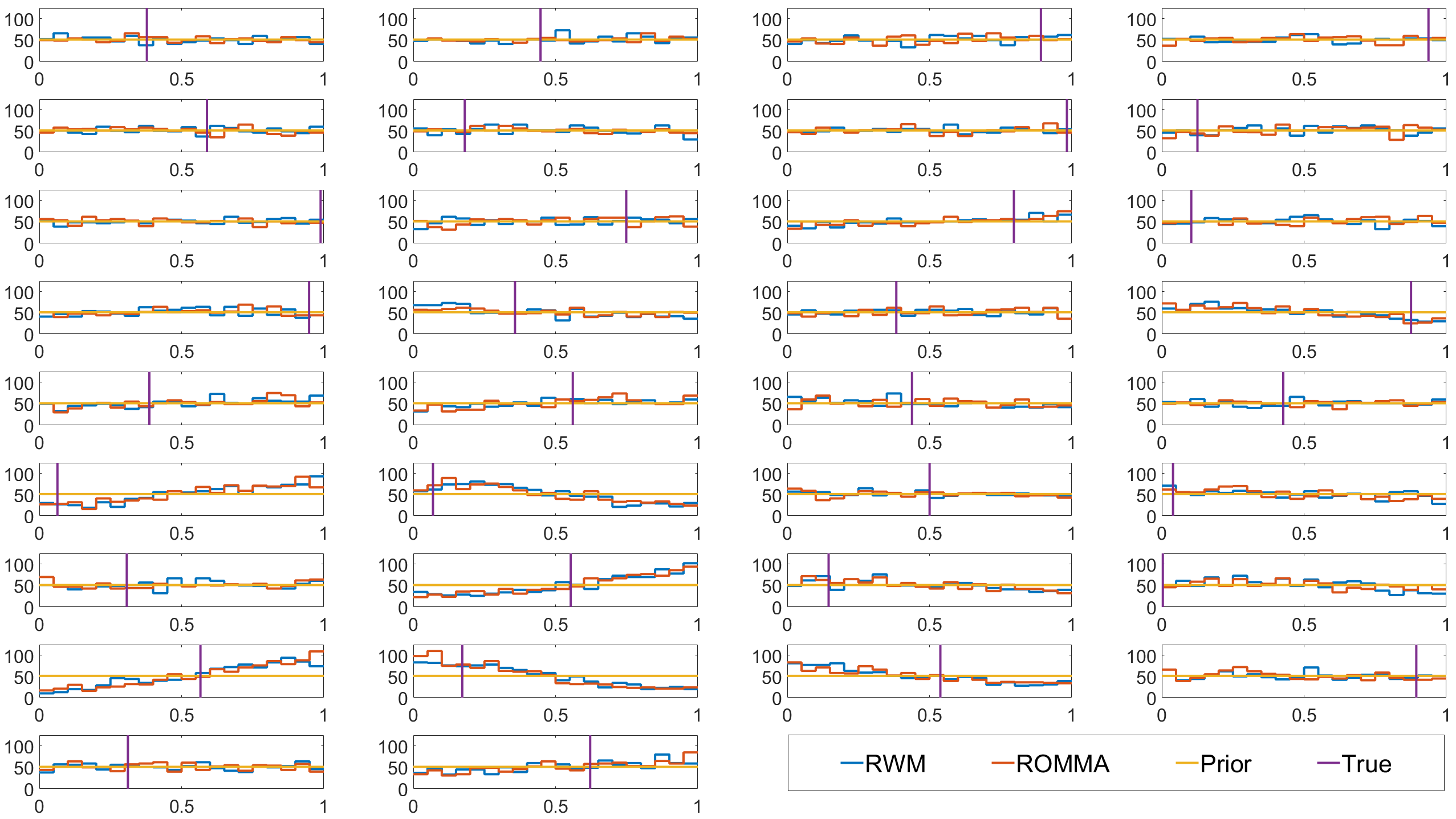}}
\caption{Case 2: Posterior marginal distributions for the leak position parameters computed using RWM and ROMMA. They are compared to the prior marginal and the true value.}
\label{fig:post_pos_marg}
\end{figure}
\vfill
\pagebreak
\hfill
\subsection{Case 2: Posterior Failure Correlation Plots}
\hfill
\begin{figure}[!h]
\centerline{\includegraphics[width=1.0\textwidth]{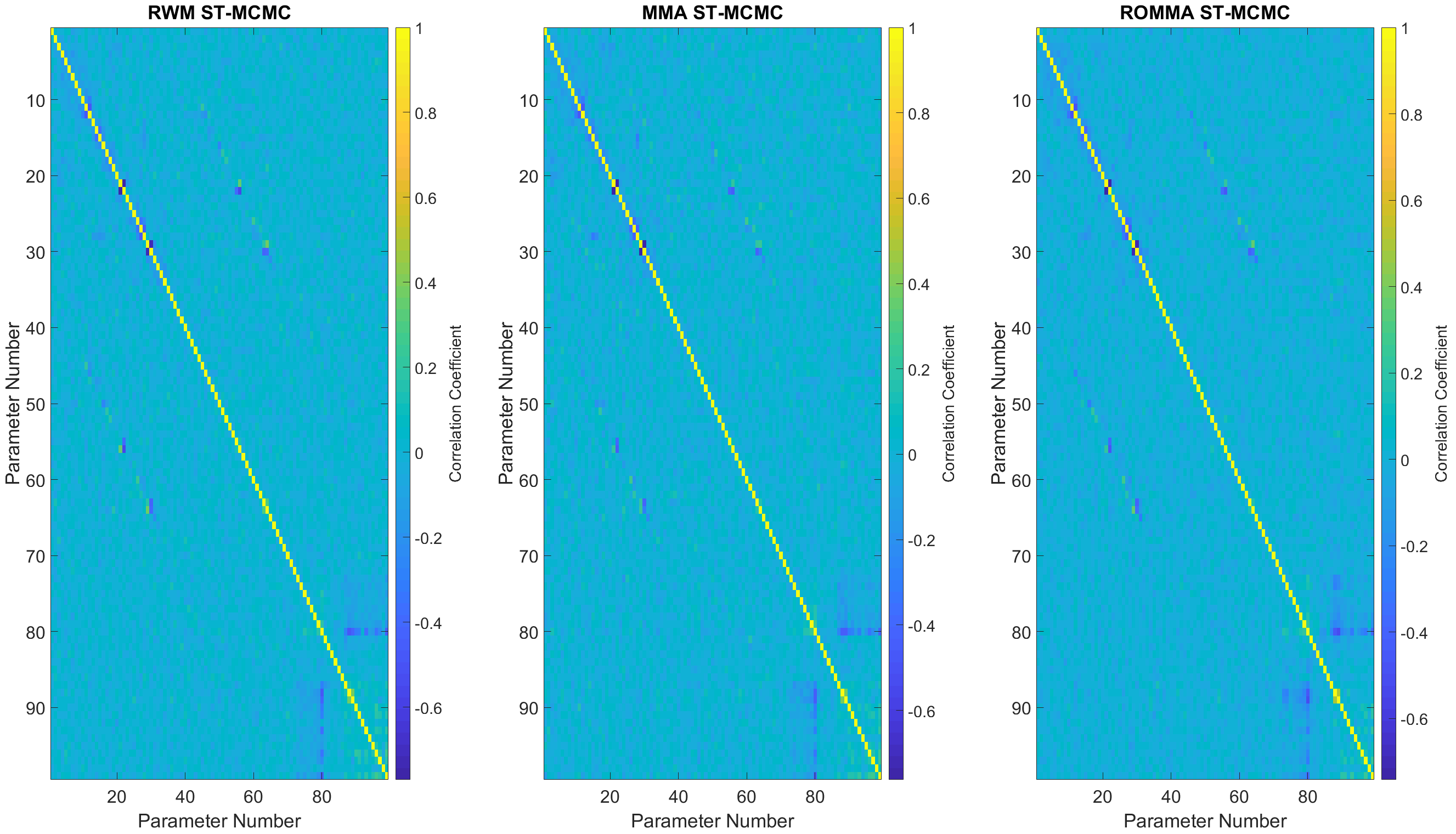}}
\caption{Case 2: Posterior failure correlation comparison between RWM, MMA, and ROMMA}
\label{fig:fail_corr_comp}
\end{figure}
\vfill
\pagebreak
\hfill
\subsection{Case 2: Posterior Failure Marginal Distributions Plots}
\hfill
\begin{figure}[!h]
\centerline{\includegraphics[width=1.0\textwidth]{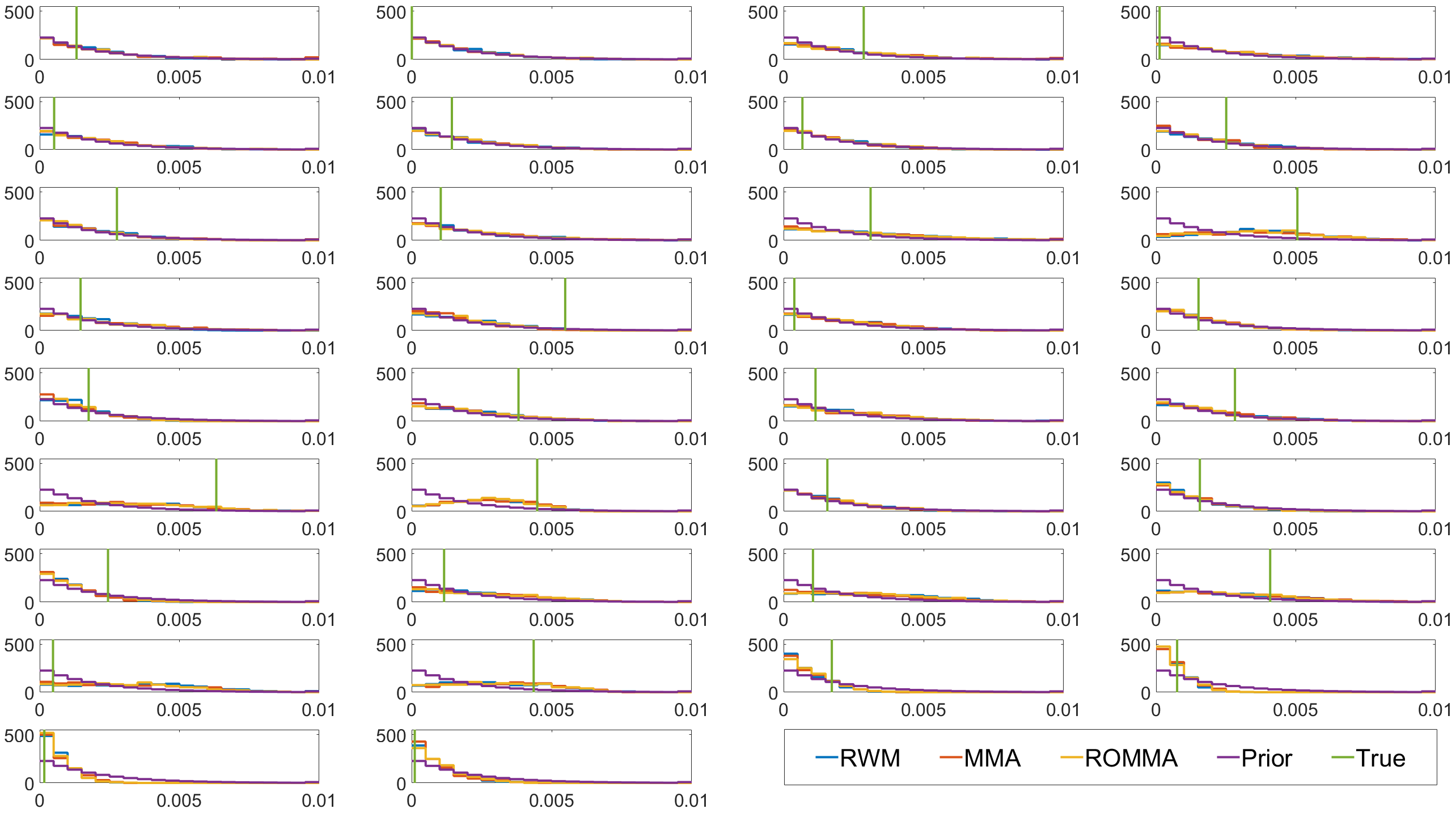}}
\caption{Case 2: Posterior failure marginal distributions for the leak size parameters computed using RWM, MMA, and ROMMA. They are compared to the prior marginal and the true value.}
\label{fig:fail_size_marg}
\end{figure}
\vfill
\pagebreak
\hfill
\begin{figure}[!h]
\centerline{\includegraphics[width=1.0\textwidth]{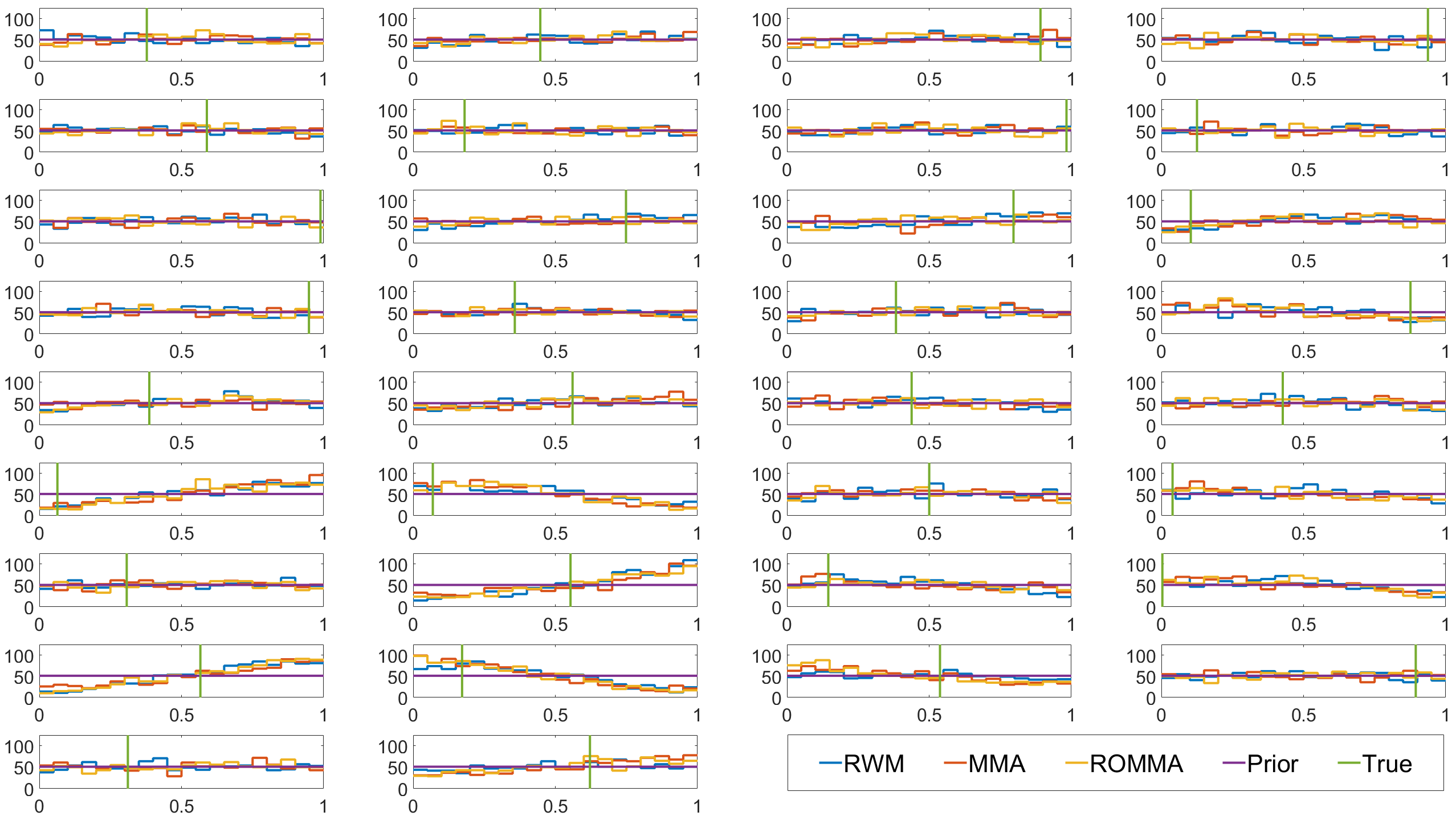}}
\caption{Case 2: Posterior failure marginal distributions for the leak position parameters computed using RWM, MMA, and ROMMA. They are compared to the prior marginal and the true value.}
\label{fig:fail_pos_marg}
\end{figure}
\vfill
\pagebreak
\hfill
\begin{figure}[!h]
\centerline{\includegraphics[width=1.0\textwidth]{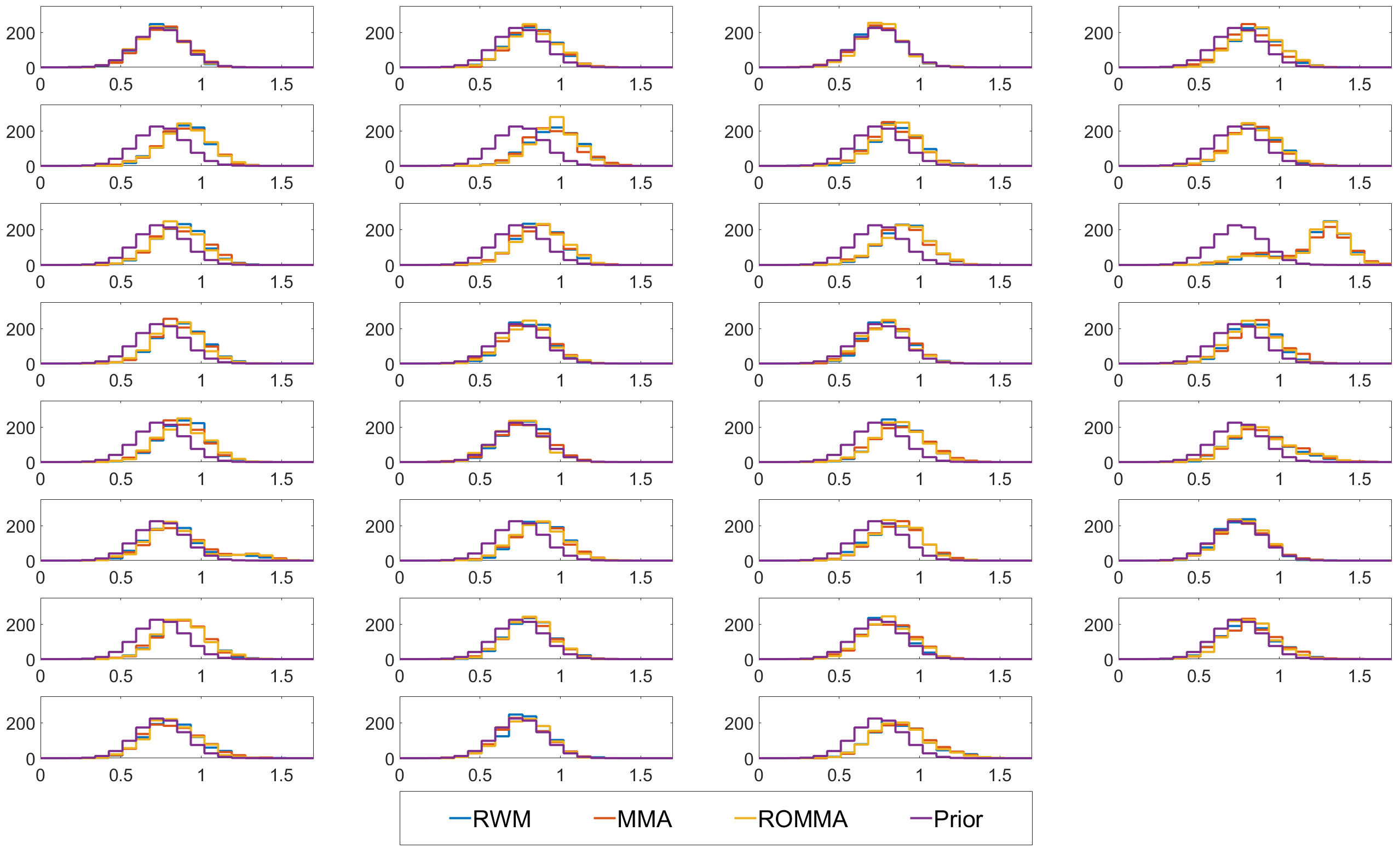}}
\caption{Case 2: Posterior failure marginal distributions for the nodal demand parameters computed using RWM, MMA, and ROMMA. They are compared to the prior marginal.}
\label{fig:fail_demand_marg}
\end{figure}
\vfill
\pagebreak
\hfill
\subsection{Case 2: Illustration of the Tri-modality of the posterior failure region}
\hfill
\begin{figure}[!h]
\centerline{\includegraphics[width=1.0\textwidth]{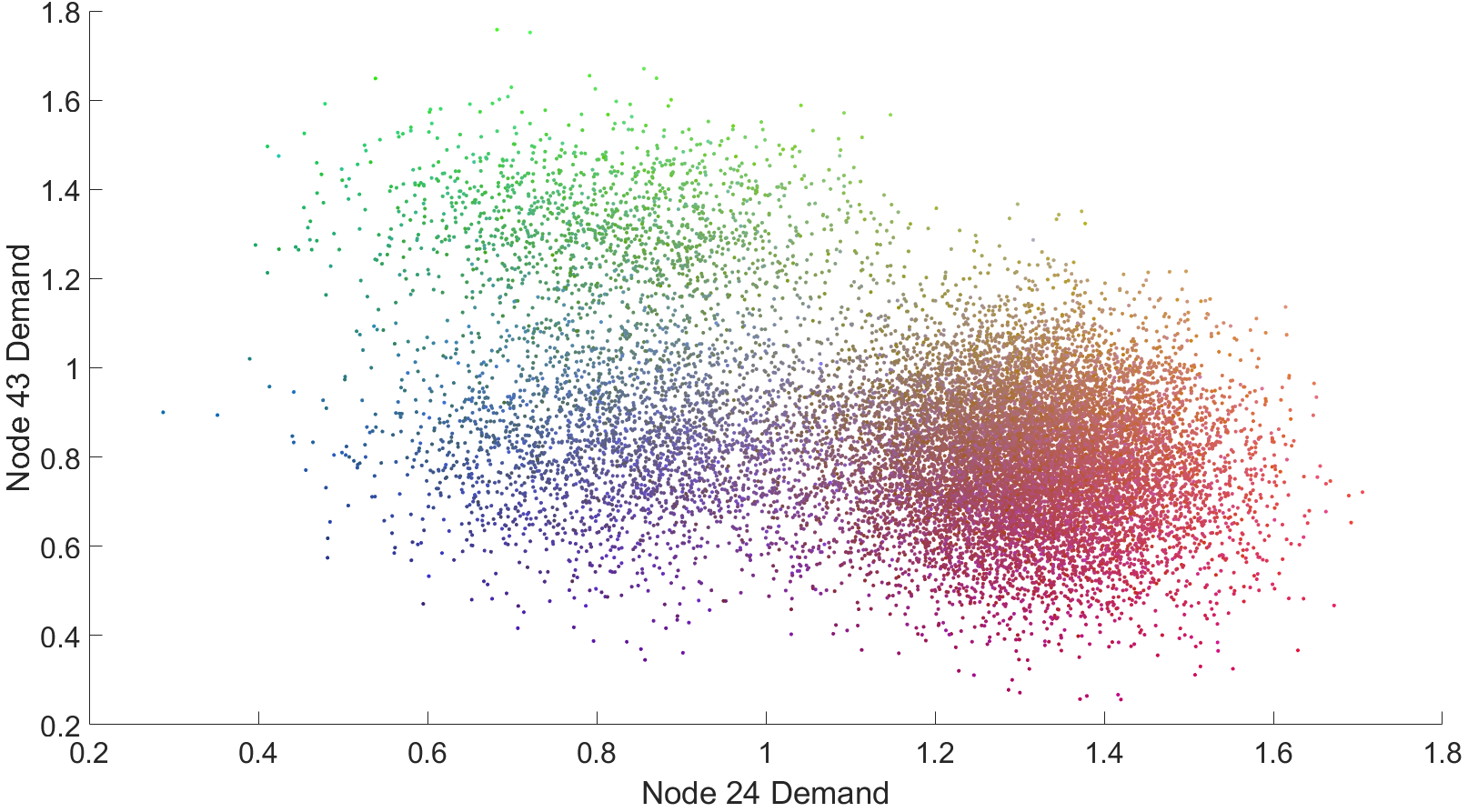}}
\caption{Case 2: Illustration of the tri-modality of the posterior failure region by plotting random points from the failure domain. The more red the point the more demand on Node 24 leads to failure. The more green the point the more demand on Node 43 leads to failure. Finally, the more blue the point the more demand on Node 64 leads to failure.}
\label{fig:fail_trimodal}
\end{figure}

\end{document}